\documentclass[trackchanges,twocolumn]{aastex701}

\newcommand{\cha}{\textit{Chandra\/}}
\def\xmm{{XMM-{\it Newton\/}}}

\def\swi{{{\it Swift}-BAT}\/}
\def\xrt{{{\it Swift}-XRT}\/}

\def\nustar{{\it NuSTAR}}

\def\myt{\texttt{MyTorus}}
\def\borus{\texttt{Borus02}}
\def\ux{\texttt{UXCLUMPY}}

\def\aap{A\&A}

\def\apjl{ApJL}
\def\apjs{ApJS}

\usepackage[latin1]{inputenc}
\usepackage[T1]{fontenc}
\usepackage[bf,small,tableposition=top]{caption}
\usepackage{subfig}
\usepackage{dirtytalk}
\usepackage{amsmath}
\usepackage{breqn}
\usepackage{placeins}
\usepackage{bm}
\usepackage{natbib}
\usepackage{longtable}
\usepackage{booktabs}
\usepackage{float}
\usepackage{hyperref}

\begin{document}

\title{Unveiling Obscured Accretion in the Local Universe}

\author[orcid=0000-0002-7825-1526,sname='Pal']{I. Pal}
\affiliation{Department of Physics and Astronomy, Clemson University, Kinard Lab of Physics, Clemson, SC 29634, USA}
\email[show]{ipal@clemson.edu}

\author[orcid=0000-0001-5544-0749]{S. Marchesi}
\affiliation{Dipartimento di Fisica e Astronomia (DIFA), Universit\`a di Bologna, via Gobetti 93/2, I-40129 Bologna, Italy}
\affiliation{Department of Physics and Astronomy, Clemson University, Kinard Lab of Physics, Clemson, SC 29634, USA}
\affiliation{INAF - Osservatorio di Astrofisica e Scienza dello Spazio di Bologna, Via Piero Gobetti, 93/3, 40129, Bologna, Italy}
\email[show]{smarche@clemson.edu}

\author[orcid=0000-0001-6564-0517]{R. Silver}
\affiliation{NASA Goddard Space Flight Center, Greenbelt, MD 20771, USA}
\email{ipal@clemson.edu}

\author[orcid=0000-0002-6584-1703]{M. Ajello}
\affiliation{Department of Physics and Astronomy, Clemson University, Kinard Lab of Physics, Clemson, SC 29634, USA}
\email{ipal@clemson.edu}

\author[orcid=0000-0002-9719-8740]{V. E. Gianolli}
\affiliation{Department of Physics and Astronomy, Clemson University, Kinard Lab of Physics, Clemson, SC 29634, USA}
\email{ipal@clemson.edu}

\author[orcid=0000-0003-3638-8943]{N. Torres-Alb\`a}
\affiliation{Department of Astronomy, University of Virginia, P.O. Box 400325, Charlottesville, VA 22904, USA}
\email{ipal@clemson.edu}

\author[orcid=0000-0003-2287-0325]{I. Cox}
\affiliation{Department of Physics and Astronomy, Clemson University, Kinard Lab of Physics, Clemson, SC 29634, USA}
\email{ipal@clemson.edu}

\author[orcid=0000-0002-7791-3671]{X. Zhao}
\affiliation{Cahill Center for Astrophysics, California Institute of Technology, 1216 East California Boulevard, Pasadena, CA 91125, USA}
\email{ipal@clemson.edu}

\author[orcid=0009-0002-6991-1534]{D. Sengupta}
\affiliation{Center for Space Science and Technology, University of Maryland Baltimore County, 1000 Hilltop Circle, Baltimore, MD 21250, USA}
\email{ipal@clemson.edu}

\author[orcid=0000-0001-7796-8907]{A. Banerjee}
\affiliation{Department of Physics and Astronomy, Clemson University, Kinard Lab of Physics, Clemson, SC 29634, USA}
\email{ipal@clemson.edu}

\author[orcid=0009-0003-3381-211X]{K. Imam}
\affiliation{Department of Physics and Astronomy, Clemson University, Kinard Lab of Physics, Clemson, SC 29634, USA}
\email{ipal@clemson.edu}

\author[orcid=0000-0001-6412-2312]{A. Pizzetti}
\affiliation{European Southern Observatory, Alonso de C\`ordova 3107, Casilla 19, Santiago,  19001, Chile}
\email{ipal@clemson.edu}


\begin{abstract}
Heavily obscured Active Galactic Nuclei (AGN), especially Compton-thick sources with line-of-sight column density ($N_{\rm H,los}$) $>$ 10$^{24}$ cm$^{-2}$, are critical to understanding supermassive black hole (SMBH) growth and the origin of the Cosmic X-ray Background (CXB). However, their observed fraction remains significantly below model predictions, due to strong absorption bias, even in the hard X-ray (i.e., above 10 keV) band. We analyze a sample of 26 nearby ($z < 0.1$) AGN from the \swi\ 150-month catalog, selected via mid-IR to X-ray diagnostics and observed with \nustar\ and soft X-ray telescopes (\xmm, \cha, or \xrt). Using self-consistent torus models (\myt, \borus, and \ux), we aim to constrain $N_{\rm H,los}$, the average torus column density, and other geometrical parameters of the obscuring medium. A comparative analysis among the three torus models showed that while estimates of $N_{\rm{H,los}}$ were generally in agreement, \borus\ tended to classify a slightly larger number of sources as Compton-thick AGN (CT-AGN). Building on this comparison, we benchmark two prediction schemes -- a mid-IR/X-ray relation and a machine-learning model -- against our broadband best-fit $N_{\rm H,los}$ measurements to assess which approach more effectively bridges the gap between predicted and measured obscuration, finding that while the former works effectively in the heavily obscured region (log$\rm{N_H} \gtrsim$ 23.5 $\rm{cm^{-2}}$), the latter provides improved accuracy, particularly for Compton-thin to moderately thick regimes (log$\rm{N_H} \lesssim$ 23.5 $\rm{cm^{-2}}$). 
\end{abstract}




\section{Introduction} 
Active Galactic Nuclei (AGN) are among the most luminous persistent sources in the Universe, with their integrated emission dominating the Cosmic X-ray Background (CXB) from 1 to $\sim$ 200--300 keV \citep{2003AJ....126..539A, 2007A&A...463...79G, 2009ApJ...696..110T}. A substantial share of CXB is expected to arise from heavily obscured AGN, in particular Compton-thick (CT-) systems with the line-of-sight (los) column density ($N_{\rm H,los}) \gtrsim10^{24}\,\mathrm{cm^{-2}}$ \citep{2014ApJ...786..104U, 2015ApJ...802...89B}. While unobscured AGN dominate the CXB below $\sim$10\,keV and are largely resolved, only a small fraction of the emission near the CXB peak at $\sim$ 30\,keV has been directly attributed to discrete sources \citep{2008ApJ...689..666A}. Population-synthesis models therefore require a sizable, still-hidden Compton-thick population of order $\sim$ 30-50\% of all AGN to reproduce both the shape and normalization of the CXB \citep{2007A&A...463...79G, 2019ApJ...871..240A}. In contrast, observational estimates in the local Universe find markedly lower CT fractions, typically $\sim$ 5-35\% (e.g., \citealp{2011ApJ...728...58B, 2015ApJ...815L..13R, 2021ApJ...922..252T, 2025ApJ...978..118B}), underscoring a persistent gap between models and data.

The shortfall is chiefly a selection problem: extreme photoelectric absorption and Compton scattering suppress the nuclear signal from the optical through the hard X-ray bands, making CT-AGN intrinsically difficult to find even with high-energy instruments \citep{2007A&A...463...79G,2008ApJ...689..666A,2019ApJ...871..240A}. The obscurer -- often dubbed the "dusty torus" -- likely sits on parsec scales \citep{2004Natur.429...47J,2008ApJ...685..160N} and is clumpy and anisotropic rather than smooth \citep{2006ApJ...648L.101E,2007ApJ...659L.111R}. In such geometries, modest changes in viewing angle or cloud statistics can shift sources across the Compton-thick boundary, diluting flux-limited samples and biasing traditional hardness-ratio or soft-X-ray selections. In this context, closing the observed-versus-predicted gap is thus necessary and it demands strategies that are less biased against obscuration. Accurately identifying and characterizing these obscured AGN is essential for tracing the full accretion history of SMBH and understanding the co-evolution of black holes and their host galaxies \citep{2015Natur.519..436T, 2017ApJ...837..149G}.

To address the observational challenges of identifying and characterizing obscured AGN, various methods have been proposed for predicting column densities ($N_{\rm H}$) using indirect indicators. Traditional approaches rely on empirical correlations--such as mid-infrared (MIR) to X-ray flux ratios--to identify obscured sources. For example, \citet{2015MNRAS.454..766A} introduced a diagnostic based on WISE 12~$\mu$m to 2-10 keV flux ratios, which is effective for sources with $\log N_{\rm H} \gtrsim 23$, but less so for less obscured AGN. Improvements have been proposed using luminosity ratios \citep{2022ApJS..261....3P} and hard X-ray spectral curvature \citep{2016ApJ...825...85K}, yet these methods still suffer from limitations in accuracy and contamination. More recently, \citet{2023AandA...675A..65S} developed a machine learning (ML) model using multiple observational features, achieving high correlation with spectroscopically measured $N_{\rm H}$ and robust classification across obscuration regimes.

Despite these advancements, there has been limited systematic testing of predictive methods against detailed spectral modeling for well-defined AGN samples. Moreover, the physical interpretation of torus parameters such as covering fraction ($c_f$), $N_{\rm H,los}$, and the average torus column density ($N_{\rm H,avg}$) remains uncertain, particularly in terms of how they relate to one another and to source properties like luminosity and accretion rate.

This work presents a focused investigation aimed at exploring some of these open questions using a well-defined subset of nearby, obscured AGN. The sample is selected using the IR--to-X-ray flux criteria of \citet{2015MNRAS.454..766A}, and is observed with the Nuclear Spectroscopic Telescope Array (\nustar; \citealt{2013ApJ...770..103H}), which provides high-sensitivity coverage in the 3-79 keV band. This enables robust measurement of the Compton hump and intrinsic continuum components, essential for accurate determination of $N_{\rm H,los}$ and other torus properties \citep{2015ApJ...805...41B, 2019ApJ...872....8M}.

We evaluate the performance of IR--X-ray diagnostics and machine learning models in predicting $N_{\rm H}$ by comparing their outputs to values derived from high-quality X-ray spectral fitting. Additionally, we investigate possible correlations between torus parameters, including the covering fraction, $N_{\rm H,los}$, and $N_{\rm H,avg}$, to explore the structure and geometry of the obscuring material. We also test the relationship between the covering fraction and key AGN properties such as intrinsic 2--10 keV luminosity and Eddington ratio.

The manuscript is arranged as follows: the selection criteria and construction of the source sample are described in Section~\ref{sec2}. Section~\ref{sec3} outlines the data reduction process, while Section~\ref{sec4} details the spectral modeling procedures. The results of the spectral fitting analysis are presented in Section~\ref{sec5}. In Section~\ref{sec6}, we compare the best-fit parameters obtained from different models, assess the agreement between predicted and measured $N_{\rm H}$ and examine correlations between torus properties. Finally, our findings and conclusions are summarized in Section~\ref{sec7}. A description of the fit results and the tabulated fit parameters are outlined in Appendix \ref{source description} and Appendix \ref{lumin:table}.

\section{Sample Selection}
\label{sec2}
\begin{figure}
\centering
     \includegraphics[scale=0.44]{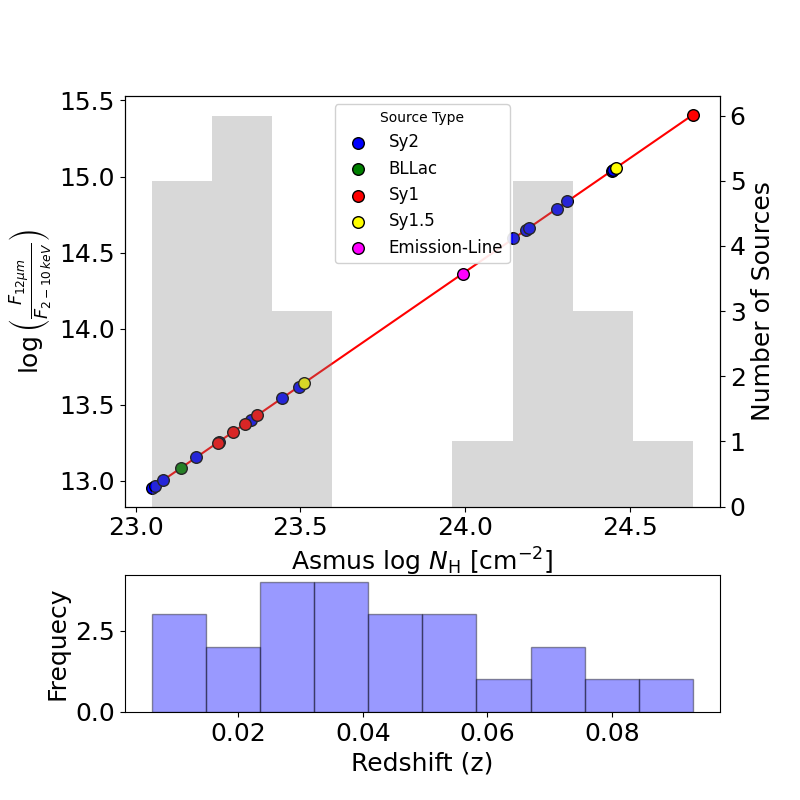}
\caption{Upper panel: Distribution of the 26 sources with predicted line-of-sight column densities ($N_{\mathrm{H,los}}$) based on the empirical relation from \citet{2015MNRAS.454..766A}. The histogram
in the background shows the number of sources in each $N_{\mathrm{H}}$ bin. The scatter points represent the logarithmic ratio of mid-infrared (12\,$\mu$m) to observed 2--10\,keV X-ray flux plotted against the predicted $N_{\mathrm{H}}$ (x-axis), with different AGN types indicated by color. The red line shows the linear regression fit, illustrating the correlation between the flux ratio and the Asmus-predicted column density. Lower panel: Distribution of source redshifts.}
\label{figure-1}
\end{figure}

Our study focuses on a sample of 26 AGN at redshift $z < 0.1$, selected from the 150-month \swi\ catalog (Imam et al., submitted). These sources were chosen to represent obscured AGN candidates based on a set of well-defined criteria. First, each target is predicted to be significantly obscured, with line-of-sight column densities $N_{\rm H,los} >  10^{23}\ \mathrm{cm}^{-2}$, as inferred from the MIR to X-ray flux correlation established by \citet{2015MNRAS.454..766A}. The details of the selected sources including their positions, type and predicted $N_{\rm H,los}$ values from \citet{2015MNRAS.454..766A} are given in Table \ref{table-1b}. Second, all selected sources have at least one archival observation with \textit{NuSTAR}, providing high-quality coverage in the 3-50~keV band. The log of observations are given in Table \ref{table-1}.
The redshift distribution and the distribution of the selected sources with the predicted $N_{\mathrm{H,los}}$ values against the mid-IR to X-ray flux ratio as obtained from \cite{2015MNRAS.454..766A} are plotted in Fig. \ref{figure-1}. 

To complement the \textit{NuSTAR} data and ensure broad energy coverage, we also associate each target with archival 0.5-10~keV observations from soft X-ray telescopes. When available, we prioritize data from \xmm, due to its superior effective area. If \xmm\ data are unavailable, we include observations from \textit{Chandra}, and when both \xmm, and \textit{Chandra} are unavailable, we incorporate data from \textit{Swift}-XRT. Therefore, the dataset includes archival observations from several X-ray observatories spanning a few years, and thus, the spectra are generally non-simultaneous. To ensure consistency, we verified that the individual spectra have consistent shapes (see Section \ref{spectral fit 25}, Fig. \ref{figure-38}) and show only modest variability; we therefore fit them together with a standard set of inter-instrument cross-normalization constants, fixing one instrument to unity and allowing the others to vary (see Section \ref{sec4}).
This multi-instrument approach allows us to construct high-quality, broadband (0.5-50~keV) spectra for each source, enabling a robust and self-consistent modeling of the obscuring material and its interaction with the AGN emission. We note that a systematic investigation of possible variations in $N_{\rm H,los}$ across different observations is not carried out in this work, as it lies beyond the scope of the present study; moreover, the available broad-band (0.5-50 keV) datasets are non-simultaneous for the majority of sources.

\begin{deluxetable*}{cccccccccc}
\caption{Properties of the 26 sources analyzed in this work. The columns are (1) serial number of the source, (2) name of the source, (3) right ascension (h:m:s), (4) declination (d:m:s), (5) redshift, (6) optical/near infrared classification of the source, (7) predicted $N_{\rm H,los}$ from \citet{2015MNRAS.454..766A}, (8) predicted $N_{\rm H,los}$ from \citealt{2023AandA...675A..65S} (9) black hole mass, and (10) Eddington ratio. Some of the information, including the right ascension, declination, $z$ and type of the sources, are from SIMBAD\footnote{\hyperref[https://simbad.cds.unistra.fr/simbad/sim-fbasic]{https://simbad.cds.unistra.fr/simbad/sim-fbasic}}. Details regarding the black hole mass references can be found in Section \ref{source description}.} \label{table-1b} 
\tablehead{\textbf{Index} & \textbf{Name} & \textbf{RA} & \textbf{DEC} & \textbf{z} & \textbf{Type} & \textbf{Asmus log $N_{\rm H,los}$} & \textbf{ML log $N_{\rm H,los}$} & $\bm{\log\left(\frac{M_{BH}}{M_{\odot}}\right)}$ & \textbf{log} {$\bm{\lambda_{\rm{Edd}}}$}} 
\startdata
&& (h:m:s) & (d:m:s) & & & ($\rm{cm^{-2}}$) & ($\rm{cm^{-2}}$) & & \\
\hline
1 & MRK 1073 & 03 15 01.43 & +42 02 08.82 & 0.023 & Sy2 & 24.450 & 23.012 & 7.78 & $-$1.63 \\
2 & UGC 5101 & 09 35 51.60 & +61 21 11.59 & 0.039 & Sy1.5 & 24.459 & 23.209 & 8.35 & $-$2.00 \\
3 & NGC 7674 & 23 27 56.70 & +08 46 44.25 & 0.029 & Sy2 & 24.311 & 23.390 & 7.73 & $-$1.90 \\
4 & IC 2227 & 08 07 07.19 & +36 14 00.46 & 0.032 & Sy2 & 24.445 & 23.597 & --- & --- \\
5 & ESO 362$-$8 & 05 11 09.09 & -34 23 36.74 & 0.016 & Sy2 & 24.193 & 23.332 & --- & --- \\
6 & ESO 406-4 & 22 42 33.37 & -37 11 07.57 & 0.029 & Sy2 & 24.279 & 23.360 & 8.02 & $-$2.49 \\
7 & 2MFGC 13496 & 16 51 05.68 & -01 27 48.23 & 0.041 & Emission Line & 23.993 & 23.641 & --- & --- \\
8 & 2MASX J03585442+1026033 & 03 58 54.44 & +10 26 02.79 & 0.031 & Sy2 & 23.058 & 23.246 & 7.75 & $-$1.20 \\
9 & M 58 & 12 37 43.59 & +11 49 05.12 & 0.005 & Sy2 & -- & -- & 8.10 & $-$3.44 \\
10 & 3C 371 & 18 06 50.68 & +69 49 28.11 & 0.049 & BL LAC & 23.139 & 22.101 & --- & --- \\
11 & IC 1198 & 16 08 36.38 & +12 19 51.60 & 0.034 & Sy1 & 23.296 & 23.048 & 7.51 & $-$0.41 \\
12 & 2MASX J09261742-8421330 & 09 26 17.43 & -84 21 33.09 & 0.064 & Sy2 & 23.331 & 22.381 & 7.10 & $-$0.06 \\
13 & UGC 12348 & 23 05 18.83 & +00 11 22.15 & 0.025 & Sy2 & 23.349 & 22.711 & --- & --- \\
14 & 2MASX J02420381+0510061 & 02 42 03.82 & +05 10 06.18 & 0.073 & Sy2 & 23.445 & 22.390 & --- & --- \\
15 & ESO 234-50 & 20 35 57.87 & -50 11 32.17 & 0.009 & Sy2 & 23.495 & 23.211 & 6.04 & $-$1.47 \\
16 & NGC 2273 & 06 50 08.67 & +60 50 44.86 & 0.006 & Sy2 & 24.184 & 23.828 & 8.22 & $-$2.68 \\
17 & FRL 265 & 06 56 29.79 & -65 33 37.73 & 0.029 & Sy1 & -- & --- & --- & --- \\
18 & MRK 231 & 12 56 14.23 & +56 52 25.24 & 0.042 & Sy1 & 24.692 & 24.343 & 7.94 & $-$1.41 \\
19 & PG 1211+143 & 12 14 17.67 & +14 03 13.18 & 0.081 & Sy1 & 23.249 & 21.986 & 8.60 & $-$1.82 \\
20 & MRK 376 & 07 14 15.08 & +45 41 55.90 & 0.056 & Sy1.5 & 23.511 & 22.480 & 8.22 & $-$0.46 \\
21 & NGC 7378 & 22 47 47.69 & -11 48 59.86 & 0.009 & Sy2 & 23.184 & 22.853 & 5.49 & $-$0.08 \\
22 & 2MASX J11462959+7421289 & 11 46 29.54 & +74 21 29.04 & 0.058 & Sy2 & 23.049 & 21.891 & 8.22 & $-$1.80 \\
23 & SWIFT J2006.5+5619 & 20 06 33.32 & +56 20 36.40 & 0.044 & Sy2 & 23.083 & 23.062 & 7.07 & $-$0.62 \\
24 & 2MASX J06363227$-$2034532 & 06 36 32.25 & -20 34 53.18 & 0.056 & Sy2 & 24.147 & 23.340 & 8.44 & $-$1.65 \\
25 & 2MASX J09034285$-$7414170 & 09 03 42.89 & -74 14 17.42 & 0.093 & Sy2 & 23.252 & 23.074 & 7.88 & $-$1.88 \\
26 & 2MASX J00091156$-$0036551 & 00 09 11.60 & -00 36 54.78 & 0.073 & Sy2 & 23.369 & 23.708 & 8.54 & $-$1.58 \\
\hline
\enddata
\end{deluxetable*}

\section{Data Reduction}
\label{sec3}
The log of all the observations analyzed in this work is provided in Table \ref{table-1}. The data reduction process is detailed in the following subsections.

\subsection{{\it NuSTAR}} 
We reduced the {\it NuSTAR} data in the 3$-$79 keV band using the standard {\it NuSTAR} data reduction software NuSTARDAS\footnote{https://heasarc.gsfc.nasa.gov/docs/nustar/analysis/nustar swguide.pdf} v0.4.12 distributed by HEASARC within HEASoft v6.34.  The calibrated, cleaned, and screened event files were generated by running the {\tt nupipeline} task using the CALDB release 20240405. To extract the source counts from both modules (FPMA and FPMB), we chose circular regions of radii between $40''$ and $100''$ centred on the source, depending on the observations, while requiring a minimum S/N of 4 per energy bin. Similarly, to extract the background counts, we selected a circular region of the same radius away from the source on the same chip to avoid contamination from source photons. We then used the  {\tt nuproducts} task to generate energy spectra, response matrix files (RMFs) and auxiliary response files (ARFs), for both the hard X-ray detectors housed inside the corresponding focal plane modules, FPMA and FPMB. The data from FPMA and FPMB are analyzed jointly, and they are not combined. 

\subsection{{\xmm}}
We carried out our analysis with the data from the pn and MOS cameras (when available) \citep{2001A&A...365L..18S}. We used SAS v1.3 for the data reduction \citep{2004ASPC..314..759G}. The event files were filtered to exclude background flares selected from time ranges where the 10-15 keV count rates in the PN camera exceeded 0.7 cts/s. The source spectra were extracted from a circular region with a radius between $30''$ and $50''$ centered on the nucleus. Background photons were selected from a source-free region of equal area on the same chip as the source. We checked for pileup using the {\tt EPATPLOT} task. We did not find any of the sources that suffered from pileup.  We constructed RMFs and ARFs using the tasks {\it RMFGEN} and {\it ARFGEN} for each observation. 

\subsection{{\it Chandra}}
The {\it Chandra} \citep{2000SPIE.4012....2W} data have been reduced using the CIAO 4.7 software and the {\it Chandra} Calibration Data Base (\texttt{caldb}) 4.6.9, adopting standard procedures; no source shows significant pile-up, as measured by the CIAO \textsc{pileup\_map} tool. We used the CIAO \texttt{specextract} tool to extract both the source and the background spectra. Source spectra have been extracted in circular regions of 4$^{\prime\prime}$, while background spectra have been extracted from annuli having inner radius $r_{\rm int}$=10$^{\prime\prime}$ and outer radius $r_{\rm out}$=25$^{\prime\prime}$: regions inside the background area have been visually inspected to avoid contamination from nearby sources.

\subsection{{\xrt}}
Eight sources in our sample (2MASX J09261742-8421330, MRK 376, NGC 7378, 2MASX J11462959+7421289, SWIFT J2006.5+5619, 2MASX J06363227-2034532, 2MASX J09034285-7414170, and 2MASX J00091156-0036551) have not been observed with either \xmm\ or \cha. For these sources, we used archival \xrt \citep{2005SSRv..120..165B} observations from the Neil Gehrels Swift Observatory \citep{2004ApJ...611.1005G} taken simultaneously / quasi-simultaneously with the \nustar observations.  In cases where multiple observations were obtained within a short time interval, the \xrt\ data were combined to produce a single 0.5--10 keV spectrum, following standard procedures of the UK Swift Science Data Centre \citep{2009MNRAS.397.1177E}. 

\section{Spectral fitting procedure}\label{sec4}
The spectral fitting procedure is performed using the XSPEC software \citep{1996ASPC..101...17A}; the Galactic absorption values are the one measured by \citet{2013MNRAS.431..394W}. We use the solar abundances from \cite{2000ApJ...542..914W}, and the \citet{1996ApJ...465..487V} photoelectric absorption cross-section. 

\begin{figure*}
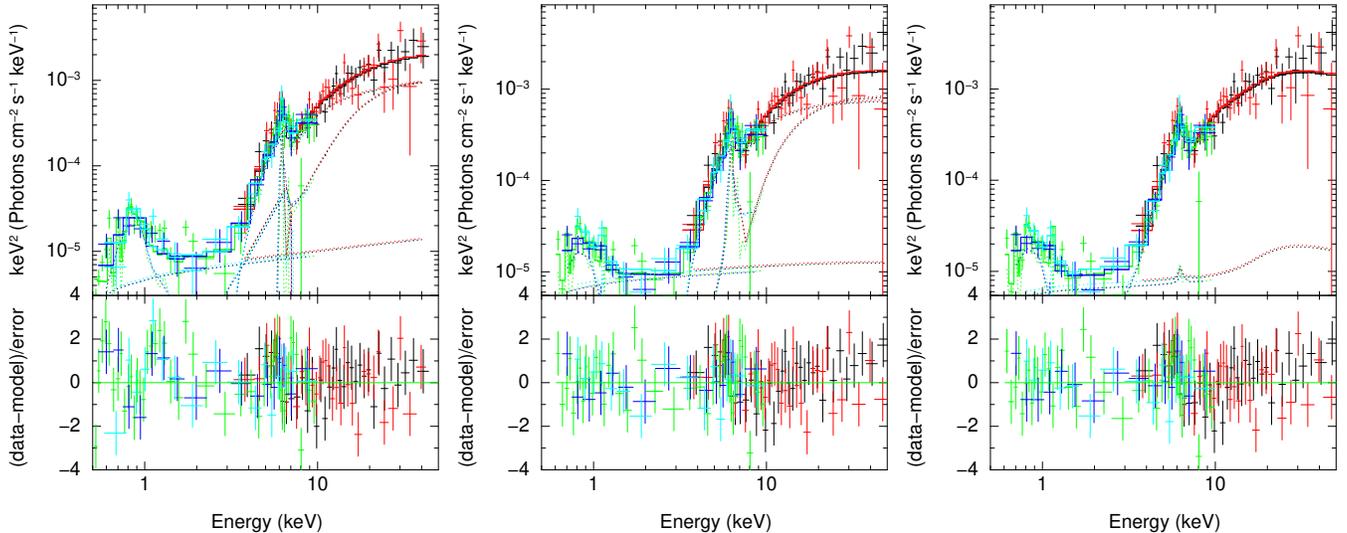

\hbox{
    \hspace{-0.7cm}
     \includegraphics[scale=0.29]{IC2227_my.eps}
     \hspace{-0.5cm}
     \includegraphics[scale=0.29]{IC2227_borus.eps}
     \hspace{-0.5cm}
     \includegraphics[scale=0.29]{IC2227_ux.eps}
     }
\caption{Unfolded spectra and model-to-data ratio for IC 2227 using \myt\ (left panel), \borus\ (middle panel), and \ux\ (right panel).}  \label{figure-A1}       
\end{figure*}

Heavily obscured AGN have complex spectra, where the contribution of the Compton scattering and of the fluorescent iron line becomes significant with respect to less obscured AGN spectra. Consequently, these sources should be treated in a self-consistent way, that allows one to properly measure the line-of-sight column densities, using models developed specifically for this purpose. To analyze these complex spectra in a more self-consistent way, we use three torus models: \myt\ \citep{2009MNRAS.397.1549M, 2012MNRAS.423.3360Y, 2015MNRAS.454..973Y}, \borus\ \citep{2018ApJ...854...42B}, and \ux\ \citep{Buchner2019}. Three best-fitted unfolded spectra with the data-to-model residuals are presented in Fig. \ref{figure-A1}, in which the \nustar, \xmm, and \cha\ observations of one of the sources (IC 2227) from our sample are jointly fitted using \myt, \borus, and \ux, respectively. The best-fit unfolded spectra, along with the corresponding data-to-model residuals for the remaining 25 sources, are shown in Appendix~\ref{spectral fit 25}, Fig.~\ref{figure-38}. All source spectra were binned with the \texttt{grppha} task to achieve a minimum of 20 counts per bin, ensuring the applicability of  $\chi^2$ statistics in the spectral analysis. For sources with only \xrt\ data, we instead adopted a minimum of 5 counts per bin and employed Cash statistics (C-stat) as the fit-statistics for spectral fitting. A brief description of each model used in this analysis is given below.

\subsection{\myt}\label{sec:myt}
The \myt\ \citep{2009MNRAS.397.1549M, 2012MNRAS.423.3360Y, 2015MNRAS.454..973Y} model is divided into three distinct components.

\begin{enumerate}
\item A multiplicative component ($\texttt{MYTZ}$) containing photoelectric absorption and Compton scattering attenuation. This component is applied to the main power law continuum. The viewing angle of $\texttt{MYTZ}$ is fixed to 90$^{\circ}$, so its $N_{H}$ corresponds to the los value.
\item A scattered continuum, also known as ``reprocessed component'' ($\texttt{MYTS}$). This component models those photons that are observed after one or more interactions with the material surrounding the SMBH. The normalization of the reprocessed component with respect to the main continuum is hereby denoted as A$_{\rm S}$. 
\item The neutral Fe fluorescent emission lines, more in detail the Fe K$\alpha$ line at 6.4\,keV and the K$\beta$ at 7.06\,keV ($\texttt{MYTL}$). We denote the normalization of these lines with respect to the main continuum as A$_{\rm L}$. 
\end{enumerate}

In \myt, the obscuring material surrounding the SMBH is assumed to have a toroidal, azimuthally symmetric shape. The torus covering factor, $c_f$, is not a free parameter and is fixed to  $c_f$=cos($\theta_{\rm OA}$)=0.5, where $\theta_{\rm OA}$=60$^\circ$ is the torus half-opening angle. The angle between the torus axis and the observer is free to vary, within the range $\theta_{\rm obs}$=[0--90]$^\circ$. In our analysis, we use \myt\ in the so-called ``decoupled mode'' \citep{2015MNRAS.454..973Y}: for the main continuum, we fix $\theta_{\rm obs}$=90$^\circ$, while for the reprocessed component we test two different scenarios, one with $\theta_{\rm obs, A_S, A_L}$ = 90$^\circ$ (near-side of the torus), the other with $\theta_{\rm obs, A_S, A_L}$ = 0$^\circ$ (far-side of the torus). The relative weights for the MYTS and MYTL components are tied together ($A_S = A_L$) and denoted as $A_{90}$ and $A_{0}$ for $\theta_{obs} = 90^{\circ}$ and $0^{\circ}$, respectively.
The \myt\ decoupled configuration allows for the separate measurement of $N_{\rm H,los}$, and the average torus column density, $N_{\rm H,avg}$, thus mimicking a clumpy torus distribution. In this configuration, while the column density associated with the transmitted component ($\texttt{MYTZ}$) is the line of sight one, the column densities associated with $\texttt{MYTS}$ and $\texttt{MYTL}$ are tied together to estimate $N_{\rm H,avg}$.

\subsection{\borus}\label{sec:borus}
\borus\ \citep{2018ApJ...854...42B} is an updated and improved version of the widely used \texttt{BNTorus} model \citep{2011MNRAS.413.1206B}. This radiative transfer code models the reprocessed emission component of an AGN X-ray spectrum, i.e., following the \myt\ nomenclature we introduced in the previous section, the ``reprocessed component'' and the neutral Fe emission lines.

In \borus, the obscuring material has a quasi--toroidal geometry, with conical polar cutouts. Both the average torus column density ($N_{\rm H,avg}$) and the torus covering factor are free parameters in the model: the torus covering factor value can vary in the range $c_f$=[0.1--1.0], corresponding to a torus opening angle range $\theta_{\rm OA}$=[84--0]$^\circ$.
In principle, the angle between the torus axis and the observer is a free parameter of this model, but in our analysis we fix it to $\theta_{\rm obs}$=60$^\circ$ and 30$^\circ$, a common assumption for the type 2 \citep{2011A&A...530A..42C, 2020ApJ...894...71Z} and type 1 \citep{2006A&A...445..451M, 2008MNRAS.386L..15D, 2010A&A...518A..47R} sources. 

Finally, since the \borus\ models itself does not take into account line-of-sight absorption, we follow the \citet{2018ApJ...854...42B} approach and derive  $N_{\rm H, los}$ in XSPEC using the components \texttt{zphabs $\times$ cabs}, to properly model Compton scattering losses out of the line of sight. In the overall fitting model, the $N_{\rm H, z}$ value is a free parameter, independent from $N_{\rm H,avg}$, and assumed to be identical in \texttt{zphabs} and \texttt{cabs}.

\begin{figure*}[htbp]
\hbox{
     \includegraphics[scale=0.47]{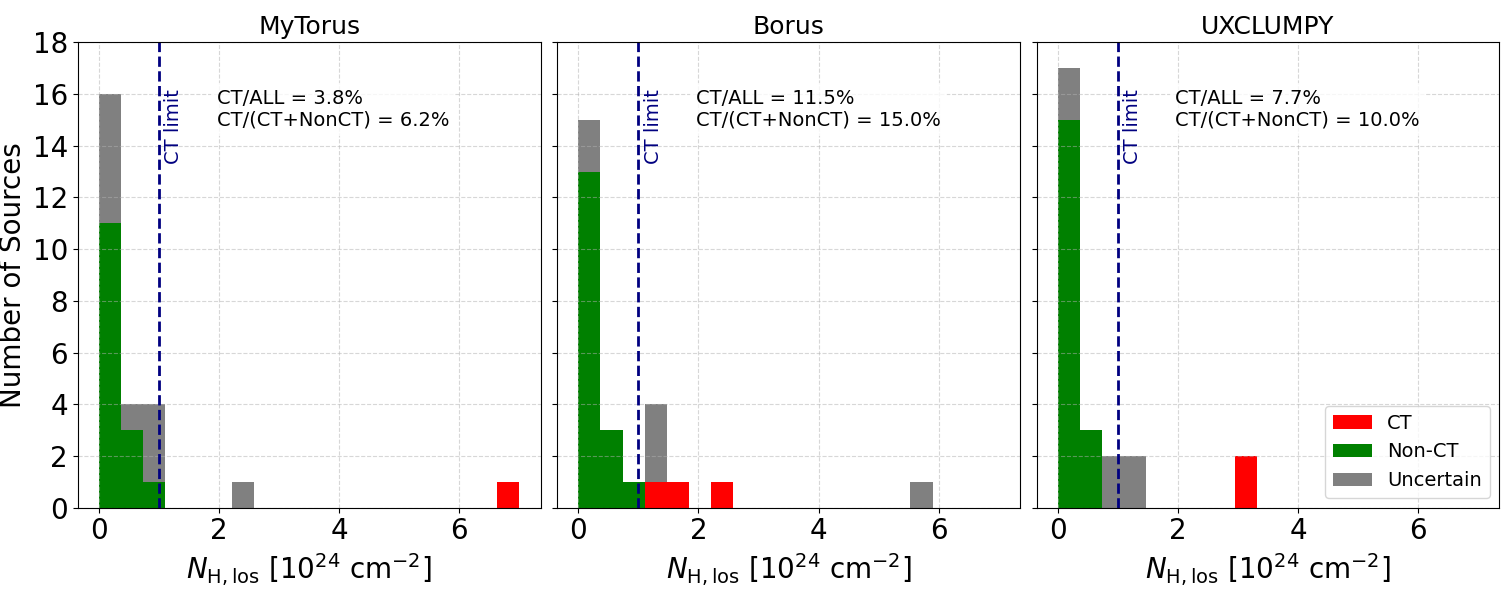}
     }
\caption{Fraction of CT-AGN (red bars) derived using \myt\ (left panel), \borus\ (middle panel), and \ux\ (right panel). In each panel, confirmed non-CT AGN are shown in green, while sources with uncertain values of $N_{\rm H,los}$ are indicated by gray bars. The blue dashed vertical line shows the CT limit.}\label{figure-4}
\end{figure*}

\subsubsection{\ux}\label{sec:uxclumpy}
Unlike \borus\ and \myt, \ux\ \citep{Buchner2019} does not assume a uniform torus. Instead, \ux\ is a physically motivated model that reproduces the X-ray data by simulating different cloud sizes and distributions. 

The first table accounts for the transmitted and reflection components, including fluorescent lines. \ux\ produces the reflection component through the cloud distribution it generates. However, for some sources that are reflection-dominated, a Compton-thick reflector near the corona can be added. This can be thought of as an inner wall that blocks the line of sight to the corona while also reflecting its emission. The second table reproduces the intrinsic continuum that leaks through the clumps of the torus.

\ux\ differs from \borus\ in that it does not include a parameter to measure the average torus column density. However, it measures other torus parameters such as the inclination angle (with a slightly larger range than \borus\; cos[$\theta_{obs}$] = 0 -- 1.00), the dispersion of the cloud distribution \texttt{TORsigma} (ranges from 6$\degr$ to 90$\degr$), and the covering factor of the inner reflector \texttt{CTKcover} (ranges from 0 to 0.6).

\subsection{Additional components to the best-fit model}
Besides using \myt, \borus, and \ux\ in the configurations described in the previous sections, we included the following components in our best-fit model:
\begin{enumerate}
\item A second power law, with photon index $\Gamma_2$=$\Gamma_1$, where $\Gamma_1$ is the photon index of the primary power law. This second power law is introduced to take into account the fraction ($f_{\rm s}$) of accreting SMBH emission which is scattered, rather than absorbed, by the gas surrounding the SMBH. We assume this component to be unabsorbed. 
\item A constant, $c_{instrument}$, allowing for a re-normalization of the \nustar\ FPMA spectrum with respect to the FPMB and the 0.5-10\,keV soft X-ray data from \xmm, \cha\ or \xrt. Such a component models both cross-calibration offsets between the instruments and potential flux variability between the different observations.
\item A \texttt{mekal} component, to account for the emission below 3\,keV caused by diffuse hot gas.
\end{enumerate}

The three torus models are implemented in \texttt{XSPEC} as follows:

\begin{dmath}\label{eq:myt}
    Model\ A = constant \times  phabs \times  [MYTZ_{90} \times zpowerlw + A_{90} \times (MYTS_{90} + MYTL_{90})
    + A_{0} \times (MYTS_{0} + MYTL_{0})  + f_s \times zpowerlw + mekal].
\end{dmath}

\begin{dmath}\label{eq:borus}
    Model\ B = constant  \times phabs \times  (borus02 + zphabs \times cabs \times zpowerlw 
    + f_s \times zpowerlaw + mekal).
\end{dmath}

\begin{dmath}\label{eq:ux}
    Model\ C = constant \times phabs \times (uxcl\_cutoff.fits +  f_s \times uxcl\_cutoff\_omni.fits + mekal).
\end{dmath}

\begin{figure*}
\hbox{
\includegraphics[scale=0.45]{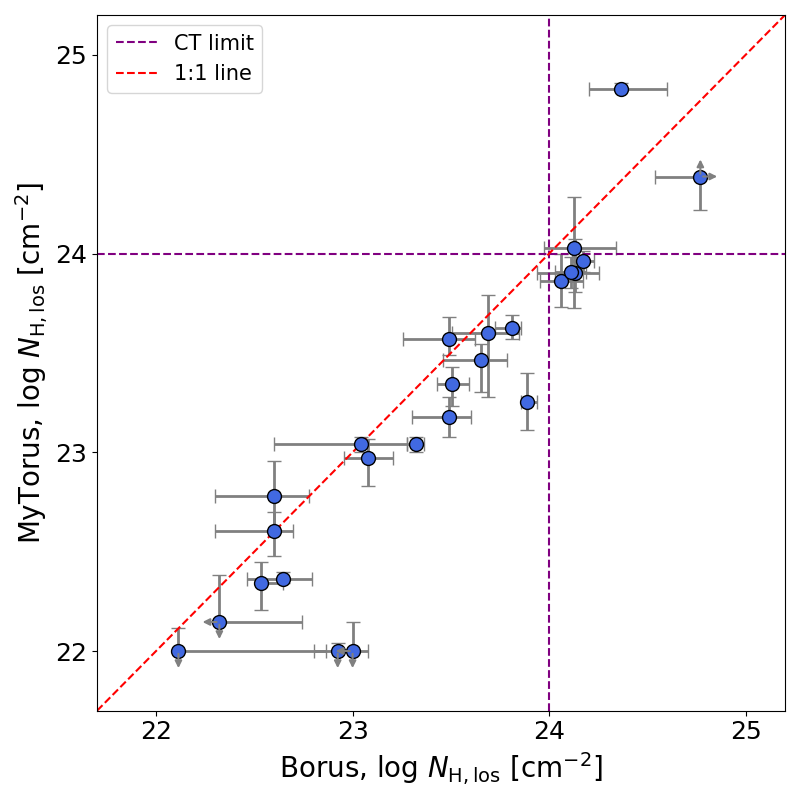}
\includegraphics[scale=0.45]{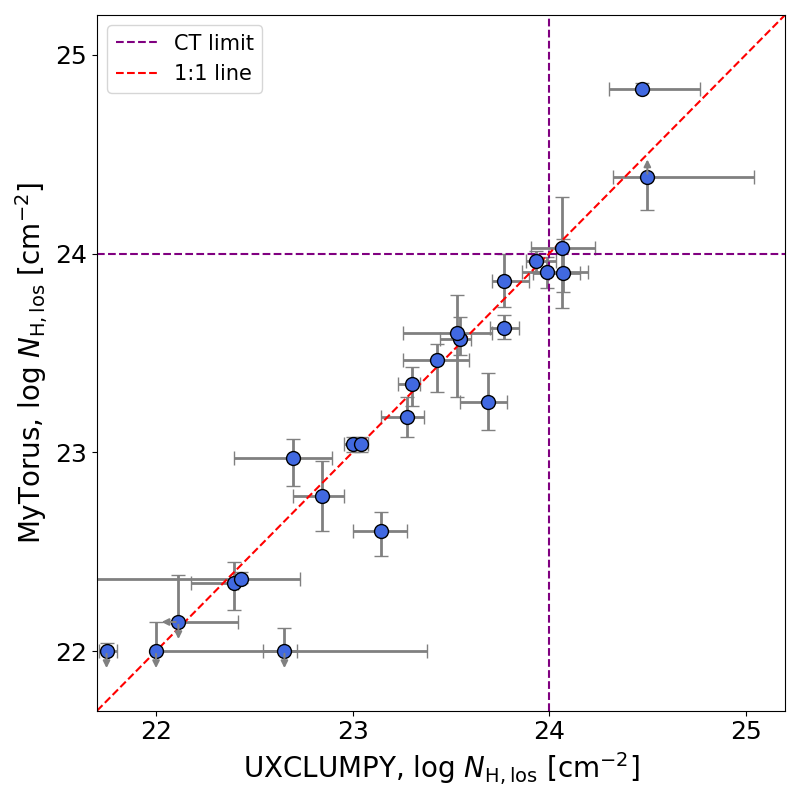}
}
\caption{Comparison between the line-of-sight column densities obtained from \myt, \borus, and \ux. The red dashed line marks the 1:1 relation between the estimated $N_{\rm H,los}$ from different torus models, while the purple dashed lines indicate the Compton-thick boundary.}
\label{figure-2}
\end{figure*}

\begin{figure*}
\hbox{
     \includegraphics[scale=0.45]{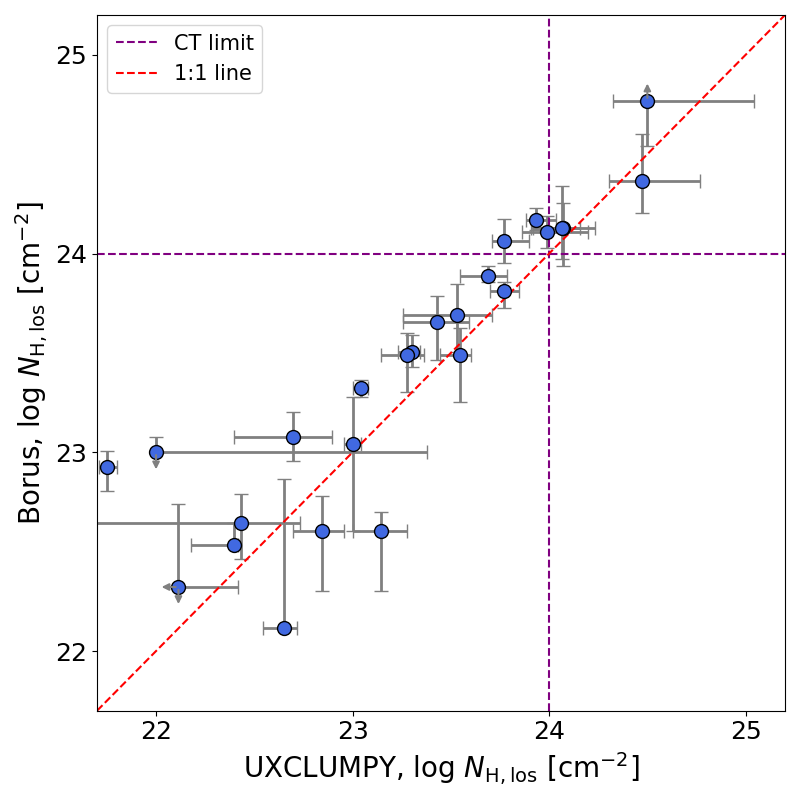}
     \includegraphics[scale=0.45]{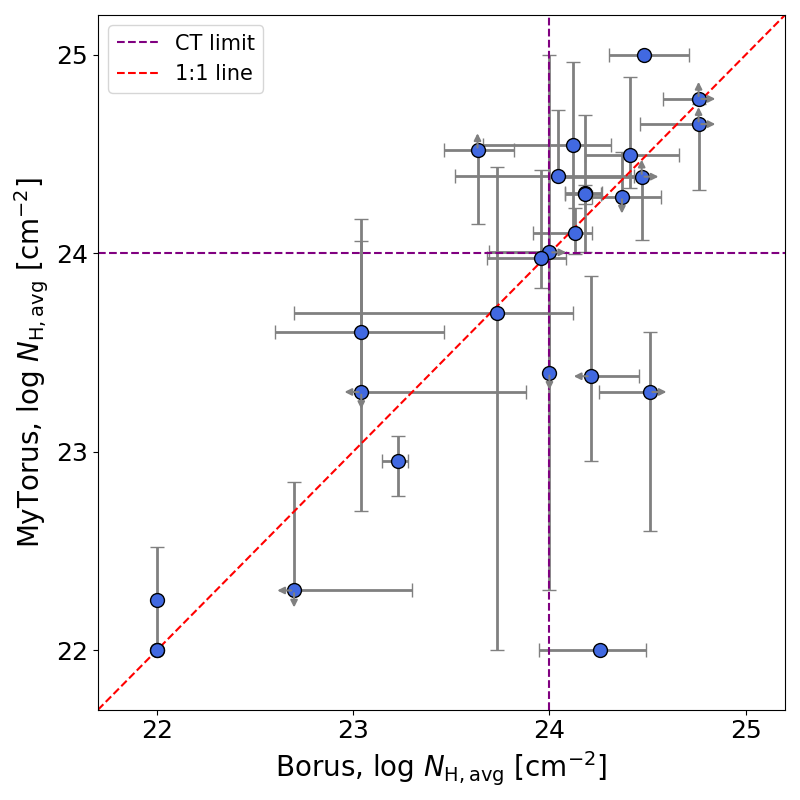}
     }
\caption{Comparison between the line-of-sight column densities obtained from \borus, and \ux\ (left panel). The right panel shows the comparison between average column densities inferred using \myt, and \borus. The red dashed line denotes the 1:1 relation between the $N_{\rm H,los}$ values estimated from different torus models, and the purple dashed lines mark the Compton-thick threshold. }  \label{figure-3}
\end{figure*}

\section{Spectral Fitting Results}
\label{sec5}
The best-fit results including the photon index ($\Gamma$) $N_{\rm H,los}$, $N_{\rm H,avg}$, $c_f$, \texttt{TORsigma}, \texttt{CTKcover}, model normalizations, the 2-10 keV flux, and the 2-10 keV absorption corrected intrinsic luminosities as obtained from the spectral analysis using the three torus models of the 26 sources are given in Table \ref{table-2}. The description of the fitting process for the sources and a case study on comparing the measured column densities in this work with the previously obtained analysis are reported in the Appendix \ref{source description}.

Our analysis shows that a small fraction of sources are true CT-AGN. To classify CT-AGN across the three torus models -- \myt, \borus, and \ux\ -- we used the line-of-sight column densities, along with their associated asymmetric errors. A source was classified as CT if both the upper and lower bounds of the error interval placed at $N_{\rm H,los} >$ $10^{24}$ $\rm{cm^{-2}}$ . Conversely, a source was classified as Non-CT if both bounds were below the threshold. If the classification could not be determined unambiguously due to the error bars crossing the threshold or due to the presence of censoring indicators it was marked as Uncertain. In Fig. \ref{figure-4}, we show the fraction of CT-AGN (red bars) obtained from the \myt\ (left), \borus\ (middle), and \ux\ (right) models. For comparison, the same panels also display confirmed non-CT AGN in green, while sources with uncertain $N_{\rm H,los}$ values are represented by gray bars. 

We found 10 (38.5\%), 6 (23.1\%), and 6 (23.1\%) sources classified as Uncertain by \myt, \borus, and \ux, respectively. Among the classified sources, \myt\ identified a single CT-AGN candidate (ESO 362$-$8; 3.8\%), \borus\ identified three (ESO 362$-$8, 2MASX J02420381+0510061, and 2MASX J06363227$-$2034532; 11.5\%), and \ux\ identified two (MRK 1073 and ESO 362$-$8; 7.7\%). The agreement across all three models highlights ESO 362$-$8 as a robust CT-AGN candidate, consistent with earlier findings \citep{2022ApJ...940..148S}. In addition to the CT classification, we examined the distribution of sources below the CT threshold (excluding the uncertain cases). We found that 10 (38.5\%), 10 (38.5\%), and 11 (42.3\%) sources in \myt, \borus, and \ux, respectively, have $10^{23} \leq N_{\rm H,los} < 10^{24}$ $\rm{cm^{-2}}$, while 5 (19.2\%), 7 (26.9\%), and 7 (26.9\%) sources fall below $10^{23}$ $\rm{cm^{-2}}$. 

\section{Discussion}
\label{sec6}
\subsection{Comparison among \myt, \borus, and \ux}
The comparison of $N_{\rm H,los}$ obtained for a sample of 26 AGN using three physically motivated X-ray torus models --- \myt, \borus, and \ux\ --- provides important insights into how different assumptions about torus geometry affect spectral fitting outcomes. Each of these models assumes distinct torus structures: \myt\ adopts a doughnut-shaped, azimuthally symmetric torus with a fixed opening angle and a uniform density distribution; \borus\ allows for a more flexible, quasi-spherical geometry with variable opening angles and covering factors; and \ux\ models the torus as a clumpy, patchy distribution of obscuring material, thought to better represent a physically realistic torus structure in many AGN unification scenarios. Despite these significant differences in geometry and parameterization, Fig. \ref{figure-2}, and \ref{figure-3} (left panel) show that the majority of the best-fit $N_{\rm H,los}$ values are in reasonably good agreement across the three models, with all of them capturing similar trends in obscuration level among sources. The spearman rank correlation performed between $N_{\rm H,los}$ values yields the coefficients ($\rho$) = 0.95, 0.97 and 0.93, respectively, for \myt\ vs. \borus, \myt\ vs. \ux, and \borus\ vs. \ux.  This suggests that the line-of-sight column density, being a robust quantity, is relatively well constrained by the data regardless of torus geometry assumptions.

However, some discrepancies still emerge. For example, Fig. \ref{figure-4} reveals that \borus\ tends to identify a larger fraction of sources as CT-AGN, defined as having $N_{\rm H,l.o.s} > 10^{24}~\text{cm}^{-2}$, compared to \myt\ and \ux. 
The scatter plots (Fig. \ref{figure-2} and left panel of Fig. \ref{figure-3}) and the $\rho$-values suggest that: while a strong correlation exists between the $N_{\rm H,los}$ obtained from \myt\ and the other two models (as shown by points clustering around the 1:1 red dashed line), several sources lie above or below the line by substantial margins, especially at lower column densities. This deviation is particularly apparent at the Compton-thin to Compton-thick transition, suggesting that model geometry and assumptions may significantly influence classification at these thresholds. The vertical and horizontal magenta lines in Fig. \ref{figure-2} and \ref{figure-3}  mark the CT threshold, showing that while most sources lie below this boundary in all models, a few are pushed into the CT-AGN regime by \borus\ or \ux.

In summary, while all three models provide broadly consistent estimates for $N_{\rm H,los}$ over a wide range of column densities, \borus\ appears more inclined to identify sources as Compton-thick. This suggests that, while geometry may not drastically alter the overall trend of $N_{\mathrm{H,los}}$, it can critically influence classification at threshold values. Our result that estimates $N_{\rm H,los}$ remain consistent across \myt, \borus, and \ux\ is echoed in previous studies (e.g., \citealt{2018ApJ...854...49M,2019ApJ...887..173L,2023A&A...678A.154T,2025ApJ...979..170P}), who also found strong agreement between \myt, and \borus\ in similar parameter regimes.

\myt, and \borus\ are widely used X-ray spectral models developed to characterize the obscuring torus in AGN. In both models, the average torus column density represents the global, line-of-sight-independent measurement of the torus matter distribution, derived primarily from modeling the reprocessed (reflected) emission components. To assess the consistency between the two models, we calculated $N_{\rm H,avg}$  values for a sample of 26 AGN using \myt, and \borus. Within the uncertainties, the results show overall agreement, supported by a Spearman rank correlation coefficient of $\rho = 0.63$ with a highly significant $p$-value of $5.3 \times 10^{-4}$, as shown in the right panel of Fig.~\ref{figure-3}.

\subsection{Clumpy Torus Scenario}
We explored the difference between $N_{\rm H,avg}$ and $N_{\rm H,los}$ as obtained from the \myt\ and \borus\ models:
\begin{equation}
    \Delta N_H = N_{\rm H,avg} - N_{\rm H,los}
\end{equation}

Asymmetric uncertainties in both quantities were propagated in logarithmic space to estimate the upper and lower bounds on $\Delta N_H$. A source was classified as having $N_{\rm H,avg}< N_{\rm H,los}$ if the lower limit of $N_{\rm H,los}$ (accounting for its upper error) exceeded the upper limit of $N_{\rm H,avg}$ (accounting for its upper error), and as $N_{\rm H,avg}> N_{\rm H,los}$ if the upper limit of $N_{\rm H,los}$ (from its lower error) was smaller than the lower limit of $N_{\rm H,avg}$. Sources for which the error ranges overlapped or included unknown (censored) bounds were conservatively assigned to an "uncertain" category.

\begin{figure*}
\hbox{
     \includegraphics[scale=0.35]{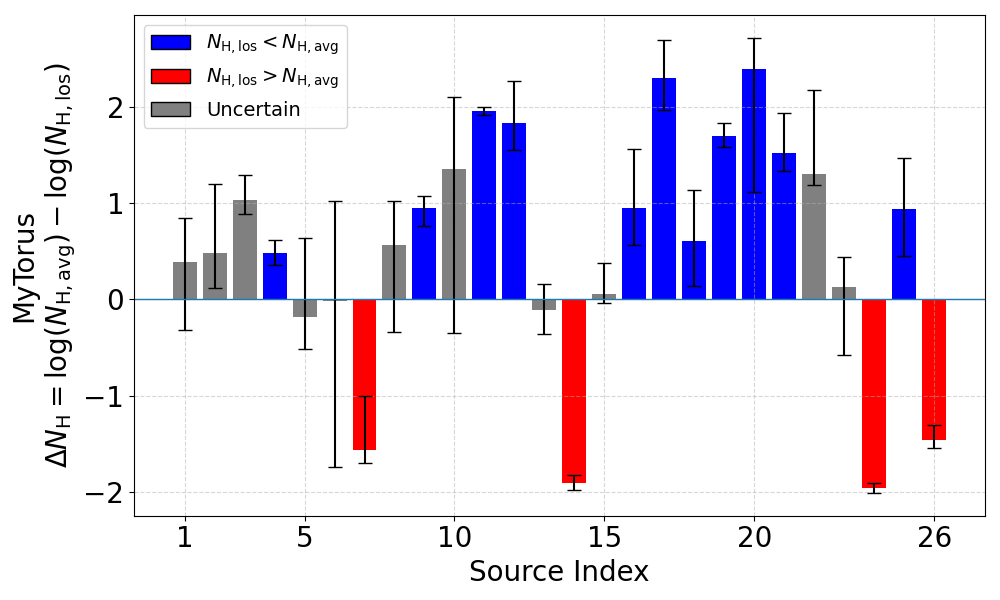}
     \includegraphics[scale=0.35]{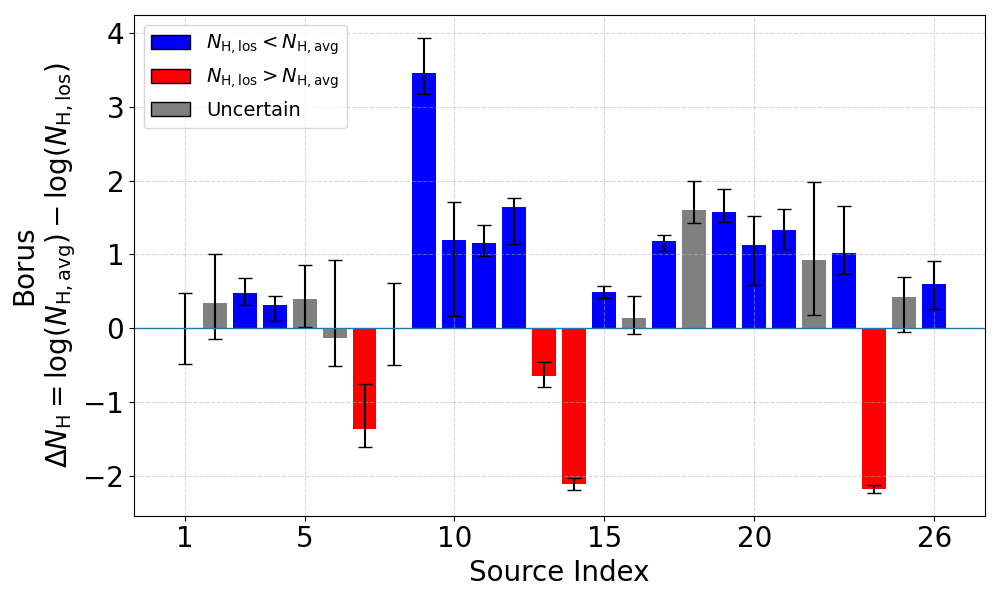}
     }
\caption{Bar plot showing the difference  $\Delta N_H = N_{\rm H,avg} - N_{\rm H,los}$ for each source obtained using \myt\ (left panel) and \borus\ (right panel), with vertical error bars representing the propagated asymmetric uncertainties in logarithmic space. Bars are color-coded by classification: red indicates sources with $N_{\rm H,los} > N_{\rm H,avg}$, blue denotes $N_{\rm H,los} < N_{\rm H,avg}$, and gray corresponds to sources with uncertain classification due to overlapping error ranges or censored (unknown) error values. The horizontal dashed line marks $\Delta N_H$=0, where the line-of-sight and average torus column densities are equal.}  \label{figure-5}
\end{figure*}

Across the sample, the classification based on $\Delta N_{\rm H} = \log N_{\rm H,avg} - \log N_{\rm H,los}$ was generally consistent between the \myt\ and \borus\ models, with no contradictions identified among the sources that could be reliably classified in both cases (see Fig. \ref{figure-5}). Specifically, for \myt, we found that $N_{\rm H,los} > N_{\rm H,avg}$ in 4 out of 26 sources (15.4\%), and $N_{\rm H,los} < N_{\rm H,avg}$ in 11 sources (42.3\%), while the rest of 11 sources (42.3\%) remained unclassified due to overlapping uncertainties or censored error bounds. Similarly, for \borus, 4 sources (15.4\%) showed $N_{\rm H,los} > N_{\rm H,avg}$, and 13 sources (50.0\%) had $N_{\rm H,los} < N_{\rm H,avg}$, with 9 sources (34.6\%) unclassified. We do not attempt to compare classifications for the unclassified cases, where uncertainties preclude meaningful interpretation. We note that in a few instances the $N_{\rm H,avg}$ values were pegged at very low limits; these cases mostly fall into the $N_{\rm H,los} > N_{\rm H,avg}$ category, and although pegged, they were still counted as classified rather than uncertain, since the inequality with respect to $N_{\rm H,los}$ is unambiguous. Only in a source \myt\ and \borus\ produced opposite interpretation: 2MASX J00091156$-$0036551 (index-26), where \myt\ yielded $N_{\rm H,avg} < N_{\rm H,los}$, \borus\ indicated the opposite, with $N_{\rm H,avg} > N_{\rm H,los}$. The discrepancy is driven by \myt\ pegging $N_{\rm H,avg}$ at a low value, while \borus\ is able to constrain it.


The predominance of $N_{\rm H,los} <$ $N_{\rm H,avg}$ may indicate that, for many AGN, our line of sight intercepts a less dense region of the torus, possibly near the edge of the torus opening, while the global structure remains more heavily obscured on average. This is naturally expected in a clumpy torus, where variations in orientation, covering factor, and anisotropic cloud distributions can lead to significant discrepancies between the global and line-of-sight column densities \citep{2008ApJ...685..160N, 2012MNRAS.423.3360Y, 2018ApJ...854...42B}. At the same time, an alternative explanation has been put forward by \citet{2023A&A...678A.154T}, who argue that cases with $N_{\rm H,los} <$ $N_{\rm H,avg}$ may instead reflect a dense, geometrically thin (possibly warped) disk-like structure that contributes strongly to reflection, even if the line of sight intersects less obscured gas. Together, these interpretations suggest that both clumpy and flattened dense reflector geometries can account for the observed trend, potentially coexisting in different sources.

The few cases where $N_{\rm H,los} >$ $N_{\rm H,avg}$ could correspond to transient intersections with dense clumps, consistent with observed variability in some AGN. A representative example is NGC 1358, which was initially identified as a candidate $N_{\rm H,los}$--variable AGN due to its elevated $N_{\rm H,los}$ relative to $N_{\rm H,avg}$ and low covering factor. Subsequent monitoring with \xmm\ and \nustar\ confirmed it as a changing-look AGN \citep{2022ApJ...935..114M}. This scenario is also supported by \cite{2021ApJ...922..252T}, who suggest that when $N_{\rm H,los} >$ $N_{\rm H,avg}$, the torus is likely clumpy rather than uniform. In this case, our line of sight happens to cross a particularly dense cloud, while the overall torus is less dense on average. This creates a "thin-reflector" situation, where the reflected X-ray emission mainly comes from the difference between dense clumps and the more diffuse gas around them.

As shown in Fig. \ref{figure-3} (right panel), the torus column densities for the majority of Compton-thin sources, as well as for the few CT-AGN in our sample extend into the $10^{24}$   $\rm{cm^{-2}}$ regime, indicating that Compton-thick reflectors are a common feature across both categories. A comparable trend has also been noted in previous works (e.g., \citealt{2019A&A...629A..16B, 2021A&A...650A..57Z}). This behavior, also highlighted by \citet{2023A&A...678A.154T}, suggests that while line-of-sight obscuration can fluctuate due to clumpy clouds, the global torus structure remains consistently Compton-thick, providing a stable reflective component.

As mentioned earlier, only one source (excluding the uncertain measurements) displayed conflicting behavior between models, suggesting that, despite differences in model assumptions, the overall trend of $N_{\rm H,avg} >$ $N_{\rm H,los}$ observed in our sample is robust. This result supports the scenario reported in \cite{2021A&A...650A..57Z}, where the authors found that $N_{\rm H,avg}$ is mostly constant across a wide range of $N_{\rm H,los}$, and is on average log$N_{\rm H,avg} \sim$ 24.1. Since most of the objects in this work have  log$N_{\rm H,los} <$ 24, it makes sense that in most of them $N_{\rm H,los}$ is smaller than  $N_{\rm H,avg}$.



\subsection{Testing the accuracy of different $\rm{N_{H,los}}$ predictors}
\cite{2015MNRAS.454..766A} leveraged the tight correlation between MIR and X-ray emission in AGN to estimate the line-of-sight column density. They compiled high-resolution 12$\mu$m MIR fluxes and 2-10 keV X-ray fluxes for 152 local AGN, finding a near-linear correlation across obscuration classes. Crucially, obscured AGN (Seyfert 2 and Compton-thick) deviate below this trend, i.e., their X-ray flux is relatively suppressed compared to MIR, because absorption reduces X-ray output. From this, \cite{2015MNRAS.454..766A} derived an empirical relation:

\begin{dmath}
    \log N_{\mathrm{H}} = (14.37 \pm 0.11) + (0.67 \pm 0.11) 
    \log \left( \frac{F_{12\,\mu\mathrm{m}}}{F_{2\text{--}10\,\mathrm{keV}}} \right)
\end{dmath}

\noindent allowing $N_{\rm H,los}$ estimates directly from the IR--X-ray flux ratio ($\frac{F_{12\mu m}}{F_{2-10keV}}$). This method works best for heavily obscured AGN (log$\rm{N_{H}} \gtrsim$ 23), but its accuracy declines for less obscured sources.

\begin{figure}
\centering
     \includegraphics[scale=0.48]{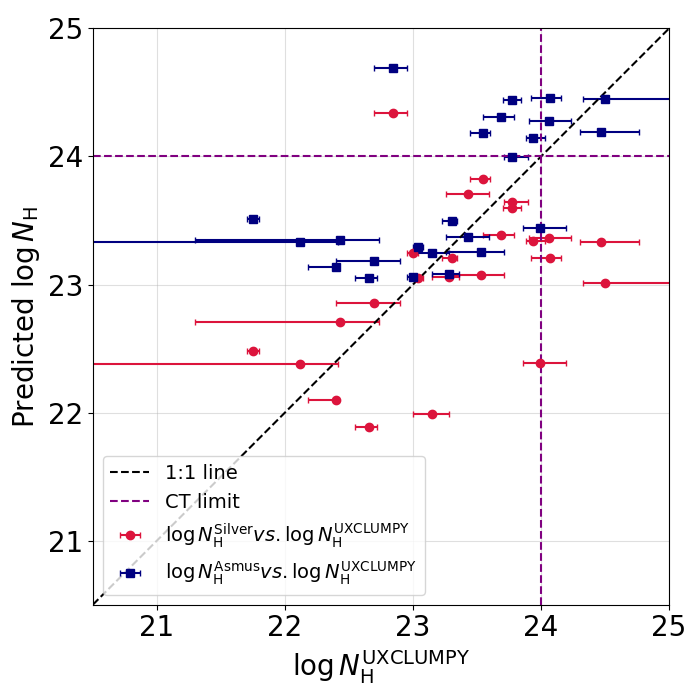}
\caption{Predicted $N_{\rm H,los}$ from \cite{2015MNRAS.454..766A} (blue), and from the machine learning method (red) by \cite{2023AandA...675A..65S}, as a function of the $N_{\rm H,los}$ obtained from the \ux\ fits. The black dashed line marks the 1:1 relation between the predicted and \ux--derived $N_{\rm H,los}$, while the purple dashed lines indicate the Compton-thick~boundary.}  \label{figure-9} 
\end{figure}
\begin{figure*}
\hbox{
     \includegraphics[scale=0.43]{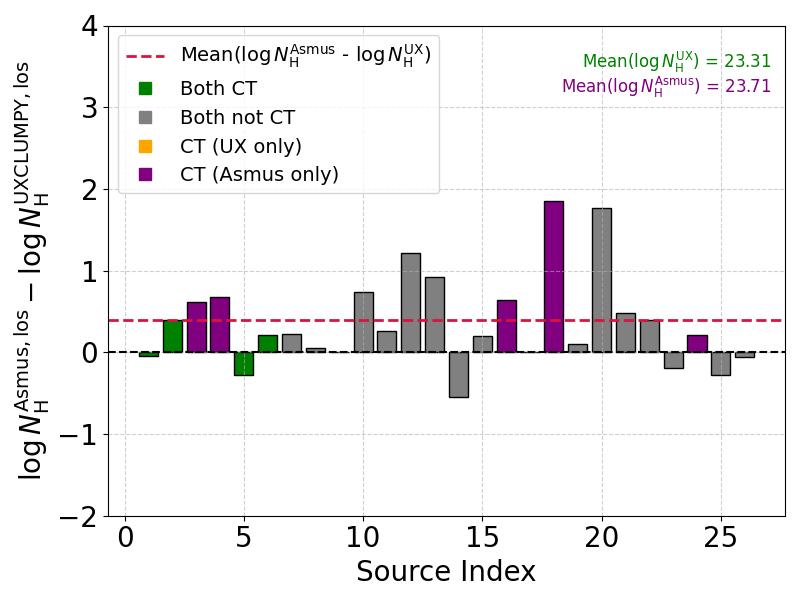}
     \includegraphics[scale=0.43]{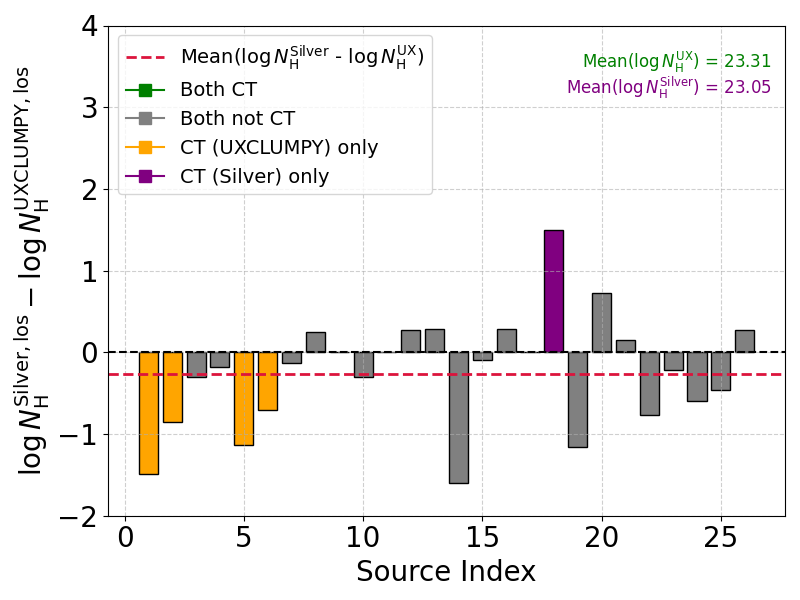}
     }
\caption{Comparison of the predicted $N_{\rm H,los}$ from  \cite{2015MNRAS.454..766A} (left panel) and from the machine learning method by \cite{2023AandA...675A..65S} (right panel),  with $N_{\rm H,los}$ obtained from \ux\ fits. Colored bars indicate differences in source classification: green marks cases where both prediction and \ux\ classify the source as CT-AGN, purple shows cases classified as CT only by the prediction, yellow as CT only by \ux, and gray where both classify the source as non-CT. Source indices follow Table \ref{table-1b}. The red dashed line indicates the offset between the mean predicted and mean derived $N_{\rm H,los}$.}  \label{figure-10} 
\end{figure*}

\cite{2023AandA...675A..65S} introduced a supervised machine-learning (ML) approach using a multiple linear regression model trained on a set of 451 \swi\ AGN with known $N_{\rm H}$ values. Their feature set included: six WISE MIR color indices (e.g., W1-W2, W2-W3), the MIR/soft X-ray flux ratio, two soft X-ray hardness ratios from \xrt, and count rates in multiple \swi\ energy bands. They demonstrated strong predictive performance, achieving a Spearman correlation of 0.86 and correctly classifying $\sim$75\% of obscuration levels-even below log$\rm{N_{H}} \lesssim$ 22.5. This method highlighted that the two soft X-ray hardness ratios and the IR--X-ray flux ratio were the most significant predictors. Their method thus refines the \cite{2015MNRAS.454..766A} relation by combining IR--X-ray information with X-ray spectral diagnostics and broad-band count rates in a statistically robust, machine-learning framework.

We compared $N_{\rm H,los}$ values derived from our X-ray spectral fitting using three independent torus models (\myt, \borus\ and \ux) with predictions from two literature methods: (1) the IR--X-ray correlation from \cite{2015MNRAS.454..766A}, and (2) the machine learning (ML) model from \cite{2023AandA...675A..65S}. The predicted $N_{\rm H,los}$ values are given in Table \ref{table-1b}.  The consistency of $N_{\rm H,los}$ values across the three torus models (see Fig. \ref{figure-2}, \ref{figure-3}) suggest the robustness of the best-fit estimates. Therefore, we compare $N_{\rm H,los}$ obtained from \ux\ model fits with the predicted values, and report this comparison in Fig. \ref{figure-9}. We exclude two sources from the analysis: M 58 (source index - 9) was omitted since its \ux\ column density was around $10^{20}\,\mathrm{cm}^{-2}$, below the lower threshold of $10^{21}\,\mathrm{cm}^{-2}$ used in the analysis of \cite{2023AandA...675A..65S}. FRL 265 (source index - 17) was also excluded from the ML prediction because it lacked a W3 measurement, preventing an estimate of $N_{\rm H,los}$. We statistically tested the agreement between the \ux\ line-of-sight column densities and the predictions from the ML and Asmus methods. Two complementary approaches were adopted:

\noindent{\bf 1. Chi-square test:} Here we tested the null hypothesis that the predicted values lie on the one-to-one relation with the \ux\ measurements (y=x). Both comparisons yield very large chi-square values relative to the degrees of freedom (ML vs \ux\: $\chi^2$ = 1044 for 24 dof; Asmus vs \ux\: $\chi^2$ = 3807 for 24 dof), corresponding to vanishingly small p-values. This indicates that neither method is formally consistent with the one-to-one line.

\noindent{\bf 2. Paired t-test:} We also compared the mean offset between predicted and measured log$\rm{N_{H}}$. For the ML method, the mean difference is -0.26 dex with a p-value of 0.087, indicating no statistically significant systematic offset. In contrast, Asmus predictions show a larger positive offset (+0.40 dex) with a highly significant p-value (p = 0.003), suggesting a systematic overprediction relative to \ux.

\begin{figure*}
\hbox{
     \includegraphics[scale=0.30]{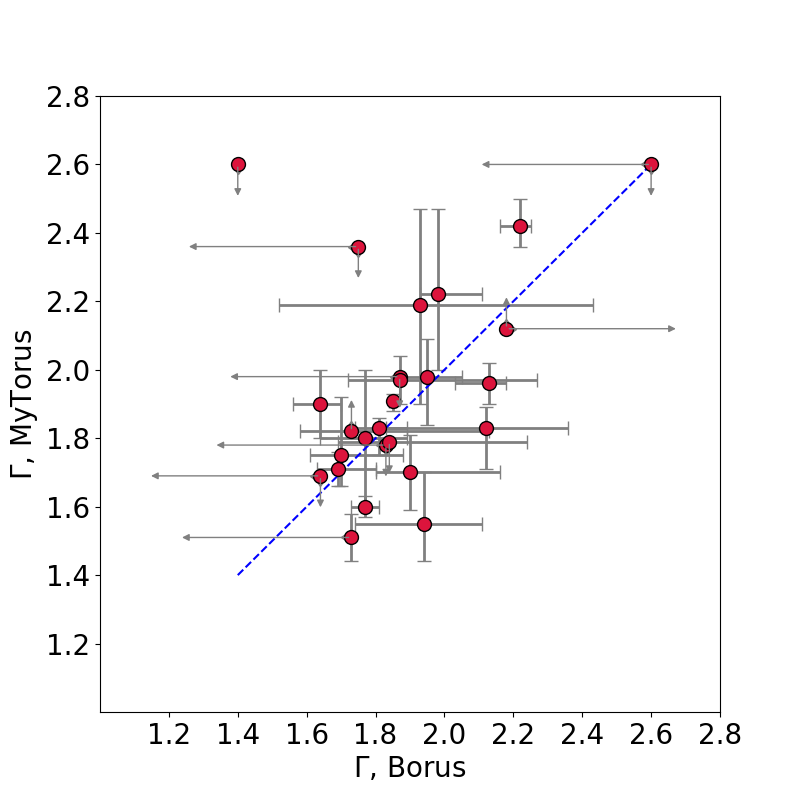}
     \includegraphics[scale=0.30]{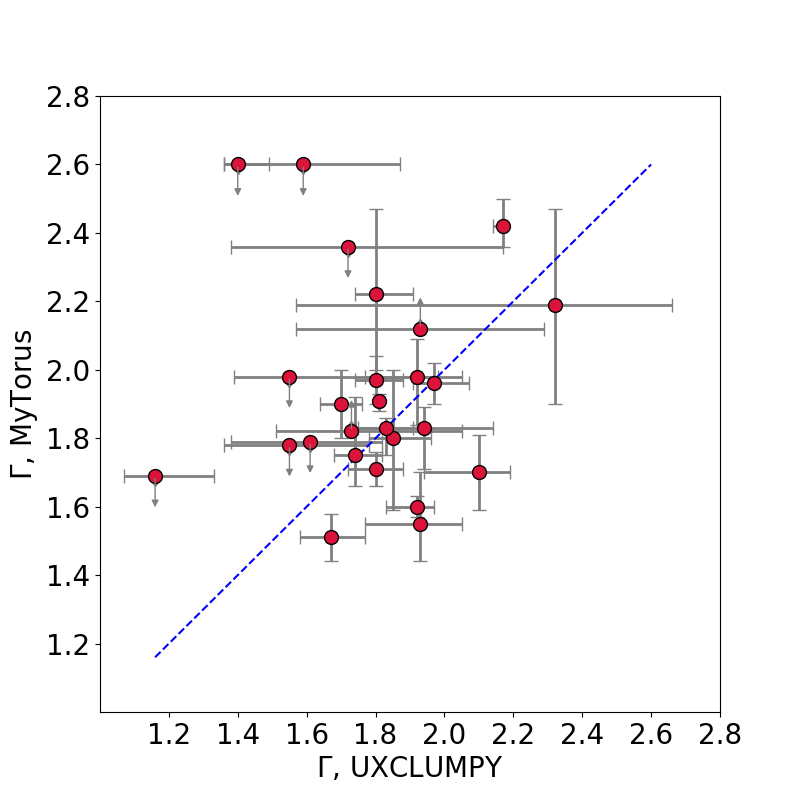}
     \includegraphics[scale=0.30]{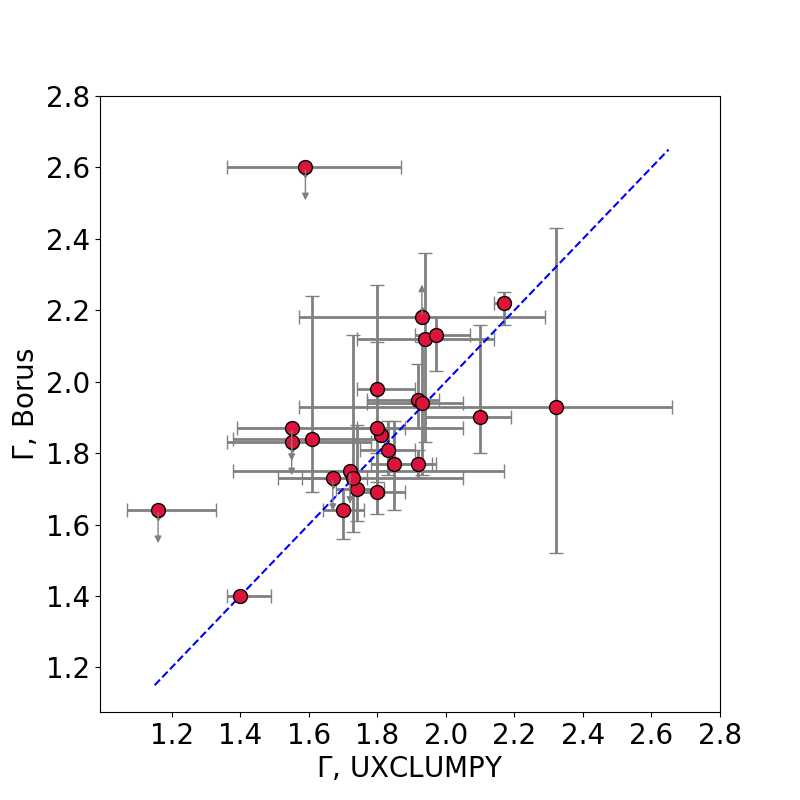}
     }
\caption{Comparison between best-fit photon index obtained from \myt, \borus, and \ux. The blue dashed line marks the 1:1 relation between the data points.}  \label{figure-11} 
\end{figure*}

\begin{figure*}
\hbox{
     \includegraphics[scale=0.45]{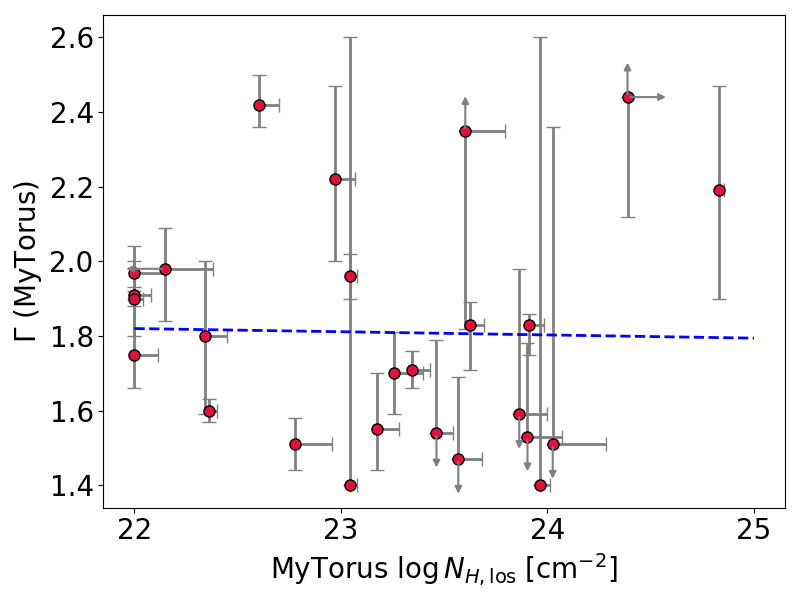}
     \includegraphics[scale=0.45]{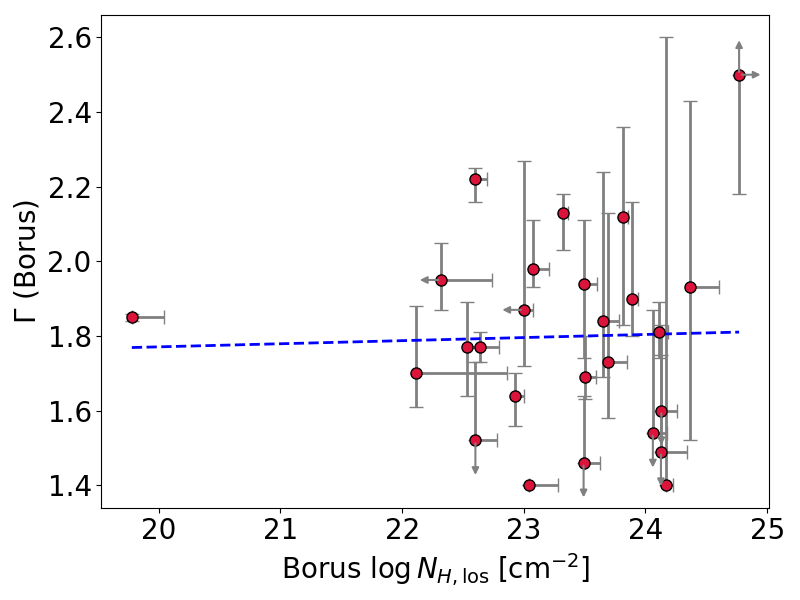}
     }
\caption{Correlation between the photon indices and the line-of-sight column densities obtained from \myt\ (left panel), and \borus\ (right panel). The blue dashed line indicates the least-squares linear fit to the data.}  \label{figure-12} 
\end{figure*}

Taken together, while both prediction methods deviate from the derived \ux\ values in a strict $\chi^2$ sense, the mean comparison indicates that the ML predictions by \cite{2023AandA...675A..65S} (mean $\log N_{\rm H,los}=23.05$) lie closer to the \ux\ mean ($23.31$) than the Asmus estimates ($23.71$). This is consistent with the t-test results, which suggest that the ML approach is statistically more compatible with the \ux\ fits, whereas the Asmus predictions exhibit a more pronounced systematic offset.


The left panel of Fig. \ref{figure-10} shows differences between the IR--X-ray $N_{\rm H}$ predictions \citep{2015MNRAS.454..766A} and the \ux\-- based fits. A large number of sources, especially those tagged as Compton-thin by \ux\ but Compton-thick  by \cite{2015MNRAS.454..766A} (purple bars), indicate that the IR--X-ray method tends to overpredict $N_{\rm H}$, particularly for sources with lower intrinsic obscuration (log$\rm{N_H} \lesssim$ 23 $\rm{cm^{-2}}$) (see Fig. \ref{figure-9}). This is consistent with the idea that MIR emission can remain strong even when the X-ray emission is weak due to moderate absorption, misleading the empirical correlation. Thus, false positives (CT, or high obscuration classification only by \citealt{2015MNRAS.454..766A}) are evident, making the method less reliable at lower $N_{\rm H,los}$.

The right panel of Fig. \ref{figure-10} shows differences between the ML-based predictions and the \ux\ fits. The results demonstrate better alignment in the lower $N_{\rm H,los}$ regime (log$\rm{N_H} \lesssim$ 23.5 $\rm{cm^{-2}}$, see Fig. \ref{figure-9}), with most sources clustering near zero. Only one source exhibits false CT classifications based solely on ML (purple bars in Fig. \ref{figure-10}), and the ML predictions show less systematic bias than the Asmus method. However, a trend of underprediction is visible at high obscuration (log$\rm{N_H} \gtrsim$ 23.5 $\rm{cm^{-2}}$), possibly due to the limited dynamic range or fewer training samples in the Compton-thick regime. As data volume and quality increase (e.g., through future missions like \texttt{Athena}, \texttt{AXIS}, and deeper IR--X-ray coverage), ML methods will become even more effective. With better representation of CT-AGN in the training set, future models can be optimized to reduce current limitations in the high-$N_{\rm H}$ regime. 


\subsection{comparison among $\Gamma$ from \myt, \borus, and \ux}
The three panels in Fig. \ref{figure-11} compare the best-fit photon indices for the same AGN sample: \myt\ vs. \borus\ (left), \myt\ vs. \ux\ (middle), and \borus\ vs. \ux\ (right). The dashed blue line indicates the 1:1 relation. In both comparisons, the points scatter around the line, showing a decent overall agreement--typical $\Gamma$ values cluster between approximately 1.6 and 2.2. However, the \myt--derived indices systematically exceed those from \ux. We assessed the consistency of the $\Gamma$ measurements with a chi--square test. For \myt\ vs.\ \borus\ we obtain $\chi^{2}=39.66$ for $\nu=26$ degrees of freedom ($\chi^{2}_{\nu}=1.53$, $p=0.042$), indicating modest but statistically significant tension at the $\sim$5\% level. \myt\ vs.\ \ux\ yields $\chi^{2}=60.84$ ($\nu=26$; $\chi^{2}_{\nu}=2.34$, $p=1.29\times10^{-4}$), demonstrating an incompatibility. In contrast, \borus\ vs.\ \ux\ gives $\chi^{2}=21.626$ ($\nu=26$; $\chi^{2}_{\nu}=0.832$, $p=0.709$), consistent with the null hypothesis of equal $\Gamma$. Overall, \borus\ and \ux\ agree best; \myt\ shows moderate tension with \borus\ and the strongest disagreement with \ux, likely indicating that model--specific assumptions and torus geometries can introduce systematic differences in $\Gamma$ estimates.

In the literature, such cross-model comparisons have been reported. For example, \cite{2018ApJ...854...49M} compared \myt\ and \borus\ across a sample of CT-AGN observed with \nustar\ and found generally good agreement in continuum parameters like $\Gamma$--though with increasing scatter at high obscuration. Other works exploring clumpy and smooth torus models \citep{2019ApJ...877...95T, 2022MNRAS.509.5485S} caution that $\Gamma$ may be biased if key geometrical parameters are poorly constrained or degenerate.

In addition to comparing photon indices values across torus models (Fig. \ref{figure-12}), we also examined the relationship between $\Gamma$ and the line-of-sight column density for each model. Our analysis shows no statistically significant correlation between these parameters, with Spearman's correlation coefficients, $r = -0.18$ (\myt), $r = -0.07$ (\borus), $r = -0.004$ (\ux), and corresponding $p$-values well above 0.05, suggesting that $\Gamma$ and $N_{\rm H,los}$ are effectively non-degenerate in all three models. These findings reinforce the idea that broadband X-ray coverage-- such as that provided by \nustar-- enables robust, independent constraints on both spectral slope and absorption. This is consistent with the conclusions of \cite{2018ApJ...854...49M}, who demonstrated that the inclusion of high--energy data can decouple $\Gamma$ and $N_{\rm H,los}$, especially in Compton-thick AGN. Similar results have also been reported by \cite{2018ApJ...854...42B, 2019ApJ...877...95T}, where modeling with geometrically flexible or clumpy torus structures reduced the degeneracy between continuum and absorption parameters. Overall, the absence of a significant $\Gamma$--$N_{\rm H,los}$ correlation in our sample supports the reliability of spectral fits across different torus models, provided that broadband X-ray spectral coverage with adequate count statistics is available.

\begin{figure*}
\centering
\hbox{
    \includegraphics[scale=0.45]{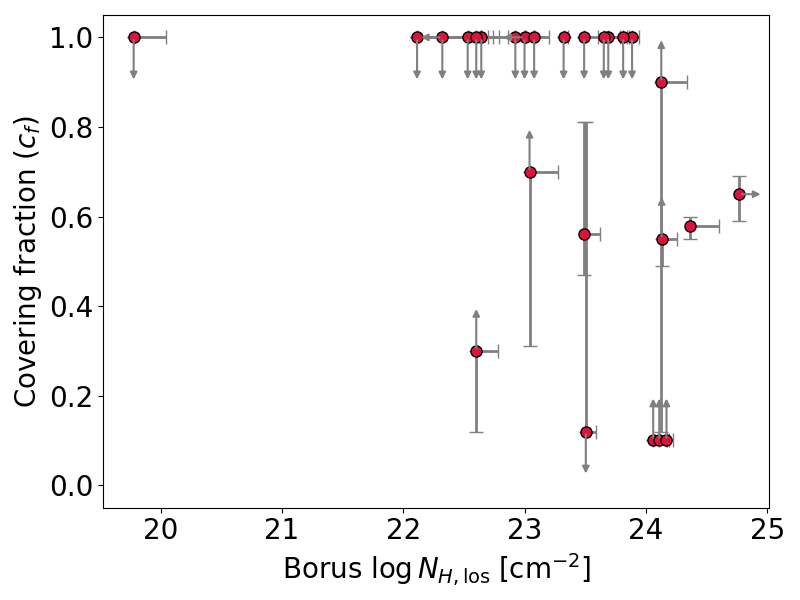}
    \includegraphics[scale=0.45]{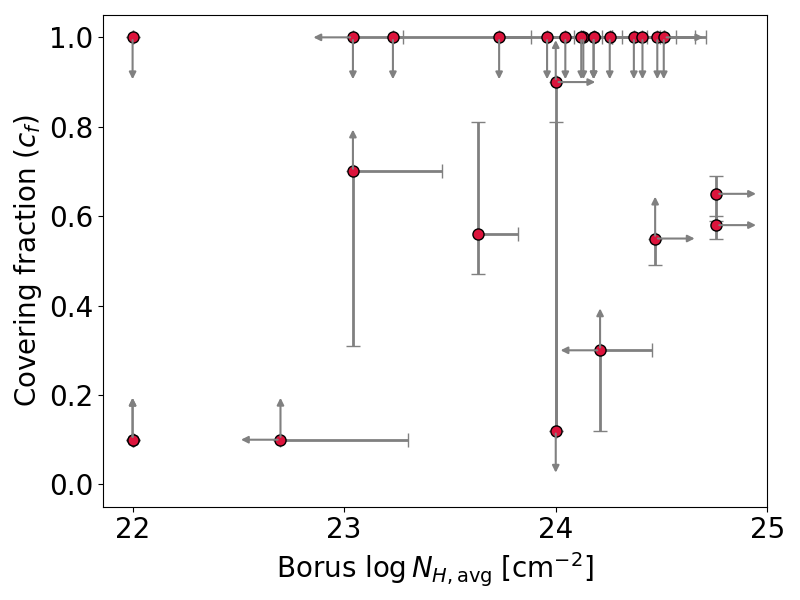}}
\caption{Left panel: Correlation between covering factor ($c_{f}$) and the line-of-sight column densities obtained from \borus. Right panel: Correlation between covering factor ($c_{f}$) and the average column densities obtained from \borus.}
\label{figure-6}
\end{figure*}

\begin{figure}[h]
\centering
     \includegraphics[scale=0.42]{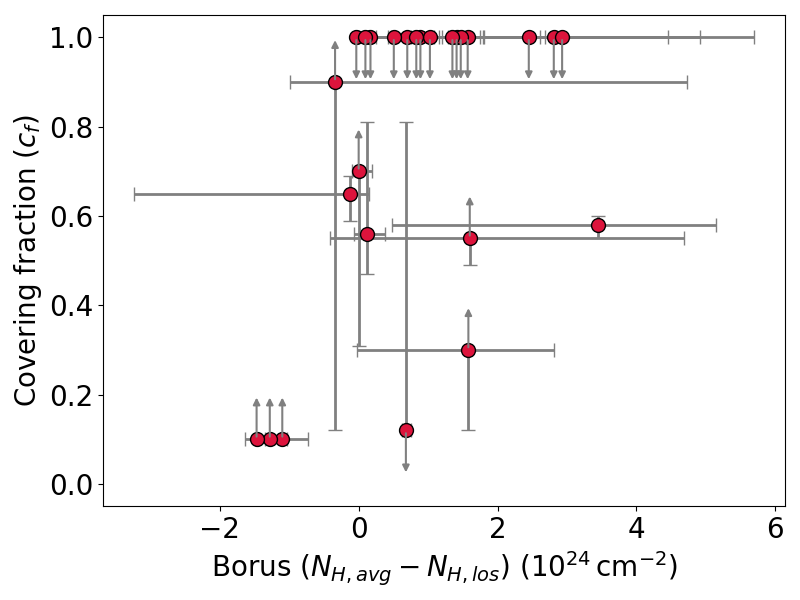}
\caption{Correlation between covering factor ($c_{f}$) and the difference between average and line-of-sight column densities as obtained from \borus.}
\label{figure-7}
\end{figure}


\subsection{Correlations Between Torus Covering Fraction and Other AGN Properties}

We investigated the possible correlations between the torus covering fraction and several AGN parameters, including the $N_{\rm H,los}$, $N_{\rm H,avg}$, their difference, the intrinsic 2--10 keV X-ray luminosity ($\rm L_{2-10 keV}$), and the Eddington ratio ($\lambda_{\mathrm{Edd}}$). Across all cases, we find no statistically significant correlations, primarily due to the limited sample size and the presence of large uncertainties and censored values in $c_f$.

As shown in Fig.~\ref{figure-6}, no clear trend is observed between $c_f$ and $N_{\rm H,los}$ or $N_{\rm H,avg}$. The data show large scatter, and covering fractions span a wide range at both low and high column densities. For $c_f$ versus $N_{\rm H,los}$, the Spearman's rank correlation coefficient is $\rho = -0.56$ with $p = 0.03$, and for $c_f$ versus $N_{\rm H,avg}$, $\rho = 0.24$ with $p = 0.24$, indicating no statistically significant correlation.
For $\Delta \rm{N_H} = \log N_{\rm H,avg} - \log N_{\rm H,los}$, shown in Fig.~\ref{figure-7}, sources with $\Delta \rm{N_H} < 0$ (i.e., $N_{\rm H,los} > N_{\rm H,avg}$) all have very low $c_f$ values pegged at the lower bound in the model, rendering these values unreliable and the apparent trend physically uninformative. For this, the Spearman's rank correlation coefficient is $\rho = 0.40$ with $p = 0.04$, dominated by sources with unconstrained $c_f$.


\begin{figure*}
\hbox{
     \includegraphics[scale=0.45]{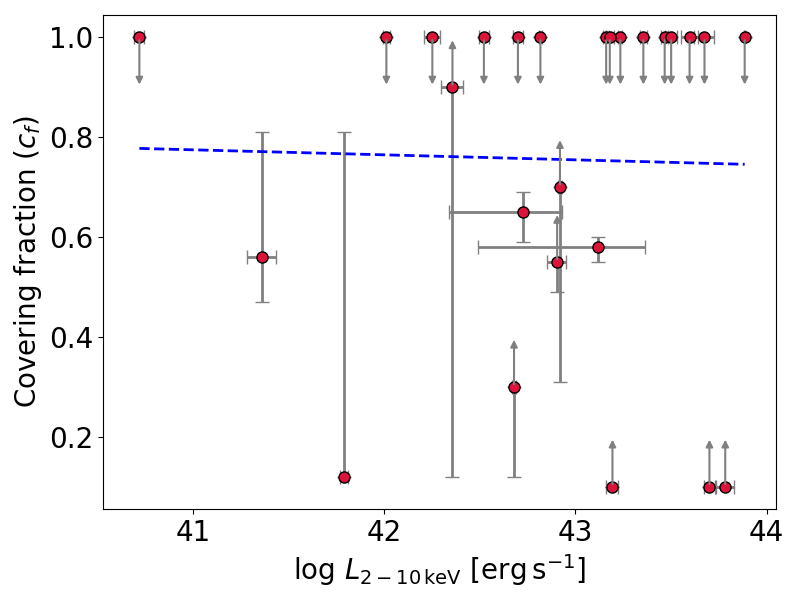}
     \includegraphics[scale=0.45]{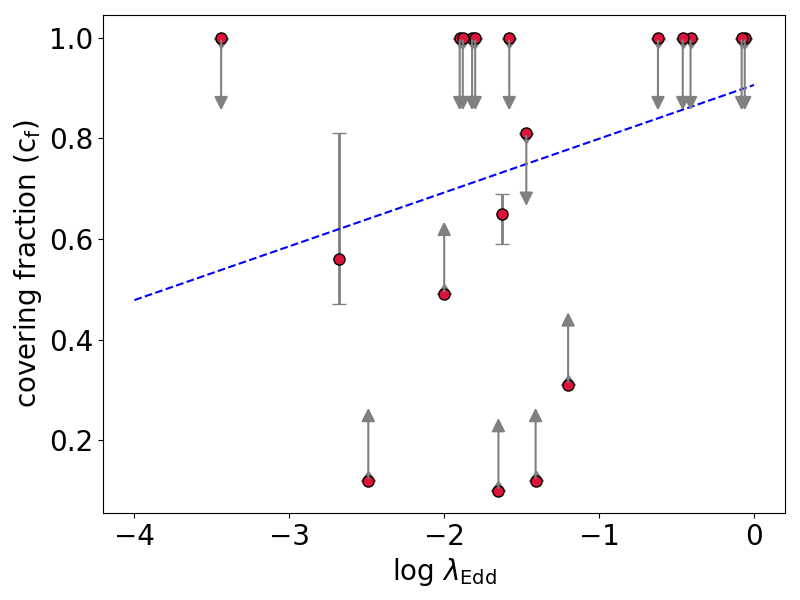}
     }
\caption{Left panel: Correlation between 2-10 keV intrinsic luminosity and the covering factor. Right panel: Correlation between the covering fraction and the Eddington ratio. The blue dashed line shows the linear regression fit to the data.}  \label{figure-8} 
\end{figure*}

Figure~\ref{figure-8} (left panel) examines the relationship between $c_f$ and the absorption-corrected 2--10 keV luminosity derived from the \borus\ model. The Spearman's rank correlation does not confirm any statistically significant correlation between these two parameters ($\rho = 0.04$, $p = 0.85$). The best-fit regression:

\begin{equation*}
    c_f = a*log \rm L_{2-10 keV} + b
\end{equation*}

\noindent yields $a = -0.01$ and $b = 1.19$. 

We also examined the correlation between $c_f$ and the Eddington ratio. The bolometric correction factors, $K_{X} = \rm{L_{ bol}/L_{2-10\,{keV}}}$, were derived following the prescription of \citet{2020A&A...636A..73D}. In their work, $K_{X}$ is parametrized as a function of the bolometric luminosity with separate relations for type~1 and type~2 AGN, calibrated using a large multi-wavelength sample. We adopted their best-fit analytic forms (eqs.~4 and~5 in \citealt{2020A&A...636A..73D}) to compute $K_{X}$ for each source, using the intrinsic $2$--$10$~keV luminosities obtained from our spectral fits. For each source, the bolometric luminosity was then determined as $\rm{L_{ bol} = K_{X} \times L_{2-10\,{\rm keV}}}$. The procedure thus accounts for the luminosity dependence of the bolometric correction, and provides separate estimates for type~1 and type~2 populations. The derived parameters, such as, the bolometric luminosities, bolometric correction factors, and Eddington luminosities for the type~1 and type~2 AGN samples are given in Table \ref{tab:type1} and \ref{tab:type2} in Appendix \ref{lumin:table}.

The Eddington luminosity is calculated as:

\begin{equation*}
  \rm{L_{\mathrm{Edd}}} = 1.3 \times 10^{38} \left(\frac{M_{\mathrm{BH}}}{M_\odot}\right)\, \mathrm{erg\,s^{-1}},  
\end{equation*}

from which the Eddington ratio is defined:

\begin{equation*}
 \lambda_{\mathrm{Edd}} = \frac{\rm{L_{\mathrm{bol}}}}{\rm{L_{\mathrm{Edd}}}}.   
\end{equation*}

\noindent Figure~\ref{figure-8} (right panel) shows the $c_f$--$\lambda_{\mathrm{Edd}}$ relation for 19 sources with available black hole mass estimates. The best-fit linear regression line has a slope of $0.11$ and an intercept = $0.91$. The Spearman's rank correlation yields $\rho = 0.20$ and $p = 0.41$.

In conclusion, none of the investigated parameters exhibit a robust or statistically significant correlation with the torus covering fraction in our sample. The results are limited by small sample size, large scatter, and poorly constrained or censored $c_f$ values. A larger sample with better $c_f$ constraints is required to test any potential correlations.

\section{Conclusion and Summary}
\label{sec7}
In this study, we performed a detailed broadband X-ray spectral analysis of 26 local ($z < 0.1$) obscured AGN selected from the Swift-BAT 150-month catalog. The sources were selected based on mid-infrared to X-ray flux diagnostics and observed with NuSTAR and complementary soft X-ray telescopes (\xmm, \cha, \xrt). We employed three physically motivated torus models --- \myt, \borus, and \ux\ --- to constrain key obscuration parameters including the line-of-sight column density, the average torus column density, and the torus covering factor.

\paragraph{} 
Despite markedly different geometric assumptions, the three torus models tested here --- \myt, \borus, and \ux --- deliver broadly consistent line-of-sight column densities for the 26 AGN. Pairwise comparisons of $N_{\rm H,los}$ are tightly correlated ($\rho=0.95$, $0.97$, and $0.93$ for \myt\ vs.\ \borus, \myt\ vs.\ \ux, and \borus\ vs.\ \ux), indicating that this quantity is largely robust to the choice of geometry (see Fig. \ref{figure-2}, and Fig. \ref{figure-3}). The places where the models start to part ways are exactly where one would expect: near the Compton-thick boundary. There, \borus\ tends to flag more CT candidates than \myt\ or \ux, underscoring how geometry and covering factor assumptions can tip borderline cases across the threshold. According to all three models, ESO 362-8 is robustly classified as a confirmed CT-AGN.

\paragraph{}
Using $\Delta N_{\rm H} = \log N_{\rm H,avg}-\log N_{\rm H,los}$ to compare line-of-sight and global obscuration, we find a clear prevalence of $N_{\rm H,avg}>N_{\rm H,los}$ among sources that can be reliably classified. In \myt, 11/26 (42.3\%) sources show $N_{\rm H,avg}>N_{\rm H,los}$ versus 4/26 (15.4\%) with the opposite trend (11/26, 42.3\% uncertain); restricting to classified objects, this corresponds to 11/15 (73.3\%). In \borus, the numbers are 13/26 (50.0\%) vs.\ 4/26 (15.4\%) (9/26, 34.6\% uncertain), i.e.\ 13/17 (76.5\%) among the classified (see Fig. \ref{figure-5}). This pattern -- together with the frequent extension of $N_{\rm H,avg}$ into the $10^{24}\,\mathrm{cm^{-2}}$ regime even for many Compton-thin lines of sight -- supports a picture in which the global torus is typically more heavily obscured than the particular sight line, as expected for a clumpy and anisotropic structure (with occasional $N_{\rm H,los}>N_{\rm H,avg}$ events plausibly linked to transient dense clouds). Overall, the two models deliver consistent classifications where constraints are robust, indicating the view that a clumpy, globally Compton-thick reflector commonly coexists with sight lines of lower column density.

\paragraph{}
We evaluated the performance of mid-IR/X-ray diagnostics from \citet{2015MNRAS.454..766A} against machine learning (ML) predictions for $N_{\mathrm{H,los}}$, based on multiwavelength observables by \citet{2023AandA...675A..65S} (Fig. \ref{figure-9}, \ref{figure-10}). We found that while the Asmus method provides a broadly reliable first-order prediction, especially for heavily obscured sources (log$\rm{N_H} \gtrsim$ 23.5 $\rm{cm^{-2}}$), the ML approach-- when trained on a diverse parameter space--achieves improved accuracy in estimating $N_{\mathrm{H}}$, specially for Compton-thin to moderately thick regions (log$\rm{N_H} \lesssim$ 23.5 $\rm{cm^{-2}}$). This highlights the potential of data-driven techniques to complement or refine empirical diagnostics in AGN classification and obscuration studies.

\paragraph{}
Across the sample, the photon indices inferred from the three torus models show overall qualitative agreement, with most sources clustering around the one-to-one relation (Fig. \ref{figure-11}). Nonetheless, \myt\ tends to return systematically steeper continua than \ux, whereas \borus\ and \ux\ are mutually more consistent. We do not find evidence for a correlation between $\Gamma$ and $N_{\rm H,los}$ in any model (Fig. \ref{figure-12}), indicating that spectral slope and line-of-sight obscuration are effectively decoupled in our broadband fits. 

\paragraph{}
Finally, we find no robust correlations between $c_f$ and $N_{\rm H,los}$, $N_{\rm H,avg}$, 2--10~keV intrinsic luminosity, or Eddington ratio; formal rank tests either are not significant or are driven by poorly constrained/censored $c_f$ values.

In summary, our analysis supports a clumpy torus structure as a dominant geometry in obscured AGN, provides new constraints on the physical parameters of the obscuring material, and demonstrates the utility of modern ML techniques in predicting AGN column densities. These findings contribute to a more nuanced understanding of obscured accretion and AGN unification, and motivate future multiwavelength follow-up studies to capture the nature of AGN torus.


\begin{acknowledgments}
IP, MA, IC acknowledge funding under NASA contract 80NSSC24K0633 and SAO contract AR4-25009X. This research made use of data obtained with the \nustar\ mission, a project led by the California Institute of Technology, managed by the Jet Propulsion Laboratory, and funded by the National Aeronautics and Space Administration (NASA). It also made use of observations from \xmm, an ESA science mission with instruments and contributions funded by ESA Member States and NASA. In addition, this work is based on data obtained with the \cha\ X-ray Observatory. The authors further acknowledge the use of public data from the {\it Swift} data archive.
\end{acknowledgments}

\facilities{\nustar, \xmm, \cha, {\it Swift}}

\software{HEASoft \citep{2014ascl.soft08004N}), XSPEC \citep{1996ASPC..101...17A}, NUSTARDAS (\hyperlink{https://heasarc.gsfc.nasa.gov/docs/nustar/analysis/}{https://heasarc.gsfc.nasa.gov/docs/nustar/analysis/}), SAS \citep{2004ASPC..314..759G}, CIAO \citep{2006SPIE.6270E..1VF}
          }


\appendix

\section{Source description}\label{source description}
In the appendix, we describe the results obtained via X-ray spectral fitting, and compare them to previous determinations (when available).

\smallskip
\noindent\textbf{MRK 1073:} MRK 1073 is a nearby ($z$ = 0.0233) Seyfert 2 AGN, with a black hole mass of log($M_{\rm{BH}}$/$M_{\odot}$) = 7.78$\pm$0.50 \citep{2011ApJ...740...94D}. The bolometric luminosity ($\rm{L_{bol}}$) and Eddington ratio ($\lambda_{\rm{Edd}}$) reported by \cite{2020ApJ...897..107Y} are 7.8$^{+4.4}_{-3.8}$ $\times$ 10$^{43}$ erg $\rm{s^{-1}}$ and 0.10$^{+0.25}_{-0.07}$, respectively. In this work, we analyze archival \nustar\ and \xmm\ observations of MRK 1073 spanning from 2002 to 2020. One \nustar\ observation was obtained in 2014, while among the three \xmm\ observations, one in 2002 and the remaining two in 2020. Previous X-ray analyses have suggested that MRK 1073 hosts a CT-AGN \citep{2005A&A...444..119G, 2011ApJ...727...19F, 2020ApJ...897..107Y, 2021MNRAS.506.5935R}. Using the \texttt{XCLUMPY} model \citep{2019ApJ...877...95T}, \cite{2020ApJ...897..107Y} analyzed the \nustar\ data along with one \xmm\ observation (ObsID: 0002942401) and reported $N_{\rm H,los}$ = 5.4$^{+7.0}_{-3.1}$ $\times$ 10$^{24}$ cm$^{-2}$ and $N_{\rm H,avg}$ = 15.0$^{+19.0}_{-9.0}$ $\times$ 10$^{24}$ cm$^{-2}$, thereby supporting the Compton-thick nature of the source. Our analysis of the joint \nustar\ and \xmm\ spectra of MRK 1073 using the \ux\ model is consistent with the CT classification reported in previous studies with $N_{\rm H,los}$ = 3.16$^{+7.84}_{-1.04}$ $\times$ 10$^{24}$ cm$^{-2}$  (see Table~\ref{table-2}). However, \myt\ and \borus\ yielded $N_{\rm H,los}$ of 2.44$^{+u}_{-0.78}$ $\times$ 10$^{24}$ cm$^{-2}$ and 5.88$^{+u}_{-2.41}$ $\times$ 10$^{24}$ cm$^{-2}$, respectively, leaving the CT classification uncertain due to the unconstrained upper limits. While, the IR--X-ray correlation analysis in \cite{2015MNRAS.454..766A} classified this source as CT with log($N_{\rm{H,l.o.s}}$/cm$^{-2}$) = 24.45, the ML algorithm \citep{2023AandA...675A..65S}, retrieved a lower value for log($N_{\rm{H,l.o.s}}$/cm$^{-2}$) = 23.01. To account for the additional nuclear emission in the soft X-ray band, we include a \texttt{mekal} component in our analysis. To model the Fe K$\beta$ emission line around $\sim$7.1 keV, we add a Gaussian component (\texttt{zgauss}) with a fixed line width of $\sigma$ = 100 eV. The best-fit line energy we obtained is 6.77$^{+0.57}_{-0.19}$ keV. Allowing the line width to vary during the fitting process does not significantly affect the fit statistics. Furthermore, the common parameters obtained from modeling the source spectra using three different torus models show good agreement with each other, strengthening the robustness of our results.

\smallskip
\vspace{0.1cm}
\noindent {\textbf{UGC 5101:}} This nearby Seyfert 2 AGN ($z$ = 0.0393) was observed by \nustar\ twice, in 2014 and 2020, respectively. It hosts a supermassive black hole with a mass of log($M_{\rm{BH}}$/$M_{\odot}$) $\sim$ 8.35 \citep{2017ApJ...850...74K}, and an Eddington ratio of $\lambda_{\rm{Edd}}$ = 0.014 \citep{2022ApJS..260...30T}. For the soft X-ray (0.5-10 keV) spectral analysis, we use archival \cha\ observation taken in 2001. Using the \texttt{XCLUMPY} model, \cite{2022ApJS..260...30T} analyzed \nustar\ and {\it Suzaku} data of this source, and reported $N_{\rm H,los}$ = 1.53$^{+0.53}_{-0.09}$ $\times$ 10$^{24}$ cm$^{-2}$ and $N_{\rm H,avg}$ = 1.72$^{+0.51}_{-0.66}$ $\times$ 10$^{24}$ cm$^{-2}$. Similarly, \cite{2017ApJ...835..179O} utilized \xmm, \nustar, and {\it Suzaku} observations, and derived $N_{\rm H,los}$ = 1.31$^{+0.31}_{-0.36}$ $\times$ 10$^{24}$ cm$^{-2}$. More recently, \cite{2022MNRAS.510.5102O} analyzed one of the \nustar\ observations (ObsID: 60001068002) and reported a line-of-sight column density of log($N_{\rm{H,l.o.s}}$/cm$^{-2}$) = 23.84$^{+0.06}_{-0.06}$. $N_{\rm H,los}$ and $N_{\rm H,avg}$ obtained from our joint spectral modeling using the \myt, \borus, and \ux\ torus models are both $\sim$ 10$^{24}$ cm$^{-2}$, indicating a Compton-thick (CT) nature. Using \myt, \borus\ and \ux\ we obtained $N_{\rm H,los}$ = 0.80$^{+0.38}_{-0.16}$ $\times$ 10$^{24}$ cm$^{-2}$, 1.35$^{+0.45}_{-0.48}$ $\times$ 10$^{24}$ cm$^{-2}$ and 1.17$^{+0.27}_{-0.34}$ $\times$ 10$^{24}$ cm$^{-2}$ respectively. Our results are in good agreement with previous studies of the source. The IR--X-ray correlation performed in \cite{2015MNRAS.454..766A} classified the source as CT with log($N_{\rm{H,l.o.s}}$/cm$^{-2}$) = 24.46, while \cite{2023AandA...675A..65S}, marked the source as non-CT with log($N_{\rm{H,l.o.s}}$/cm$^{-2}$) = 23.21.

\smallskip
\vspace{0.1cm}
\noindent {\textbf{NGC 7674:}} We used the \nustar, \xmm\ and \cha\ observations taken in 2014, 2004 and 2020, respectively, of this source for our analysis. This nearby AGN ($z$ = 0.02903) is classified as a type 2 Seyfert with log($M_{\rm{BH}}$/$M_{\odot}$) $\sim$ 7.73 \citep{2020ApJ...901..161K}. Fitting \myt\ in decoupled patchy mode, using \nustar\ and \xmm\ observations, \cite{2020ApJ...901..161K} reported $N_{\rm H,los}$ = 0.31$^{+0.05}_{-0.07}$ $\times$ 10$^{24}$ cm$^{-2}$ and $N_{\rm H,avg}$ = 5.7$^{+1.0}_{-2.3}$ $\times$ 10$^{24}$ cm$^{-2}$. Similar results are reported in \cite{2023MNRAS.524.4670J}, who used the \borus\ model to fit the \nustar\ and \xmm\ data, reported log($N_{\rm{H,l.o.s}}$/cm$^{-2}$) = 23.61$^{+0.25}_{-0.10}$ and log($N_{\rm{H,avg}}$/cm$^{-2}$) = 24.57$^{+25.50}_{-1.02}$. Using the \texttt{XCLUMPY} model to fit the {\it Suzaku} and \nustar\ data \cite{2020ApJ...897....2T} reported $N_{\rm H,los}$ = 0.24$^{+0.22}_{-0.10}$ $\times$ 10$^{24}$ cm$^{-2}$. Fitting the \nustar\ spectrum alone with the \myt\ model in decoupled mode, \cite{2017MNRAS.467.4606G} also suggested a Compton-thin line of sight obscurer with  $N_{\rm H,los}$ = 0.13$^{+0.03}_{-0.03}$ $\times$ 10$^{24}$ cm$^{-2}$ for NGC 7674. The previous results agree well with our spectral analysis with all three torus models suggested a clumpy nature of the obscuring medium, with $N_{\rm H,los}$ $<$ 10$^{24}$ cm$^{-2}$ and $N_{\rm H,avg}$ $\gtrsim$ 10$^{24}$ cm$^{-2}$ (see Table \ref{table-2}). The three torus models yielded broadly consistent results, with \myt\ giving 0.18$^{+0.38}_{-0.16}$ $\times$ 10$^{24}$ cm$^{-2}$, \borus\ yielding 0.77$^{+0.45}_{-0.48}$ $\times$ 10$^{24}$ cm$^{-2}$ and \ux\ returning 0.49$^{+0.12}_{-0.14}$ $\times$ 10$^{24}$ cm$^{-2}$. According to the IR--X-ray correlation presented in \cite{2015MNRAS.454..766A} the source was identified as CT-AGN with log($N_{\rm{H,l.o.s}}$/cm$^{-2}$) = 24.31, in contrast to the findings of \citep{2023AandA...675A..65S}, which classified it as non-CT with log($N_{\rm{H,l.o.s}}$/cm$^{-2}$) = 23.01. The spectra of NGC 7674 showed an excess at $\sim$ 7 keV that possibly corresponds to Fe XXVI. We modeled it by adding a Gaussian emission line. The soft X-ray emission is modeled with a \texttt{mekal} component.

\smallskip
\vspace{0.1cm}
\noindent {\textbf{IC 2227:}} The nearby Seyfert 2 galaxy IC 2227 (z = 0.0323) was recently targeted by \nustar\ (2022), \xmm\ (2022), and \cha\ (2021). In this work, we make use of all these observations to fit the source spectra with three different torus model. From an analysis of \nustar\ and \xmm\ observations using \myt\ (decoupled Face-on), \borus, and \ux, \cite{2022ApJ...940..148S} reported $N_{\rm H,los}$ values of 0.6$^{+0.11}_{-0.07}$ $\times$ 10$^{24}$ cm$^{-2}$, 0.64$^{+0.05}_{-0.06}$ $\times$ 10$^{24}$ cm$^{-2}$, and 0.63$^{+0.10}_{-0.04}$ $\times$ 10$^{24}$ cm$^{-2}$, respectively. The $N_{\rm H,avg}$ values obtained from the \myt\ and \borus\ fits were 6.06$^{+u}_{-2.84}$ $\times$ 10$^{24}$ cm$^{-2}$ and 31.62$^{+u}_{-26.00}$ $\times$ 10$^{24}$ cm$^{-2}$, respectively. The best-fitted parameters obtained from our analysis are in close agreement with the previous results with $N_{\rm H,los}$ = 0.42$^{+0.07}_{-0.05}$ $\times$ 10$^{24}$ cm$^{-2}$ using \myt, 0.65$^{+0.07}_{-0.12}$ $\times$ 10$^{24}$ cm$^{-2}$ from \borus\  and 0.59$^{+0.11}_{-0.09}$ $\times$ 10$^{24}$ cm$^{-2}$ using \ux. From spectral modeling with \myt\ and \borus\, $N_{\rm H,avg}$ estimated to be 1.26$^{+0.42}_{-0.27}$ $\times$ 10$^{24}$ cm$^{-2}$ and 1.35$^{+0.30}_{-0.53}$ $\times$ 10$^{24}$ cm$^{-2}$, suggesting a patchy torus scenario with a Compton-thin $N_{\rm H,los}$ and $N_{\rm H,avg}$ $\gtrsim$ 10$^{24}$ cm$^{-2}$ (see Table \ref{table-2}). Based on the IR--X-ray correlation, \cite{2015MNRAS.454..766A} placed the source in the CT class with log($N_{\rm{H,l.o.s}}$/cm$^{-2}$) = 24.45, whereas the ML algorithm in \cite{2023AandA...675A..65S} identified it as non-CT, finding log($N_{\rm{H,l.o.s}}$/cm$^{-2}$) = 23.60. The soft X-ray emission in the spectra of IC 2227 is modeled with a \texttt{mekal} component with $\rm{kT_{e}}$ $\sim$ 0.6 keV. 

\smallskip
\vspace{0.1cm}
\noindent {\textbf{ESO 362-8:}} This nearby ($z$ = 0.0150) Sy 2 AGN has been observed once with \nustar\ in 2021 and twice with \xmm in 2006 and 2021, respectively. We used all the observations to fit the source spectra using the three torus models. From the analysis of the \nustar\ and one \xmm\ observation (Obs ID - 0890440101), \cite{2022ApJ...940..148S} reported $N_{\rm H,los}$ = 2.78$^{+u}_{-0.65}$ $\times$ 10$^{24}$ cm$^{-2}$, 3.96$^{+u}_{-1.30}$ $\times$ 10$^{24}$ cm$^{-2}$ and 3.93$^{+u}_{-1.41}$ $\times$ 10$^{24}$ cm$^{-2}$ using \myt\ (decoupled Face-on), \borus\ and \ux\ respectively. The authors retrieved $N_{\rm H,avg}$ of 9.91$^{+u}_{-5.66}$ $\times$ 10$^{24}$ cm$^{-2}$ and 10.00$^{+0.15}_{-4.75}$ $\times$ 10$^{24}$ cm$^{-2}$, using \myt\ and \borus. Our analysis also revealed the CT nature of the source with both $N_{\rm H,los}$ and $N_{\rm H,avg}$ $\gtrsim$ 10$^{24}$ cm$^{-2}$ (see Table \ref{table-2}. The best-fitted parameters obtained from the three models are in agreement, and retrieved a relatively soft spectrum with $\Gamma$ $\sim$ 2.00. Using \myt, \borus, and \ux, we derived $N_{\rm H,los}$ = 6.74$^{+u}_{-6.20}$ $\times$ 10$^{24}$ cm$^{-2}$, 2.31$^{+1.69}_{-0.72}$ $\times$ 10$^{24}$ cm$^{-2}$ and 2.96$^{+2.88}_{-0.95}$ $\times$ 10$^{24}$ cm$^{-2}$, respectively. While \cite{2015MNRAS.454..766A} used IR--X-ray scaling relations to classify the source as CT with log($N_{\rm{H,l.o.s}}$/cm$^{-2}$) = 24.19, an analysis by \cite{2023AandA...675A..65S} assigned it to the non-CT category, estimating log($N_{\rm{H,l.o.s}}$/cm$^{-2}$) = 23.33. Fitting the soft X-ray emission with \texttt{mekal} produced a plasma temperature ($\rm{kT_{e}}$) $\sim$ 0.5 keV (see Table \ref{table-2}).

\smallskip
\vspace{0.1cm}
\noindent {\textbf{ESO 406-4:}}
This nearby Sy 2 AGN ($z$ = 0.0297), with log($M_{\rm{BH}}$/$M_{\odot}$) $\sim$ 8.02$\pm$0.02 \citep{2022ApJS..261....2K}, is a well-studied source. In this work, we make use of observations from \nustar, \xmm, and \cha. The dataset includes two \nustar\ observations taken in 2016 and 2020, one \xmm\ observation from 2021, and an earlier \cha\ observation from 2012. The obscuration properties of this source have been debated: some studies suggest a CT-AGN classification based on spectral modeling with the torus model of \cite{2011MNRAS.413.1206B}, yielding log($N_{\rm{H,l.o.s}}$/cm$^{-2}$) = 24.74$^{+u}_{-0.55}$ \citep{2015ApJ...815L..13R}, or based on spectral curvature estimates \citep{2016ApJ...825...85K}. More recent works, however, including \cite{2016A&A...594A..73A, 2022ApJS..260...30T, 2023A&A...676A.103S}, do not confirm the CT nature of the obscurer. Using the torus model of \cite{2011MNRAS.413.1206B}, and fitting the \xrt\ and \texttt{BAT} spectra, \cite{2016A&A...594A..73A} reported a Compton-thin obscurer with $N_{\rm H,los}$ = 0.32$^{+0.20}_{-0.11}$ $\times$ 10$^{24}$ cm$^{-2}$. Applying the \texttt{XCLUMPY} model to the Swift and \nustar\ data, \cite{2022ApJS..260...30T} reported $N_{\rm H,los}$ = 0.54$^{+5.82}_{-0.14}$ $\times$ 10$^{24}$ cm$^{-2}$. Similarly, \cite{2023A&A...676A.103S} reported $N_{\rm H,los}$ = 0.73$^{+0.27}_{-0.14}$ $\times$ 10$^{24}$ cm$^{-2}$ using \myt\ and 0.79$^{+0.04}_{-0.12}$ $\times$ 10$^{24}$ cm$^{-2}$ using \borus, based on the two \nustar\ and one \cha\ observations. In the present work, we include for the first time the \xmm\ observation in a joint fit with the three torus models. We find consistent results across \myt, \borus, and \ux, with line-of-sight column densities of 1.06$^{+0.87}_{-0.53}$ $\times$ 10$^{24}$ cm$^{-2}$, 1.34$^{+0.85}_{-0.40}$ $\times$ 10$^{24}$ cm$^{-2}$, and 1.16$^{+0.56}_{-0.35}$ $\times$ 10$^{24}$ cm$^{-2}$, respectively. Although the best-fit $N_{\rm H,los}$ values suggest a CT nature, the large uncertainties prevent a definitive confirmation. Likewise, the large error bars on $N_{\rm H,avg}$ hinder any firm conclusion on the clumpy nature of the torus, with $N_{\rm H,avg}$ estimated at 1.01$^{+8.98}_{-0.99}$ $\times$ 10$^{24}$ cm$^{-2}$ and 1.00$^{+u}_{-0.51}$ $\times$ 10$^{24}$ cm$^{-2}$ from the \myt\ and \borus\ fits, respectively. Based on the IR--X-ray correlation, \cite{2015MNRAS.454..766A} placed the source in the CT class with log($N_{\rm{H,l.o.s}}$/cm$^{-2}$) = 24.28; however, employing a ML technique, \cite{2023AandA...675A..65S} concluded that the source belongs to the non-CT class with log($N_{\rm{H,l.o.s}}$/cm$^{-2}$) = 23.360.

\smallskip
\vspace{0.1cm}
\noindent {\textbf{2MFGC 13496:}} This nearby ($z$ = 0.0407) emission line galaxy is observed by \nustar, \xmm\ and \cha\ between 2015 and 2019. The nature of the obscuration for this galaxy is studied for the first time in this work. Using the three models, we find that $N_{\rm H,los}$ $\lesssim$ 10$^{24}$ cm$^{-2}$, thus allowing us to flag this source as Compton-thin. In our fits, the estimated line-of-sight column densities spanned a narrow range: \myt\ indicated 0.73$^{+0.27}_{-0.19}$ $\times$ 10$^{24}$ cm$^{-2}$, \borus\ estimated 1.15$^{+0.34}_{-0.25}$ $\times$ 10$^{24}$ cm$^{-2}$, and \ux\ found 0.59$^{+0.20}_{-0.08}$ $\times$ 10$^{24}$ cm$^{-2}$.
However, the $N_{\rm H,avg}$ obtained from the \myt\ and \borus\ fits is even smaller, being 2.00$^{+5.00}_{-u}$ $\times$ 10$^{22}$ cm$^{-2}$ and 5.00$^{+15.00}_{-u}$ $\times$ 10$^{22}$ cm$^{-2}$, respectively, thus supporting a clumpy torus scenario. The IR--X-ray correlation presented in \cite{2015MNRAS.454..766A}, reported log($N_{\rm{H,l.o.s}}$/cm$^{-2}$) = 23.99, while \cite{2023AandA...675A..65S} estimated log($N_{\rm{H,l.o.s}}$/cm$^{-2}$) = 23.64 using their ML algorithm.

\smallskip
\vspace{0.1cm}
\noindent {\textbf{2MASX J03585442+1026033:}} This nearby Seyfert 2 galaxy ($z$ = 0.031) with log($M_{\rm{BH}}$/$M_{\odot}$) $\sim$ 7.75 \citep{2015A&A...579A..90H} has been observed by \nustar\ once in 2014, twice by \xmm\ in 2002 and 2003, and once by \cha\ in 2008. From an analysis of the two \xmm\ and one \cha\ observations using an absorbed power-law, \cite{2015A&A...579A..90H} classified the source as a Compton-thin AGN with $N_{\rm H,los} \sim$ 10$^{22}$ cm$^{-2}$. We analyzed all the observations and our analysis using all three models features the Compton-thin nature of the source with  $N_{\rm H,los}$ = 0.11$^{+0.01}_{-0.01}$ $\times$ 10$^{24}$ cm$^{-2}$ using \myt, $N_{\rm H,los}$ = 0.11$^{+0.08}_{-0.07}$ $\times$ 10$^{24}$ cm$^{-2}$ using \borus, and $N_{\rm H,los}$ = 0.11$^{+0.01}_{-0.01}$ $\times$ 10$^{24}$ cm$^{-2}$ from \ux\ fits. $N_{\rm H,avg}$ obtained from the \myt\ and \borus\ analyses has the similar set of values ($N_{\rm H,avg}$ = 0.40$^{+0.75}_{-0.35}$ $\times$ 10$^{24}$ cm$^{-2}$ and 0.11$^{+0.18}_{-0.07}$ $\times$ 10$^{24}$ cm$^{-2}$, respectively) as of $N_{\rm H,los}$, indicating an uniform density torus. Both the IR--X-ray correlation in \cite{2015MNRAS.454..766A} and the analysis of \cite{2023AandA...675A..65S} classified the source as non-CT, with 
log($N_{\rm{H,l.o.s}}$/cm$^{-2}$) = 23.06 and 23.25, respectively. The soft part of the spectra is modeled using a \texttt{mekal} component with $\rm{kT_{e}}$ $\sim$ 0.30 keV (see Table \ref{table-2}).  

\smallskip
\vspace{0.1cm}
\noindent{\textbf{M 58:}} This well studied LINER Type 1.9 AGN ($z$ = 0.005), with log($M_{\rm{BH}}$/$M_{\odot}$) $\sim$ 8.10,  has been observed once by \nustar\ in 2016 and twice by \xmm\ and \cha, respectively between 2000 and 2016. We have analyzed all the observations using three torus models. Previous analyses have suggested a very low line-of-sight column density. For instance, from the spectral analysis of the \nustar\ and one \xmm\ observation (ObsID: 0790840201) using \borus, \cite{2023MNRAS.524.4670J} reported log$N_{\rm H,l.o.s/cm^{-2}}$ = 20.00 (fixed) and log$N_{\rm H,avg/cm^{-2}}$ = 23.80$^{+0.17}_{-0.12}$. On the other hand, using an absorbed power-law with reflection component, \cite{2022MNRAS.510.5102O}, reported log$N_{\rm H,los/cm^{-2}}$ = 22.34$^{+0.09}_{-0.11}$. In our analysis, we similarly find evidence for a Compton-thin obscurer with $N_{\rm H,los}$ $\lesssim$ 10$^{22}$ cm$^{-2}$. Finally, we note that for this source we do not need an additional scattered component to fit the spectra, a result consistent with the almost unobscured nature of the source. 

\smallskip
\vspace{0.1cm}
\noindent{\textbf{3C 371:}} This is the only BL LAC source analyzed in this work. This source, located at $z$ = 0.0495, was observed once by \nustar\ in 2022, and once by \cha\ in 2000. The three torus models yielded broadly consistent results, with \myt\ giving $N_{\rm H,los}$ = 0.022$^{+0.006}_{-0.006}$ $\times$ 10$^{24}$ cm$^{-2}$, \borus\ yielding $N_{\rm H,los}$ = 0.034$^{+0.010}_{-0.018}$ $\times$ 10$^{24}$ cm$^{-2}$, and \ux\ estimated $N_{\rm H,los}$ = 0.025$^{+0.001}_{-0.010}$ $\times$ 10$^{24}$ cm$^{-2}$. the $N_{\rm H,avg}$ obtained from the \myt\ and \borus\ fits are 0.50$^{+0.010}_{-0.088}$ $\times$ 10$^{24}$ cm$^{-2}$ and 0.54$^{+0.010}_{-0.088}$ $\times$ 10$^{24}$ cm$^{-2}$, respectively. While the derived column densities differ--log$N_{\rm H,los/cm^{-2}}$ = 23.14 from \cite{2015MNRAS.454..766A} and 22.10 from \cite{2023AandA...675A..65S}-- both studies ultimately classified the source as Compton-thin. The soft part of the spectra is modeled using a \texttt{mekal} component with $\rm{kT_{e}}$ $\sim$ 0.35 keV (see Table \ref{table-2}).

\smallskip
\vspace{0.1cm}
\noindent{\textbf{IC 1198:}} IC 1198, a nearby ($z$ = 0.034) Sy 1 galaxy, is been studied for the first time in this work to characterize its torus properties. It has a black hole mass of log($M_{\rm{BH}}$/$M_{\odot}$) $\sim$ 7.51 \citep{2022ApJS..261....5M}.  This source is been observed once each by \nustar\ and \xmm\  in 2017 and 2022, respectively. The spectral fit with the three models, \myt,\ \borus,\ and \ux\ agrees well and features a Compton-thin nature of the source with $N_{\rm H,los}$ = 0.11$^{+0.01}_{-0.01}$ $\times$ 10$^{24}$ cm$^{-2}$ using \myt, 0.21$^{+0.02}_{-0.02}$ $\times$ 10$^{24}$ cm$^{-2}$ using \borus\ and 0.11$^{+0.01}_{-0.01}$ $\times$ 10$^{24}$ cm$^{-2}$ from the \ux\ fits. However, from the \myt\ and \borus\ analysis we obtain a much larger $N_{\rm H,avg}$ of $\gtrsim$ 10$^{24}$ cm$^{-2}$, indicating a clumpy nature of the torus in IC 1198. Consistently, \cite{2015MNRAS.454..766A} and \cite{2023AandA...675A..65S} both identified the source as Compton-thin with log$N_{\rm H,los/cm^{-2}}$ = 23.30 and 23.05, respectively.

\smallskip
\vspace{0.1cm}
\noindent{\textbf{2MASX J09261742-8421330:}} This nearby ($z$ = 0.0643) Sy 2 galaxy has been observed once by \nustar\ in 2015. The source has never been observed by either \xmm,\ or \cha. For the soft X-ray band (0.5-10 keV), we therefore used the simultaneous \xrt\ observation with \nustar.  This source has a black hole mass of log($M_{\rm{BH}}$/$M_{\odot}$) = 7.10$\pm$0.02 \citep{2022ApJS..261....5M}. From the broadband spectral analysis of the \swi\ observations \cite{2017ApJS..233...17R}, reported log$N_{\rm H,los/cm^{-2}}$ = 22.40$^{+0.08}_{-0.12}$. Consistently with previous results reported in \cite{2017ApJS..233...17R}, we also find a Compton-thin line of sight obscuration with $N_{\rm H,los}$ = 1.4$^{+1.0}_{-u}$ $\times$ 10$^{22}$ cm$^{-2}$ using \myt, 2.1$^{+3.4}_{-u}$ $\times$ 10$^{22}$ cm$^{-2}$ using \borus, and 1.3$^{+1.3}_{-u}$ $\times$ 10$^{22}$ cm$^{-2}$ from the \ux\ fits. From \myt\ and \borus\, we obtain a larger value of  $N_{\rm H,avg}$ $\sim$ 0.9 $\times$ 10$^{24}$ cm$^{-2}$, indicating a possible clumpy nature of the torus.  While the derived column densities differ-- log$N_{\rm H,los/cm^{-2}}$ = 23.33 from \cite{2015MNRAS.454..766A} and 22.38 from \cite{2023AandA...675A..65S}-- both studies ultimately classified the source as non-CT AGN.

\smallskip
\vspace{0.1cm}
\noindent{\textbf{UGC 12348:}} This nearby ($z$ = 0.0253) Seyfert 2 galaxy is analyzed for the first time in this study to explore the nature of its obscuration. It was observed simultaneously by \nustar\ and \xmm\ in 2014, and later by \cha\ in 2020. Our spectral analysis using all three datasets shows consistent values of $N_{\rm H,los}$ and $N_{\rm H,avg}$, suggesting a uniform torus structure. In our fits, the estimated line-of-sight column densities spanned a narrow range: \myt\ indicated $N_{\rm H,los}$ = 0.023$^{+0.002}_{-0.001}$ $\times$ 10$^{24}$ cm$^{-2}$, , \borus\ suggested $N_{\rm H,los}$ = 0.044$^{+0.018}_{-0.015}$ $\times$ 10$^{24}$ cm$^{-2}$, and \ux\ found $N_{\rm H,los}$ = 0.027$^{+0.027}_{-0.025}$ $\times$ 10$^{24}$ cm$^{-2}$. Therefore, all models agree on a Compton-thin obscuration ($N_{\rm H,los}$ $<$ 10$^{24}$ cm$^{-2}$). The IR--X-ray correlation analysis by \cite{2015MNRAS.454..766A} and the more recent study by \cite{2023AandA...675A..65S} arrived at the same classification of the source as Compton-thin, with estimated column densities of log$N_{\rm H,los/cm^{-2}}$ = 23.35 and 22.71, respectively. To model the soft X-ray emission, we included a \texttt{mekal} component with $\rm{kT_{e}}$ $\sim$ 0.70 keV.

\smallskip
\vspace{0.1cm}
\noindent{\textbf{2MASX J02420381+0510061:}} This nearby Seyfert 2 AGN ($z = 0.073$) was observed simultaneously with \nustar\ and \xmm\ in 2017, and later by \cha\ in 2022. Previously \cite{2017ApJS..233...17R} reported $N_{\rm H,los} = 3.16^{+2.46}_{-0.65} \times 10^{23}$ $\rm{cm^{-2}}$ using an absorbed cut-off power-law model. In our analysis, all three torus models yield a Compton-thin line-of-sight column density around $10^{23}$ $\rm{cm^{-2}}$, with $N_{\rm H,los}$ = 8.1$^{+1.5}_{-1.4}$ $\times$ 10$^{23}$ cm$^{-2}$. from \myt, $N_{\rm H,los}$ = 12.9$^{+2.5}_{-2.2}$ $\times$ 10$^{23}$ cm$^{-2}$ from \borus, and $N_{\rm H,los}$ = 9.8$^{+6.0}_{-2.5}$ $\times$ 10$^{24}$ cm$^{-2}$ from \ux. Despite yielding different numerical estimates-- log$N_{\rm H,los/cm^{-2}}$ = 23.45 in \cite{2015MNRAS.454..766A} and 22.39 in \cite{2023AandA...675A..65S}-- both analyses classified the source as non-CT candidate.

\smallskip
\vspace{0.1cm}
\noindent{\textbf{ESO 234-50:}} This Seyfert 2 AGN ($z = 0.009$) was observed by \nustar\ in May 2021 and by \cha\ in March 2021. The black hole mass is estimated to be log($M_{\rm{BH}}$/$M_{\odot}$) $\sim 6.04\pm0.07$ \cite{2022ApJS..261....2K}. From the broad-band spectral analysis of the Swift observations, \cite{2017ApJS..233...17R} reported $N_{\rm H,los} = 1.20^{+0.38}_{-0.41} \times 10^{23}$ $\rm{cm^{-2}}$. Our results for $N_{\rm H,los}$ - 0.22$^{+0.05}_{-0.05}$ $\times$ 10$^{24}$ cm$^{-2}$ from \myt, 0.32$^{+0.07}_{-0.05}$ $\times$ 10$^{24}$ cm$^{-2}$ from \borus, and 0.20$^{+0.02}_{-0.03}$ $\times$ 10$^{24}$ cm$^{-2}$ from \ux- are in agreement with this value. The IR--X-ray correlation analysis \citep{2015MNRAS.454..766A} and the more recent study by \cite{2023AandA...675A..65S} arrived at the same classification of the source as Compton-thin, with estimated column densities of
log$N_{\rm H,los/cm^{-2}}$ = 23.50 and 23.21, respectively.

\smallskip
\vspace{0.1cm}
\noindent{\textbf{NGC 2273:}} This nearby Seyfert 2 AGN ($z = 0.0061$) has multiple archival observations: one by \nustar\ in 2014, three by \xmm\ in 2003, and one by \cha\ in 2017. We used all of them in our joint spectral analysis. The black hole mass is log($M_{\rm{BH}}$/$M_{odot}$) $\sim 8.22\pm0.03$ \citep{2022ApJS..261....2K}. A \nustar-only analysis by \cite{2022MNRAS.510.5102O}, using a partial covering continuum model with a reflection component, classified the source as Compton-thick with $N_{\rm H,los} \sim 2 \times 10^{25}$ $\rm{cm^{-2}}$. \cite{2022MNRAS.510.4909W} reported a cold absorber with $N_{\rm H} < 0.16 \times 10^{22}$ $\rm{cm^{-2}}$ and a warm absorber with $N_{\rm H} = 15.69^{+5.74}_{-8.36} \times 10^{22}$ $\rm{cm^{-2}}$. In our fits, the estimated line-of-sight column densities spanned a narrow range: \myt\ indicated $N_{\rm H,los}$ = 0.37$^{+0.11}_{-0.06}$ $\times$ 10$^{24}$ cm$^{-2}$, \borus\ suggested $N_{\rm H,los}$ = 0.31$^{+0.11}_{-0.13}$ $\times$ 10$^{24}$ cm$^{-2}$, and \ux\ found $N_{\rm H,los}$ = 0.35$^{+0.05}_{-0.07}$ $\times$ 10$^{24}$ cm$^{-2}$. To investigate the discrepancy with \cite{2022MNRAS.510.5102O}, we fitted the \nustar\ spectrum from 2014 and the \xmm\ spectra from 2003 separately with the \ux\ model to check for possible variability in $N_{\rm H,los}$. The \cha\ data did not contain enough counts to allow a separate analysis. The individual \nustar\ and \xmm\ fits with \ux\ could not constrain most spectral parameters, including $N_{\rm H,los}$. The \nustar-only fit gave $N_{\rm H,los} = 5.02^{+u}_{-4.65} \times 10^{25}$ cm$^{-2}$, while the \xmm-only fit yielded $N_{\rm H,los} = 4.47^{+u}_{-2.68} \times 10^{24}$ cm$^{-2}$. Although $N_{\rm H,los}$ is poorly constrained with the \ux\ model and the lower limit prevents us from drawing firm conclusions about variability between observations, the discrepant values nonetheless make NGC 2273 a promising candidate for future studies of line-of-sight absorption variability with improved data quality. Based on the IR--X-ray correlation, \cite{2015MNRAS.454..766A} placed the source in the CT class with log$N_{\rm H,los/cm^{-2}}$ = 24.18, a later analysis by \cite{2023AandA...675A..65S}, estimated log$N_{\rm H,los/cm^{-2}}$ = 23.83. The soft X-ray component of NGC 2273 was modeled using a \texttt{mekal} component with $kT_e \sim 0.30$ keV (see Table \ref{table-2}). 

\smallskip
\vspace{0.1cm}
\noindent{\textbf{FRL 265:}} This type 1 Seyfert galaxy ($z$ = 0.0295) was observed by \nustar\ in 2019. We did not use the 2022 \xmm\ data due to contamination from extended emission. This is the first study of the circumnuclear matter in this source. The three torus models yielded broadly consistent results, with \myt\ producing $N_{\rm H,los}$ = 0.010$^{+0.004}_{-u}$ $\times$ 10$^{24}$ cm$^{-2}$, \borus\ yielding a upper-limit of $N_{\rm H,los} <$ 0.06  $\times$ 10$^{24}$ cm$^{-2}$, and \ux\ estimating $N_{\rm H,los}$ = 0.01$^{+0.23}_{-0.06}$ $\times$ 10$^{24}$ cm$^{-2}$. Using \myt\ and \borus, we obtained $N_{\rm H,avg}$ = 2.00$^{+2.95}_{-0.99}$ $\times$ 10$^{24}$ cm$^{-2}$ and 1.51$^{+0.34}_{-0.31}$ $\times$ 10$^{24}$ cm$^{-2}$, respectively. A clear difference between $N_{\rm H,los}$ and $N_{\rm H,avg}$ (from \myt\ and \borus) suggests we observe the AGN through a less dense region, supporting a clumpy-torus model. 

\smallskip
\vspace{0.1cm}
\noindent{\textbf{MRK 231:}} Located at $z = 0.042$, this Seyfert 1 galaxy is known for strong winds and outflows \citep{2014ApJ...788..123L,2024ApJS..274....8Y}. It has been observed multiple times by \nustar\ and \xmm\ (see Table \ref{table-1}). To avoid the complexity of the joint fit due to variable nature of soft X-ray source spectrum below 10 keV \citep{2004A&A...420...79B}, we used only \nustar\ data for spectral analysis. However, results from joint \nustar\ and \xmm\ fits are consistent. Analysis of contemporaneous \cha\ and \nustar\ broadband spectra \cite{2014ApJ...785...19T} revealed strong X-ray emission associated with a powerful circumnuclear starburst. The direct AGN emission is absorbed and scattered by a patchy, Compton-thin torus with $N_{\rm H} \sim 1.2^{+0.3}_{-0.3} \times 10^{23}$ cm$^{-2}$. In contrast, \cite{2004A&A...420...79B} concluded, based on their modeling of the 0.5-50 keV broadband spectrum, that MRK 231 is a CT quasar with a column density of $\sim 1.8$-$2.6 \times 10^{24}$ cm$^{-2}$. However, they also suggested that the variability observed below 10 keV is consistent with changes in the partial covering fraction of the Compton-thick absorber. Our analysis of only \nustar\ observations agrees well with the recent findings of \cite{2014ApJ...785...19T} with $N_{\rm H,los}$ = 0.06$^{+0.03}_{-0.02}$ $\times$ 10$^{24}$ cm$^{-2}$ from \myt, 0.10$^{+0.02}_{-0.02}$ $\times$ 10$^{24}$ cm$^{-2}$ from \borus, and 0.07$^{+0.02}_{-0.02}$ $\times$ 10$^{24}$ cm$^{-2}$ from \ux. However, The IR--X-ray correlation analysis (\cite{2015MNRAS.454..766A}) and the more recent study by \cite{2023AandA...675A..65S} arrived at the same classification of the source as CT-AGN, with estimated column densities of log$N_{\rm H,los/cm^{-2}}$ = 24.69 and log$N_{\rm H,los/cm^{-2}}$ = 24.34, respectively.

\smallskip
\vspace{0.1cm}
\noindent{\textbf{PG 1211+143:}} This bright, nearby ($z = 0.081$) narrow-line Seyfert 1 galaxy has multiple observations from \nustar\ and \xmm, that were taken in 2014. \cite{2018ApJ...854...28R} reported variable soft X-ray outflows using \xmm\ data, so we excluded the \xmm\ observations from our analysis. Reverberation mapping results from \cite{2004ApJ...613..682P} and \cite{2020A&A...640A..39C} estimate log($M_{\rm{BH}}$/$M_{\odot}$) $\sim 8.6$. The three torus models yielded broadly consistent results, with \myt\ giving $N_{\rm H,los}$ = 0.04$^{+0.01}_{-0.01}$ $\times$ 10$^{24}$ cm$^{-2}$, \borus\ yielding $N_{\rm H,los}$ = 0.04$^{+0.01}_{-0.02}$ $\times$ 10$^{24}$ cm$^{-2}$, and \ux\ estimating $N_{\rm H,los}$ = 0.14$^{+0.05}_{-0.04}$ $\times$ 10$^{24}$ cm$^{-2}$ from \ux. While \cite{2015MNRAS.454..766A} used IR--X-ray scaling relations to classify the source as Compton-thin with log$N_{\rm H,los/cm^{-2}}$ = 23.25, a later analysis by \cite{2023AandA...675A..65S} estimating a lower value for log$N_{\rm H,los/cm^{-2}}$ = 21.99.

\smallskip
\vspace{0.1cm}
\noindent{\textbf{MRK 376:}} MRK 376 is a nearby ($z = 0.056$) Seyfert 1.5 galaxy with a black hole mass log($M_{\rm{BH}}$/$M_{\odot}$) $\sim 8.22$ \citep{2022ApJS..261....2K}. This source was observed by \nustar\ and \xrt\ in 2015. \cite{2017ApJS..233...17R} reported a Compton-thin $N_{\rm H,los} \sim 8.71^{+5.41}_{-3.47} \times 10^{22}$ $\rm{cm^{-2}}$, consistent with our findings. We also estimate a Compton-thick $N_{\rm H,avg} \sim 2.00 \times 10^{24}$ $\rm{cm^{-2}}$, suggesting a patchy torus. Both the IR--X-ray correlation in \cite{2015MNRAS.454..766A} and the analysis of \cite{2023AandA...675A..65S} classified the source as Compton-thin, with log$N_{\rm H,los/cm^{-2}}$ = 23.51 and 22.48, respectively. The soft X-ray emission is modeled with a \texttt{mekal} component ($kT_e \sim 0.2$ keV).

\smallskip
\vspace{0.1cm}
\noindent{\textbf{NGC 7378:}} This type 2 AGN at $z = 0.009$ was observed by \nustar\ in December, 2018. For the soft X-ray analysis (0.5--10 keV), we used the \xrt\ observations taken quasi-simultaneously with \nustar\ in between December and January of 2018. \cite{2022ApJS..261....2K} estimated log($M_{\rm{BH}}$/$M_{\odot}$) $= 5.49\pm0.16$, making this an AGN powered by an intermediate mass black hole . We find $N_{\rm H,los} \sim 10^{23}$ $\rm{cm^{-2}}$ and $N_{\rm H,avg} \sim 3 \times 10^{24}$ $\rm{cm^{-2}}$, indicating a patchy torus. No previous $N_{\rm H}$ measurements are reported in the literature for this target. Consistently, \cite{2015MNRAS.454..766A} and \cite{2023AandA...675A..65S} both identified the source as Compton-thin AGN, with log$N_{\rm H,los/cm^{-2}}$ = 23.18 and 22.85, respectively.

\smallskip
\noindent{\textbf{2MASX J11462959+7421289:}} This nearby Seyfert 2 AGN ($z = 0.058$) was observed quasi-simultaneously by \nustar\ and \xrt\ in 2013. \cite{2022ApJS..261....2K} reported log($M_{\rm{BH}}$/$M_{\odot}$) $= 8.22\pm0.03$. No previous $N_{\rm H}$ measurements are available. Our analysis suggests a quasi--unobscured nature for this target, with $N_{\rm H,los} \sim 10^{22}$ $\rm{cm^{-2}}$. From spectral modeling with \myt, \borus, and \ux, the line-of-sight column densities were estimated to be 0.010$^{+0.003}_{-u}$ $\times$ 10$^{24}$ cm$^{-2}$, 0.013$^{+0.060}_{-0.010}$ $\times$ 10$^{24}$ cm$^{-2}$, and 0.045$^{+0.007}_{-0.010}$ $\times$ 10$^{24}$ cm$^{-2}$, respectively. The IR--X-ray correlation in \cite{2015MNRAS.454..766A} classified the source as Compton-thin AGN with log$N_{\rm H,los/cm^{-2}}$ = 23.05, while \cite{2023AandA...675A..65S} also reported it as Compton-thin with a lower log$N_{\rm H,los/cm^{-2}}$ = 21.89.

\smallskip
\noindent{\textbf{SWIFT J2006.5+5619:}} This Seyfert 2 AGN ($z = 0.044$) was observed by \nustar\  and by \xrt\ in 2014. Its black hole mass is log($M_{\rm{BH}}$/$M_{\odot}$) $\sim 7.07\pm0.11$ \cite{2022ApJS..261....2K}. Previous analyses by \cite{2017ApJS..233...17R} and \cite{2021A&A...653A.162P} found $N_{\rm H,los}$ values around $1.8$--$3.0 \times 10^{23}$ $\rm{cm^{-2}}$, consistent with our result of $N_{\rm H,los}$ as estimated to be 0.15$^{+0.04}_{-0.03}$ $\times$ 10$^{24}$ cm$^{-2}$, 0.31$^{+0.09}_{-0.11}$ $\times$ 10$^{24}$ cm$^{-2}$, and 0.19$^{+0.04}_{-0.05}$ $\times$ 10$^{24}$ cm$^{-2}$, respectively, using \myt, \borus\ and \ux. Consistently, \cite{2015MNRAS.454..766A} and \cite{2023AandA...675A..65S} both identified the source as Compton-thin AGN, with log$N_{\rm H,los/cm^{-2}}$ = 23.08 and 23.06, respectively.

\smallskip
\noindent{\textbf{2MASX J06363227-2034532:}} This Seyfert 2 AGN at $z = 0.056$ was observed quasi-simultaneously by \nustar\ and \xrt\ in Spetember, 2023. The black hole mass is log($M_{\rm{BH}}$/$M\_{\odot}$) $\sim 8.44\pm0.15$ \cite{2022ApJS..261....2K}. All three models indicate a Compton-thick nature with $N_{\rm H,los} \sim 1.0 \times 10^{24}$ $\rm{cm^{-2}}$ (Table \ref{table-2}), matching \cite{1999ApJ...526...60S} who found $N_{\rm H,los} = 7.9^{+12.9}_{-4.3} \times 10^{23}$ $\rm{cm^{-2}}$ from \textit{ASCA} data. The soft X-rays are modeled using a \texttt{mekal} component ($kT_e \sim 0.20$ keV). According to the IR--X-ray correlation presented in \cite{2015MNRAS.454..766A}, the source was identified as CT-AGN with log$N_{\rm H,los/cm^{-2}}$ = 24.15, in contrast to the findings of \cite{2023AandA...675A..65S}, which classified it as Compton-thin with log$N_{\rm H,los/cm^{-2}}$ = 23.34.

\smallskip
\noindent{\textbf{2MASX J09034285-7414170:}} \nustar\ observed this nearby ($z = 0.093$) Seyfert 2 AGN in 2015, while in the soft band the source has been targeted multiple times by \xrt\ in 2015. The black hole mass is log($M_{\rm{BH}}$/$M_{\odot}$) $= 7.88\pm0.17$ \cite{2022ApJS..261....2K}. In our fits, the estimated line-of-sight column densities spanned a narrow range: \myt\ indicated $N_{\rm H,los}$ = 0.40$^{+0.22}_{-0.21}$ $\times$ 10$^{24}$ cm$^{-2}$, \borus\ suggested $N_{\rm H,los}$ = 0.49$^{+0.21}_{-0.17}$ $\times$ 10$^{24}$ cm$^{-2}$, and \ux\ estimated $N_{\rm H,los}$ = 0.34$^{+0.17}_{-0.16}$ $\times$ 10$^{24}$ cm$^{-2}$. These findings are consistent with \cite{2017ApJS..233...17R} who reported $N_{\rm H,los} \sim 3.89 \times 10^{23}$ $\rm{cm^{-2}}$ for this source. Our results suggest a patchy torus with $N_{\rm H,los} \sim 4.0 \times 10^{23}$ $\rm{cm^{-2}}$ and $N_{\rm H,avg} \sim 2.0 \times 10^{24}$ $\rm{cm^{-2}}$ from modeling the source spectra with \myt\ and \borus. Consistently, \cite{2015MNRAS.454..766A} and \cite{2023AandA...675A..65S} both identified the source as Compton-thin AGN, with log$N_{\rm H,los/cm^{-2}}$ = 23.25 and 23.07, respectively.

\smallskip
\noindent{\textbf{2MASX J00091156-0036551:}} This Seyfert 2 AGN ($z = 0.073$) was observed by \nustar\ in 2015. \xrt\ observed this source simultaneously with \nustar\ once in 2015.  The black hole mass is log($M_{\rm{BH}}$/$M_{\odot}$) $= 8.54\pm0.11$ \cite{2022ApJS..261....2K}. Our spectral analysis gives $N_{\rm H,los} \sim 3.0 \times 10^{23}$ $\rm{cm^{-2}}$, consistent with \cite{2017ApJS..233...17R} who found $N_{\rm H,los} \sim 4.07 \times 10^{23}$ $\rm{cm^{-2}}$. The classification of this source is consistent across studies: \cite{2015MNRAS.454..766A} inferred log$N_{\rm H,los/cm^{-2}}$ = 23.37, whereas \cite{2023AandA...675A..65S} identified it as Compton-thin with log$N_{\rm H,los/cm^{-2}}$ = 23.71.

\clearpage
\begin{center}
\startlongtable
\begin{deluxetable*}{cccccc}
\tablecaption{Summary of the X-ray observations analyzed in this work, for the 26 sources in our sample.  The columns are (1) serial number of source, (2) source name, (3) telescopes used for observation, (4) observation ID, (5) exposure time in sec., and (6) date of observation.}\label{table-1}
\tablehead{Index & Name & Telescope & Obs ID & Exposure (s) & Obs Date}
\startdata
1 & MRK 1073 & {\it NuSTAR} & 60001161002 & 22199 & 2014-10-08 \\
&& {\it XMM-Newton} & 0002942401 & 7868 & 2002-01-28 \\
&&& 0862510301 & 14900 & 2020-08-20 \\
&&& 0862510701 & 13800 & 2020-08-31 \\
2 & UGC 5101 & {\it NuSTAR} & 60001068004 & 23214 & 2020-04-17 \\
&& {\it NuSTAR} & 60001068002 & 18291 & 2014-03-21 \\
&& {\it Chandra} & 2033 & 49930 & 2001-05-28 \\
3 & NGC 7674 & {\it NuSTAR} & 60001151002 & 51997 & 2014-09-30 \\
&& {\it XMM-Newton} & 0200660101 & 10420 & 2004-06-02 \\
&& {\it Chandra} & 23715 & 2090 & 2020-09-29 \\
4 & IC 2227 & \textit{NuSTAR} & 60701049002 & 52172 & 2022-03-28 \\
&& \xmm\ & 0890440201 & 38000 & 2022-03-27 \\
&& \textit{Chandra} & 23723 & 4380 & 2021-04-10 \\
5 & ESO 362$-$8 & \textit{NuSTAR} & 60701048002 & 48531 & 2021-10-05 \\
&& \xmm\ & 0890440101 & 38000 & 2021-10-05 \\
&&& 0307001401 & 18308 & 2006-02-13 \\
6 & ESO 406-4 & \textit{NuSTAR} & 60201039002 & 36317 & 2016-05-25 \\
&&& 60161799002 & 23719 & 2020-06-26 \\
&& {\it XMM-Newton} & 0883210401 & 24000 & 2021-11-21 \\
&& {\it Chandra} & 14050 & 5090 & 2012-06-07 \\
7 & 2MFGC 13496 & \textit{NuSTAR} & 60001145002 & 22866 & 2015-03-03 \\
&& {\it XMM-Newton} & 0743010301 & 22800 & 2015-03-04 \\
&& \textit{Chandra} & 22059 & 10090 & 2019-01-27 \\
8 & 2MASX J03585442+1026033 & \textit{NuSTAR} & 60061042002 & 27331 & 2014-09-25 \\
&& {\it XMM-Newton} & 0064600301 & 11914 & 2003-02-05 \\
&&& 0064600101 & 28904 & 2002-09-07 \\
&& \textit{Chandra} & 10234 & 32130 & 2008-12-24 \\
9 & M 58 & \textit{NuSTAR} & 60201051002 & 117843 & 2016-12-06 \\
&& \xmm\ & 0790840201 & 23002 & 2016-12-06 \\
&&& 0112840101 & 23669 & 2003-06-12 \\
&& \textit{Chandra} & 406 & 2940 & 2000-02-23 \\
&&& 807 & 35640 & 2000-05-02 \\
&&& 9558 & 49450 & 2008-02-29 \\
10 & 3C 371 & \textit{NuSTAR} & 60801044002 & 17981 & 2022-06-12 \\
&& \textit{Chandra} & 841 & 10250 & 2000-03-21 \\
11 & IC 1198 & \textit{NuSTAR} & 60361014002 & 26973 & 2017-05-07 \\
&& \xmm\ & 0903041301 & 32200 & 2022-08-29 \\
12 & 2MASX J09261742-8421330 & \textit{NuSTAR} & 60160360002 & 35127 & 2015-11-30 \\
&& \xrt\ & 00081046001 & 5500 & 2015-11-30 \\
13 & UGC 12348 & \textit{NuSTAR} & 60001147002 & 26702 & 2014-12-09 \\
&& \xmm\ & 0743010501 & 35700 & 2014-12-09 \\
&& \textit{Chandra} & 23719 & 2090 & 2020-11-25 \\
14 & 2MASX J02420381+0510061 & \textit{NuSTAR} & 60363003002 & 19800 & 2017-07-13 \\
&& \xmm\ & 0802200401 & 23000 & 2017-07-13 \\
15 & ESO 234-50 & \textit{NuSTAR} & 60760005002 & 32271 & 2021-05-22 \\
&& \textit{Chandra} & 23814 & 20060 & 2021-03-14 \\
16 & NGC 2273 & {\it NuSTAR} & 60001064002 & 23230 & 2014-03-23 \\
&& {\it XMM-Newton} & 0140950701 & 12551 & 2003-03-17 \\
&&& 0140951001 & 13016 & 2003-09-05 \\
&&& 0140950901 & 7857 & 2003-04-20 \\
&& {\it Chandra} & 19377 & 10070 & 2017-08-17 \\
17 & FRL 265 & \textit{NuSTAR} & 60360009002 & 29420 & 2019-06-10 \\
&& \xmm\ & 0903040501 & 23000 & 2022-04-23 \\
18 & MRK 231 & \textit{NuSTAR} & 80302608002 & 82057 & 2017-10-19 \\
&&& 60002025002 & 41061 & 2012-08-26 \\
&&& 90102001002 & 31787 & 2015-04-02 \\
&&& 90102001006 & 30521 & 2015-05-28 \\
&&& 60002025004 & 28555 & 2013-05-09 \\
&&& 90102001004 & 28379 & 2015-04-19 \\
19 & PG 1211+143 & \textit{NuSTAR} & 60001100002 & 111437 & 2014-02-18 \\
&&& 91001637002 & 83886 & 2024-12-01 \\
&&& 6000110000 & 74886 & 2014-07-07 \\
&&& 60001100005 & 64429 & 2014-04-09 \\
&&& 60001100004 & 48948 & 2014-04-08 \\
20 & MRK 376 & \textit{NuSTAR} & 60160288002 & 24174 & 2015-04-07 \\
&& \xrt\ & 00081006001 & 2300 & 2015-04-07 \\
&&& 00081006002 & 3200 & 2015-04-08 \\
21 & NGC 7378 & \textit{NuSTAR} & 60464202002 & 20301 & 2018-12-22 \\
&& \xrt\ & 00087196008 & 3700 & 2018-01-01 \\
&&& 00087196009 & 1300 & 2018-01-11 \\
&&& 00088600001 & 3300 & 2018-12-22 \\
&&& 00088600002 & 2800 & 2018-22-23 \\
22 & 2MASX J11462959+7421289 & \textit{NuSTAR} & 60061214002 & 22831 & 2013-12-10 \\
&& \xrt\ & 00080060001 & 5900 & 2013-12-07 \\
23 & SWIFT J2006.5+5619 & \textit{NuSTAR} & 60061355002 & 21370 & 2014-06-30 \\
&& \xrt\ & 00080690001 & 5400 & 2014-07-01 \\
24 & 2MASX J06363227-2034532 & \textit{NuSTAR} & 60061069002 & 21436 & 2023-09-06 \\
&& \xrt\ & 00080375001 & 5900 & 2023-09-08 \\
25 & 2MASX J09034285-7414170 & \textit{NuSTAR} & 60160346002 & 17038 & 2015-11-26 \\
&& \xrt\ & 00081040001 & 820 & 2015-11-26 \\
&&& 00081040002 & 3600 & 2015-11-27 \\
&&& 00081040003 & 1600 & 2015-12-02 \\
26 & 2MASX J00091156-0036551 & \textit{NuSTAR} & 60061002002 & 23533 & 2015-08-01 \\
&& \xrt\ & 00080001001 & 4100 & 2015-08-01 \\
\hline
\enddata
\end{deluxetable*} 
\end{center}

\clearpage

\begin{center}
\startlongtable
\begin{deluxetable*}{cccc}
\tablecaption{Best-fit parameters as obtained from the spectral analysis of the 26 sources using \myt, \borus\ and \ux. The details of the fit parameters are given at the end of the table.}\label{table-2}
\tablehead{Parameter & \myt\ & \borus\ & \ux}
\startdata
\hline
\multicolumn{4}{c}{\textbf{MRK 1073}}\\
\hline
$\chi^2$/dof & 113/88 & 114/90 & 110/86  \\
kT & 0.66$^{+0.08}_{-0.05}$ & 0.66$^{+0.08}_{-0.08}$ & 0.66$^{+0.04}_{-0.04}$  \\
$\Gamma$ & 2.44$^{+u}_{-0.32}$ & 2.50$^{+u}_{-0.32}$ & 1.93$^{+0.36}_{-0.36}$  \\
norm $\times$ $10^{-2}$ & 0.11$^{+0.14}_{-0.07}$ & 3.50$^{+7.15}_{-2.04}$ & 0.44$^{+1.10}_{-0.35}$ \\
$c_f$ & ... & 0.65$^{+0.04}_{-0.06}$ & ... \\
CTKcover & ... & ... & 0.60$^{+u}_{-0.19}$ \\
TORsigma & ... & ... & 27.83$^{+26.95}_{-}$ \\
$\rm{N_{H,los}}$ & 2.44$^{+u}_{-0.78}$ & 5.88$^{+u}_{-2.41}$ & 3.16$^{+7.84}_{-1.04}$ \\
$\rm{N_{H,avg}}$ & 5.99$^{+u}_{-0.37}$ & 5.75$^{+u}_{-1.95}$ & ...  \\
$f_s$ $\times$ 10$^{-2}$ & 2.35$^{+4.77}_{-1.27}$ & 0.15$^{+0.98}_{-0.12}$ &  4.66$^{+9.66}_{-2.65}$ \\
$A_{90}$ & 1.00* & ... & ... \\
$A_{0}$ & 1.00* & ... & ... \\
c$_{nus}$ & 1.14$^{+0.29}_{-0.23}$ & 1.17$^{+0.28}_{-0.23}$ & 1.15$^{+0.28}_{-0.23}$  \\
c$_{xmm1}$ & 0.92$^{+0.28}_{-0.40}$ & 0.81$^{+0.48}_{-0.38}$ & 0.90$^{+0.38}_{-0.31}$  \\
c$_{xmm2}$ & 0.86$^{+0.51}_{-0.38}$ & 0.76$^{+0.45}_{-0.36}$ & 0.85$^{+0.37}_{-0.29}$  \\
c$_{xmm3}$ & 0.78$^{+0.48}_{-0.35}$ & 0.68$^{+0.41}_{-0.32}$ & 0.77$^{+0.34}_{-0.28}$  \\
\hline
F$\rm_{2-10\,keV}$ & 1.08$^{+0.14}_{-0.14}$ $\times$ 10$^{-13}$ & 0.95$^{+0.11}_{-0.11}$ $\times$ 10$^{-13}$ & 0.92$^{+0.12}_{-0.12}$ $\times$ 10$^{-13}$  \\
L$\rm_{2-10\,keV}$ & 5.35$^{+3.17}_{-3.17}$ $\times$ 10$^{42}$ & 5.88$^{+17.31}_{-u}$ $\times$ 10$^{43}$ & 1.53$^{+2.54}_{-1.32}$ $\times$ 10$^{43}$ \\
\hline
\multicolumn{4}{c}{\textbf{UGC 5101}}\\
\hline
$\chi^2$/dof & 78/74 & 70/75 & 78/73  \\
kT & ... & ... & ...  \\
$\Gamma$ & 1.53$^{+0.25}_{-u}$ & 1.60$^{+0.23}_{-u}$ & 1.55$^{+0.23}_{-0.19}$  \\
norm $\times$ $10^{-2}$ & 0.05$^{+0.07}_{-0.02}$ & 0.08$^{+0.18}_{-0.05}$ & 0.11$^{+0.10}_{-0.04}$ \\
$c_f$ & ... & 0.55$^{+u}_{-0.06}$ & ... \\
CTKcover & ... & ... & 0.05$^{+0.47}_{-u}$ \\
TORsigma & ... & ... & 10.62$^{+16.59}_{-u}$ \\
$\rm{N_{H,los}}$ & 0.80$^{+0.38}_{-0.16}$ & 1.35$^{+0.45}_{-0.48}$ & 1.17$^{+0.27}_{-0.34}$ \\
$\rm{N_{H,avg}}$ & 2.43$^{+u}_{-1.27}$ & 2.95$^{+u}_{-1.95}$ & ...  \\
$f_s$ $\times$ 10$^{-2}$ & 6.55$^{+7.13}_{-4.16}$ & 3.15$^{+4.36}_{-1.97}$ &  5.00* \\
$A_{90}$ & 1.00* & ... & ... \\
$A_{0}$ & 1.00* & ... & ... \\
c$_{nus1}$ & 0.88$^{+0.18}_{-0.15}$ & 0.88$^{+0.17}_{-0.15}$ & 1.00$^{+0.19}_{-0.16}$  \\
c$_{nus2}$ & 0.96$^{+0.17}_{-0.14}$ & 0.96$^{+0.17}_{-0.15}$ & 0.94$^{+0.17}_{-0.14}$  \\
c$_{nus2}$ & 0.95$^{+0.17}_{-0.14}$ & 0.96$^{+0.17}_{-0.14}$ & 0.93$^{+0.17}_{-0.14}$  \\
c$_{chan}$ & 0.32$^{+0.15}_{-0.11}$ & 0.34$^{+0.15}_{-0.07}$ & 0.34$^{+0.17}_{-0.14}$  \\
\hline
F$\rm_{2-10\,keV}$ & 2.50$^{+0.17}_{-0.17}$ $\times$ 10$^{-13}$ & 2.67$^{+0.18}_{-0.18}$ $\times$ 10$^{-13}$ & 2.36$^{+0.16}_{-0.16}$ $\times$ 10$^{-13}$  \\
L$\rm_{2-10\,keV}$ & 8.03$^{+0.95}_{-0.95}$ $\times$ 10$^{42}$ & 1.41$^{+0.16}_{-0.16}$ $\times$ 10$^{43}$ & 2.01$^{+0.10}_{-0.08}$ $\times$ 10$^{43}$ \\
\hline
\multicolumn{4}{c}{\textbf{NGC 7674}}\\
\hline
$\chi^2$/dof & 190/132 & 218/133 & 213/133  \\
kT & 0.32$^{+0.10}_{-0.04}$ & 0.20$^{+0.07}_{-0.02}$ & 0.24$^{+0.03}_{-0.02}$  \\
$\Gamma$ & 1.70$^{+0.11}_{-0.11}$ & 1.90$^{+0.26}_{-0.10}$ & 2.10$^{+0.09}_{-0.16}$  \\
norm $\times$ $10^{-2}$ & 0.02$^{+0.07}_{-0.02}$ & 0.10$^{+0.18}_{-0.05}$ & 0.18$^{+0.11}_{-0.09}$ \\
$c_f$ & ... & 1.00(p) & ... \\
CTKcover & ... & ... & 0.60$^{+u}_{-0.03}$ \\
TORsigma & ... & ... & 84.00$^{+u}_{-3.16}$ \\
$\rm{N_{H,los}}$ & 0.18$^{+0.38}_{-0.16}$ & 0.77$^{+0.45}_{-0.48}$ & 0.49$^{+0.12}_{-0.14}$ \\
$\rm{N_{H,avg}}$ & 1.93$^{+1.32}_{-U}$ & 2.34$^{+1.37}_{-0.69}$ & ...  \\
$f_s$ $\times$ 10$^{-2}$ & 23.68$^{+7.13}_{-4.16}$ & 3.37$^{+4.36}_{-1.97}$ &  238.99$^{+201.76}_{-92.78}$ \\
$A_{90}$ & $<$11.32 & ... & ... \\
$A_{0}$ & 7.41$^{+4.39}_{-3.95}$ & ... & ... \\
c$_{nus1}$ & 0.97$^{+0.18}_{-0.15}$ & 0.98$^{+0.17}_{-0.15}$ & 0.97$^{+0.08}_{-0.08}$  \\
c$_{xmm}$ & 0.67$^{+0.15}_{-0.11}$ & 0.57$^{+0.15}_{-0.07}$ & 0.66$^{+0.10}_{-0.11}$  \\
\hline
F$\rm_{2-10\,keV}$ & 6.32$^{+0.29}_{-0.29}$ $\times$ 10$^{-13}$ & 6.71$^{+0.29}_{-0.29}$ $\times$ 10$^{-13}$ & 7.08$^{+0.29}_{-0.29}$ $\times$ 10$^{-13}$  \\
L$\rm_{2-10\,keV}$ & 1.79$^{+0.16}_{-0.16}$ $\times$ 10$^{42}$ & 9.71$^{+0.50}_{-0.50}$ $\times$ 10$^{42}$ & 7.62$^{+0.43}_{-0.42}$ $\times$ 10$^{42}$ \\
\hline
\multicolumn{4}{c}{\textbf{IC 2227}}\\
\hline
$\chi^2$/dof & 192/141 & 160/136 & 163/136  \\
kT & 0.65$^{+0.06}_{-0.06}$ & 0.20$^{+0.07}_{-0.08}$ & 0.62$^{+0.06}_{-0.08}$  \\
$\Gamma$ & 1.83$^{+0.06}_{-0.12}$ & 2.12$^{+0.24}_{-0.29}$ & 1.94$^{+0.20}_{-0.20}$  \\
norm $\times$ $10^{-2}$ & 0.05$^{+0.01}_{-0.01}$ & 0.24$^{+0.11}_{-0.06}$ & 0.15$^{+0.12}_{-0.06}$ \\
$c_f$ & ... & 1.00(p) & ... \\
CTKcover & ... & ... & 0.60$^{+u}_{-0.33}$ \\
TORsigma & ... & ... & $<$30.20 \\
$\rm{N_{H,los}}$ $\times$ $10^{24}$ & 0.42$^{+0.07}_{-0.05}$ & 0.65$^{+0.07}_{-0.12}$ & 0.59$^{+0.11}_{-0.09}$ \\
$\rm{N_{H,avg}}$  $\times$ $10^{24}$ & 1.26$^{+0.42}_{-0.27}$ & 1.35$^{+0.30}_{-0.53}$ & ...  \\
$f_s$ $\times$ 10$^{-2}$ & 2.26$^{+0.68}_{-0.54}$ & 0.90$^{+0.90}_{-0.41}$ &  2.54$^{+2.02}_{-1.20}$ \\
$A_{90}$ & 6.49$^{+4.77}_{-3.94}$ & ... & ... \\
$A_{0}$ & $<$0.94 & ... & ... \\
c$_{nus}$ & 1.04$^{+0.12}_{-0.11}$ & 1.04$^{+0.12}_{-0.11}$ & 1.04$^{+0.12}_{-0.11}$  \\
c$_{xmm}$ & 0.87$^{+0.11}_{-0.09}$ & 0.90$^{+0.12}_{-0.11}$ & 0.90$^{+0.12}_{-0.11}$  \\
\hline
F$\rm_{2-10\,keV}$ & 3.65$^{+0.16}_{-0.16}$ $\times$ 10$^{-13}$ & 3.66$^{+0.16}_{-0.16}$ $\times$ 10$^{-13}$ & 3.75$^{+0.16}_{-0.16}$ $\times$ 10$^{-13}$  \\
L$\rm_{2-10\,keV}$ & 5.01$^{+0.30}_{-0.30}$ $\times$ 10$^{42}$ & 6.45$^{+0.38}_{-0.38}$ $\times$ 10$^{42}$ & 1.00$^{+0.37}_{-0.16}$ $\times$ 10$^{43}$ \\
\hline
\multicolumn{4}{c}{\textbf{ESO 362-8}}\\
\hline
$\chi^2$/dof & 107/115 & 108/114  & 110/114  \\
kT & 0.54$^{+0.07}_{-0.06}$ & 0.51$^{+0.06}_{-0.05}$ & 0.51$^{+0.07}_{-0.06}$  \\
$\Gamma$ & 2.19$^{+0.28}_{-0.29}$ & 1.93$^{+0.50}_{-0.41}$ & 2.32$^{+0.34}_{-0.75}$  \\
norm $\times$ $10^{-2}$ & 1.04$^{+1.14}_{-0.99}$ & 0.17$^{+1.11}_{-0.13}$ & 1.09$^{+4.00}_{-0.98}$ \\
$c_f$ & ... & 0.58$^{+0.02}_{-0.03}$ & ... \\
CTKcover & ... & ... & 0.60(p) \\
TORsigma & ... & ... & 56.35$^{+20.49}_{-56.32}$ \\
$\rm{N_{H,los}}$ $\times$ $10^{24}$ & 6.74$^{+0.43}_{-5.20}$ & 2.31$^{+1.69}_{-0.72}$ & 2.96$^{+2.88}_{-0.95}$ \\
$\rm{N_{H,avg}}$  $\times$ $10^{24}$ & 4.47$^{+u}_{-2.40}$ & 5.75$^{+u}_{-2.87}$ & ...  \\
$f_s$ $\times$ 10$^{-2}$ & 0.10$^{+0.24}_{-0.06}$ & 0.48$^{+0.95}_{-0.27}$ &  1.21$^{+5.06}_{-0.89}$ \\
$A_{0}$ & 1.00(f) & ... & ... \\
$A_{90}$ & 0.18$^{+1.46}_{-0.07}$ & ... & ... \\
c$_{nus}$ & 1.06$^{+0.25}_{-0.20}$ & 1.05$^{+0.24}_{-0.20}$ & 1.07$^{+0.25}_{-0.21}$  \\
c$_{xmm1}$ & 0.95$^{+0.52}_{-0.34}$ & 1.18$^{+0.48}_{-0.36}$ & 1.04$^{+0.39}_{-0.30}$  \\
c$_{xmm2}$ & 0.92$^{+0.48}_{-0.22}$ & 1.16$^{+0.44}_{-0.38}$ & 1.02$^{+0.37}_{-0.28}$  \\
\hline
F$\rm_{2-10\,keV}$ & 9.25$^{+0.79}_{-0.79}$ $\times$ 10$^{-14}$ & 8.00$^{+0.71}_{-0.71}$ $\times$ 10$^{-14}$ & 8.52$^{+0.75}_{-0.75}$ $\times$ 10$^{-14}$  \\
L$\rm_{2-10\,keV}$ & 1.31$^{+1.00}_{-1.00}$ $\times$ 10$^{43}$ & 2.52$^{+0.61}_{-0.61}$ $\times$ 10$^{43}$ & 9.61$^{+1.24}_{-1.23}$ $\times$ 10$^{42}$ \\
\hline
\multicolumn{4}{c}{\textbf{ESO 406-4}}\\
\hline
$\chi^2$/dof & 58/65 & 59/66  & 59/66  \\
kT & 0.46$^{+0.11}_{-0.13}$ & 0.45$^{+0.11}_{-0.13}$ & 0.46$^{+0.11}_{-0.16}$  \\
$\Gamma$ & 1.51$^{+0.85}_{-u}$ & 1.49$^{+0.26}_{-u}$ & 1.72$^{+0.45}_{-0.34}$  \\
norm $\times$ $10^{-2}$ & 0.02$^{+0.22}_{-0.02}$ & 0.02$^{+0.09}_{-0.01}$ & 0.05$^{+18}_{-0.04}$ \\
$c_f$ & ... & 0.90$^{+u}_{-0.78}$ & ... \\
CTKcover & ... & ... & $>$0.03 \\
TORsigma & ... & ... & 34.38$^{+u}_{-23.83}$ \\
$\rm{N_{H,los}}$ $\times$ $10^{24}$ & 1.06$^{+0.87}_{-0.53}$ & 1.34$^{+0.85}_{-0.40}$ & 1.16$^{+0.56}_{-0.35}$ \\
$\rm{N_{H,avg}}$  $\times$ $10^{24}$ & 1.01$^{+8.98}_{-0.99}$ & 1.00$^{+u}_{-0.51}$ & ...  \\
$f_s$ $\times$ 10$^{-2}$ & 5.74$^{+10.33}_{-5.39}$ & 7.42$^{+4.27}_{-4.41}$ &  11.18$^{+66.18}_{-7.96}$ \\
$A_{0}$ & 1.00$^{+53.35}_{-u}$ & ... & ... \\
$A_{90}$ & $>$2.16 & ... & ... \\
c$_{nus1}$ & 0.91$^{+0.34}_{-0.25}$ & 0.91$^{+0.34}_{-0.25}$ & 0.91$^{+0.32}_{-0.25}$  \\
c$_{nus2}$ & 2.27$^{+0.65}_{-0.44}$ & 2.27$^{+0.65}_{-0.43}$ & 2.26$^{+0.64}_{-0.43}$  \\
c$_{xmm}$ & 0.64$^{+0.32}_{-0.23}$ & 0.63$^{+0.30}_{-0.22}$ & 0.67$^{+0.31}_{-0.23}$  \\
\hline
F$\rm_{2-10\,keV}$ & 1.49$^{+0.12}_{-0.97}$ $\times$ 10$^{-13}$ & 1.50$^{+0.22}_{-1.50}$ $\times$ 10$^{-13}$ & 1.45$^{+1.82}_{-0.11}$ $\times$ 10$^{-13}$  \\
L$\rm_{2-10\,keV}$ & 2.28$^{+0.30}_{-0.30}$ $\times$ 10$^{42}$ & 1.79$^{+0.30}_{-0.30}$ $\times$ 10$^{42}$ & 3.94$^{+0.25}_{-0.24}$ $\times$ 10$^{42}$ \\
\hline
\multicolumn{4}{c}{\textbf{2MFGC 13496}}\\
\hline
$\chi^2$/dof & 50/36 & 50/37  & 56/37  \\
kT & -- & -- & -- \\
$\Gamma$ & 1.59$^{+0.39}_{-u}$ & 1.54$^{+0.33}_{-u}$ & 1.55$^{+0.50}_{-0.16}$  \\
norm $\times$ $10^{-2}$ & 0.09$^{+0.29}_{-0.04}$ & 0.11$^{+0.24}_{-0.05}$ & 0.07$^{+0.22}_{-0.04}$ \\
$c_f$ & ... & 0.10(f) & ... \\
CTKcover & ... & ... & 0.25$^{+0.26}_{-u}$ \\
TORsigma & ... & ... & 6.99$^{+44.20}_{-u}$ \\
$\rm{N_{H,los}}$ $\times$ $10^{24}$ & 0.73$^{+0.27}_{-0.19}$ & 1.15$^{+0.34}_{-0.25}$ & 0.59$^{+0.20}_{-0.08}$ \\
$\rm{N_{H,avg}}$  $\times$ $10^{24}$ & 0.02$^{+0.05}_{-u}$ & 0.05$^{+0.15}_{-u}$ & ...  \\
$f_s$ $\times$ 10$^{-2}$ & 0.59$^{+1.03}_{-0.50}$ & 0.40$^{+0.66}_{-0.18}$ &  1.72$^{+2.51}_{-1.17}$ \\
$A_{0}$ & 1.01$^{+3.75}_{-u}$ & ... & ... \\
$A_{90}$ & $<$1.63 & ... & ... \\
c$_{nus}$ & 1.16$^{+0.20}_{-0.17}$ & 1.16$^{+0.20}_{-0.17}$ & 1.16$^{+0.19}_{-0.16}$  \\
c$_{xmm}$ & 0.93$^{+0.23}_{-0.10}$ & 0.97$^{+0.24}_{-0.20}$ & 0.90$^{+0.20}_{-0.18}$  \\
\hline
F$\rm_{2-10\,keV}$ & 3.74$^{+0.29}_{-1.94}$ $\times$ 10$^{-13}$ & 3.74$^{+0.46}_{-2.69}$ $\times$ 10$^{-13}$ & 3.83$^{+0.37}_{-2.77}$ $\times$ 10$^{-13}$  \\
L$\rm_{2-10\,keV}$ & 1.56$^{+0.12}_{-0.12}$ $\times$ 10$^{43}$ & 2.08$^{+0.15}_{-0.15}$ $\times$ 10$^{43}$ & 1.38$^{+0.16}_{-0.19}$ $\times$ 10$^{43}$ \\
\hline
\multicolumn{4}{c}{\textbf{2MASX J03585442+1026033}}\\
\hline
$\chi^2$/dof & 357/318 & 368/320  & 361/320  \\
kT & 0.33$^{+0.26}_{-0.06}$ & 0.30$^{+0.08}_{-0.05}$ & 0.30$^{+0.10}_{-0.05}$ \\
$\Gamma$ & 1.40$^{+1.20}_{-u}$ & 1.40(p) & 1.40$^{+0.09}_{-0.04}$  \\
norm $\times$ $10^{-2}$ & 0.06$^{+0.01}_{-0.01}$ & 0.06$^{+0.01}_{-0.01}$ & 0.24$^{+0.06}_{-0.06}$ \\
$c_f$ & ... & 0.70$^{+u}_{-0.39}$ & ... \\
CTKcover & ... & ... & 0.014$^{+0.030}_{-u}$ \\
TORsigma & ... & ... & 1.80$^{+0.84}_{-0.72}$ \\
$\rm{N_{H,los}}$ $\times$ $10^{24}$ & 0.11$^{+0.01}_{-0.01}$ & 0.11$^{+0.08}_{-0.07}$ & 0.10$^{+0.01}_{-0.01}$ \\
$\rm{N_{H,avg}}$  $\times$ $10^{24}$ & 0.40$^{+0.75}_{-0.35}$ & 0.11$^{+0.18}_{-0.07}$ & ...  \\
$f_s$ $\times$ 10$^{-2}$ & 1.49$^{+0.36}_{-0.28}$ & 1.49$^{+0.37}_{-0.53}$ & 1.00(f) \\
$A_{0}$ & 1.19$^{+1.42}_{-u}$ & ... & ... \\
$A_{90}$ & $<$1.17 & ... & ... \\
c$_{nus}$ & 0.97$^{+0.05}_{-0.06}$ & 0.98$^{+0.06}_{-0.06}$ & 0.97$^{+0.06}_{-0.06}$  \\
c$_{xmm1}$ & 1.21$^{+0.08}_{-0.07}$ & 1.18$^{+0.07}_{-0.07}$ & 1.21$^{+0.08}_{-0.07}$  \\
c$_{xmm2}$ & 0.54$^{+0.07}_{-0.07}$ & 0.53$^{+0.07}_{-0.06}$ & 0.54$^{+0.07}_{-0.07}$  \\
c$_{chan}$ & 0.79$^{+0.06}_{-0.05}$ & 0.76$^{+0.05}_{-0.05}$ & 0.78$^{+0.06}_{-0.05}$  \\
\hline
F$\rm_{2-10\,keV}$ & 2.29$^{+0.41}_{-0.57}$ $\times$ 10$^{-12}$ & 2.36$^{+0.05}_{-0.07}$ $\times$ 10$^{-12}$ & 2.29$^{+0.20}_{-0.09}$ $\times$ 10$^{-12}$  \\
L$\rm_{2-10\,keV}$ & 8.31$^{+0.19}_{-0.19}$ $\times$ 10$^{42}$ & 8.89$^{+0.20}_{-0.20}$ $\times$ 10$^{42}$ & 3.43$^{+0.14}_{-0.25}$ $\times$ 10$^{43}$ \\
\hline
\multicolumn{4}{c}{\textbf{M 58}}\\
\hline
$\chi^2$/dof & 2466/2095 & 2497/2095  & 2643/2097  \\
kT & 0.65$^{+0.03}_{-0.03}$ & 0.61$^{+0.04}_{-0.05}$ & 0.61$^{+0.04}_{-0.04}$ \\
$\Gamma$ & 1.91$^{+0.02}_{-0.03}$ & 1.85$^{+0.01}_{-0.01}$ & 1.81$^{+0.01}_{-0.01}$  \\
norm $\times$ $10^{-2}$ & 0.03$^{+0.01}_{-0.02}$ & 0.23$^{+0.06}_{-0.05}$ & 0.29$^{+0.02}_{-0.01}$ \\
$c_f$ & ... & 1.00(p) & ... \\
CTKcover & ... & ... & -- \\
TORsigma & ... & ... & 2.92$^{+1.65}_{-1.92}$ \\
$\rm{N_{H,los}}$ $\times$ $10^{24}$ & 0.010$^{+0.002}_{-u}$ & 0.00006$^{+0.00005}_{-0.00004}$ & 0.0001$^{+0.0000}_{-u}$ \\
$\rm{N_{H,avg}}$  $\times$ $10^{24}$ & 0.09$^{+0.03}_{-0.03}$ & 0.17$^{+0.02}_{-0.03}$ & ...  \\
$f_s$ $\times$ 10$^{-2}$ & ... & ... & ... \\
$A_{0}$ & 49.90$^{+105.27}_{-20.86}$ & ... & ... \\
$A_{90}$ & 1.00(f) & ... & ... \\
c$_{nus}$ & 1.00$^{+0.02}_{-0.02}$ & 1.00$^{+0.02}_{-0.03}$ & 1.00$^{+0.02}_{-0.02}$  \\
c$_{xmm}$ & 0.51$^{+0.01}_{-0.01}$ & 0.50$^{+0.01}_{-0.02}$ & 0.49$^{+0.01}_{-0.01}$  \\
c$_{xmm}$ & 0.80$^{+0.02}_{-0.02}$ & 0.79$^{+0.03}_{-0.01}$ & 0.78$^{+0.02}_{-0.02}$  \\
c$_{chan}$ & 0.55$^{+0.02}_{-0.02}$ & 0.54$^{+0.03}_{-0.02}$ & 0.53$^{+0.02}_{-0.02}$  \\
\hline
F$\rm_{2-10\,keV}$ & 7.91$^{+0.06}_{-1.73}$ $\times$ 10$^{-12}$ & 7.88$^{+0.08}_{-0.10}$ $\times$ 10$^{-12}$ & 8.06$^{+0.74}_{-0.14}$ $\times$ 10$^{-12}$ \\
L$\rm_{2-10\,keV}$ & 5.25$^{+0.33}_{-0.33}$ $\times$ 10$^{40}$ & 4.10$^{+0.02}_{-0.02}$ $\times$ 10$^{41}$ & 5.43$^{+0.05}_{-0.09}$ $\times$ 10$^{41}$ \\
\hline
\multicolumn{4}{c}{\textbf{3C 371}}\\
\hline
$\chi^2$/dof & 203/173 & 203/176  & 203/176  \\
kT & 0.37$^{+0.49}_{-0.14}$ & 0.37(p) & 0.33(f) \\
$\Gamma$ & 1.80$^{+0.20}_{-0.21}$ & 1.77$^{+0.12}_{-0.13}$ & 1.85$^{+0.11}_{-0.07}$  \\
norm $\times$ $10^{-2}$ & 0.06$^{+0.03}_{-0.03}$ & 0.05$^{+0.02}_{-0.02}$ & 0.13$^{+0.08}_{-0.03}$ \\
$c_f$ & ... & 1.00(p) & ... \\
CTKcover & ... & ... & 0.60(p) \\
TORsigma & ... & ... & 1.91$^{+3.21}_{-u}$ \\
$\rm{N_{H,los}}$ $\times$ $10^{24}$ & 0.022$^{+0.006}_{-0.006}$ & 0.034$^{+0.010}_{-0.018}$ & 0.025$^{+0.001}_{-0.010}$ \\
$\rm{N_{H,avg}}$  $\times$ $10^{24}$ & 0.50$^{+2.22}_{-0.49}$ & 0.54$^{+0.78}_{-0.49}$ & ...  \\
$f_s$ $\times$ 10$^{-2}$ & 41.30$^{+25.72}_{-14.42}$ & 41.40$^{+17.85}_{-11.12}$ & 99.30$^{+44.74}_{-87.14}$ \\
$A_{0}$ & 3.11$^{+78.22}_{-20.86}$ & ... & ... \\
$A_{90}$ & $<$76.13 & ... & ... \\
c$_{nus}$ & 1.03$^{+0.11}_{-0.10}$ & 1.03$^{+0.11}_{-0.10}$ & 1.03$^{+0.11}_{-0.10}$  \\
c$_{chan}$ & 0.80$^{+0.10}_{-0.09}$ & 0.80$^{+0.10}_{-0.09}$ & 0.80$^{+0.08}_{-0.10}$  \\
\hline
F$\rm_{2-10\,keV}$ & 2.55$^{+0.39}_{-0.19}$ $\times$ 10$^{-12}$ & 2.43$^{+0.06}_{-0.34}$ $\times$ 10$^{-12}$ & 2.53$^{+0.22}_{-1.11}$ $\times$ 10$^{-12}$  \\
L$\rm_{2-10\,keV}$ & 1.45$^{+0.05}_{-0.05}$ $\times$ 10$^{43}$ & 1.36$^{+0.04}_{-0.04}$ $\times$ 10$^{43}$ & 2.40$^{+0.06}_{-0.07}$ $\times$ 10$^{43}$ \\
\hline
\multicolumn{4}{c}{\textbf{IC 1198}}\\
\hline
$\chi^2$/dof & 438/338 & 425/340  & 443/340  \\
kT & -- & -- & -- \\
$\Gamma$ & 1.96$^{+0.06}_{-0.06}$ & 2.13$^{+0.05}_{-0.10}$ & 1.97$^{+0.10}_{-0.06}$  \\
norm $\times$ $10^{-2}$ & 0.25$^{+0.03}_{-0.03}$ & 0.40$^{+0.06}_{-0.08}$ & 1.25$^{+0.12}_{-0.10}$ \\
$c_f$ & ... & 1.00(p) & ... \\
CTKcover & ... & ... & $<$0.03 \\
TORsigma & ... & ... & 1.87$^{+0.78}_{-0.84}$ \\
$\rm{N_{H,los}}$ $\times$ $10^{24}$ & 0.11$^{+0.01}_{-0.01}$ & 0.21$^{+0.02}_{-0.02}$ & 0.11$^{+0.01}_{-0.01}$ \\
$\rm{N_{H,avg}}$  $\times$ $10^{24}$ & 10.00(p) & 3.02$^{+2.11}_{-1.02}$ & ...  \\
$f_s$ $\times$ 10$^{-2}$ & 3.64$^{+0.61}_{-0.50}$ & 2.31$^{+0.56}_{-0.32}$ & 1.00(f) \\
$A_{0}$ & 3.11$^{+5.89}_{-u}$ & ... & ... \\
$A_{90}$ & 0.99$^{+1.00}_{-0.67}$ & ... & ... \\
c$_{nus}$ & 1.03$^{+0.06}_{-0.05}$ & 1.02$^{+0.06}_{-0.05}$ & 1.03$^{+0.06}_{-0.05}$  \\
c$_{xmm}$ & 0.30$^{+0.02}_{-0.02}$ & 0.30$^{+0.02}_{-0.02}$ & 0.31$^{+0.02}_{-0.02}$  \\
\hline
F$\rm_{2-10\,keV}$ & 3.42$^{+0.10}_{-0.10}$ $\times$ 10$^{-12}$ & 3.47$^{+0.09}_{-0.11}$ $\times$ 10$^{-12}$ & 3.47$^{+0.91}_{-0.01}$ $\times$ 10$^{-12}$  \\
L$\rm_{2-10\,keV}$ & 1.72$^{+0.04}_{-0.04}$ $\times$ 10$^{43}$ & 2.27$^{+0.06}_{-0.06}$ $\times$ 10$^{43}$ & 8.88$^{+0.05}_{-0.06}$ $\times$ 10$^{43}$ \\
\hline
\multicolumn{4}{c}{\textbf{2MASX J09261742-8421330}}\\
\hline
C-stat/dof & 265/265 & 264/266 & 267/266   \\
kT & -- & -- & -- \\
$\Gamma$ & 1.98$^{+0.11}_{-0.14}$ & 1.95$^{+0.10}_{-0.08}$ & 1.92$^{+0.06}_{-0.15}$  \\
norm $\times$ $10^{-2}$ & 0.25$^{+0.10}_{-0.05}$ & 0.21$^{+0.07}_{-0.03}$ & 0.29$^{+0.05}_{-0.08}$ \\
$c_f$ & ... & 1.00(p) & ... \\
CTKcover & ... & ... & $<$0.50 \\
TORsigma & ... & ... & 83.42$^{+u}_{-77.44}$ \\
$\rm{N_{H,los}}$ $\times$ $10^{24}$ & 0.014$^{+0.010}_{-u}$ & 0.021$^{+0.034}_{-u}$ & 0.013$^{+0.013}_{-u}$ \\
$\rm{N_{H,avg}}$  $\times$ $10^{24}$ & 0.95$^{+1.67}_{-0.28}$ & 0.91$^{+0.31}_{-0.43}$ & ...  \\
$f_s$ $\times$ 10$^{-2}$ & -- & -- & -- \\
$A_{0}$ & 2.62$^{+3.45}_{-u}$ & ... & ... \\
$A_{90}$ & $<$1.46 & ... & ... \\
c$_{nus}$ & 0.99$^{+0.04}_{-0.04}$ & 0.99$^{+0.04}_{-0.04}$ & 0.99$^{+0.04}_{-0.04}$  \\
c$_{xrt}$ & 0.90$^{+0.06}_{-0.11}$ & 0.91$^{+0.13}_{-0.10}$ & 0.90$^{+0.11}_{-0.14}$  \\
\hline
F$\rm_{2-10\,keV}$ & 5.61$^{+0.06}_{-4.89}$ $\times$ 10$^{-12}$ & 5.38$^{+0.16}_{-0.16}$ $\times$ 10$^{-12}$ & 5.31$^{+-0.04}_{-5.24}$ $\times$ 10$^{-12}$  \\
L$\rm_{2-10\,keV}$ & 4.73$^{+0.56}_{-0.56}$ $\times$ 10$^{43}$ & 1.57$^{+0.04}_{-0.04}$ $\times$ 10$^{43}$ & 8.32$^{+0.45}_{-0.46}$ $\times$ 10$^{43}$ \\
\hline
\multicolumn{4}{c}{\textbf{UGC 12348}}\\
\hline
$\chi^2$/dof & 760/772 & 751/774 & 733/773   \\
kT & 0.71$^{+0.11}_{-0.14}$ & 0.70(f) & 0.70(f) \\
$\Gamma$ & 1.60$^{+0.03}_{-0.03}$ & 1.77$^{+0.04}_{-0.04}$ & 1.92$^{+0.05}_{-0.09}$  \\
norm $\times$ $10^{-2}$ & 0.10$^{+0.01}_{-0.01}$ & 0.16$^{+0.02}_{-0.01}$ & 0.26$^{+0.03}_{-0.04}$ \\
$c_f$ & ... & 1.00(p) & ... \\
CTKcover & ... & ... & 0.6(p) \\
TORsigma & ... & ... & 23.64$^{+40.36}_{-08.30}$ \\
$\rm{N_{H,los}}$ $\times$ $10^{24}$ & 0.023$^{+0.002}_{-0.001}$ & 0.044$^{+0.018}_{-0.015}$ & 0.027$^{+0.027}_{-0.025}$ \\
$\rm{N_{H,avg}}$  $\times$ $10^{24}$ & 0.018$^{+0.015}_{-0.008}$ & 0.01(p) & ...  \\
$f_s$ $\times$ 10$^{-2}$ & 1.00(f) & 1.01$^{+0.30}_{-0.30}$ & 3.10$^{+0.06}_{-0.01}$ \\
$A_{0}$ & 16.92$^{+13.16}_{-6.07}$ & ... & ... \\
$A_{90}$ & $<$3.69 & ... & ... \\
c$_{nus}$ & 1.03$^{+0.05}_{-0.05}$ & 1.02$^{+0.05}_{-0.05}$ & 1.03$^{+0.05}_{-0.05}$  \\
c$_{xmm}$ & 1.06$^{+0.05}_{-0.05}$ & 1.05$^{+0.05}_{-0.05}$ & 1.04$^{+0.05}_{-0.05}$  \\
c$_{chan}$ & 0.54$^{+0.07}_{-0.07}$ & 0.46$^{+0.05}_{-0.05}$ & 0.46$^{+0.07}_{-0.06}$  \\
\hline
F$\rm_{2-10\,keV}$ & 4.21$^{+1.34}_{-0.03}$ $\times$ 10$^{-12}$ & 4.56$^{+0.15}_{-0.14}$ $\times$ 10$^{-12}$ & 4.61$^{+0.06}_{-2.26}$ $\times$ 10$^{-12}$  \\
L$\rm_{2-10\,keV}$ & 6.56$^{+0.09}_{-0.09}$ $\times$ 10$^{42}$ & 8.30$^{+0.10}_{-0.10}$ $\times$ 10$^{42}$ & 1.09$^{+0.25}_{-0.69}$ $\times$ 10$^{43}$ \\
\hline
\multicolumn{4}{c}{\textbf{2MASX J02420381+0510061}}\\
\hline
$\chi^2$/dof & 188/218 & 187/217 & 192/218   \\
kT & -- & -- & -- \\
$\Gamma$ & 1.83$^{+0.03}_{-0.08}$ & 1.81$^{+0.08}_{-0.07}$ & 1.83$^{+0.08}_{-0.08}$  \\
norm $\times$ $10^{-2}$ & 0.16$^{+0.03}_{-0.05}$ & 0.20$^{+0.04}_{-0.04}$ & 7.44$^{+1.13}_{-1.04}$ \\
$c_f$ & ... & 0.1(p) & ... \\
CTKcover & ... & ... & 0.6(p) \\
TORsigma & ... & ... & 0.15$^{+1.48}_{-0.06}$ \\
$\rm{N_{H,los}}$ $\times$ $10^{24}$ & 0.81$^{+0.15}_{-0.14}$ & 1.29$^{+0.25}_{-0.22}$ & 0.98$^{+0.60}_{-0.25}$ \\
$\rm{N_{H,avg}}$  $\times$ $10^{24}$ & 0.01(p) & 0.01(p) & ...  \\
$f_s$ $\times$ 10$^{-2}$ & 17.78$^{+8.43}_{-5.39}$ & 12.27$^{+6.07}_{-2.88}$ & 1.00(f) \\
$A_{0}$ & 1.00 (f) & ... & ... \\
$A_{90}$ & 1.00(f) & ... & ... \\
c$_{nus}$ & 1.00$^{+0.14}_{-0.12}$ & 1.00$^{+0.14}_{-0.12}$ & 0.99$^{+0.13}_{-0.12}$  \\
c$_{xmm}$ & 5.03$^{+0.71}_{-0.60}$ & 5.13$^{+0.74}_{-0.62}$ & 4.89$^{+0.68}_{-0.57}$  \\
\hline
F$\rm_{2-10\,keV}$ & 1.15$^{+0.06}_{-0.18}$ $\times$ 10$^{-12}$ & 1.14$^{+0.03}_{-0.18}$ $\times$ 10$^{-12}$ & 1.16$^{+0.07}_{-0.08}$ $\times$ 10$^{-12}$  \\
L$\rm_{2-10\,keV}$ & 6.08$^{+0.67}_{-0.67}$ $\times$ 10$^{43}$ & 8.80$^{+0.95}_{-0.94}$ $\times$ 10$^{43}$ & 4.88$^{+1.25}_{-2.96}$ $\times$ 10$^{44}$ \\
\hline
\multicolumn{4}{c}{\textbf{ESO 234-50}}\\
\hline
$\chi^2$/dof & 137/132 & 136/132 & 137/131   \\
kT & -- & -- & -- \\
$\Gamma$ & 1.71$^{+0.05}_{-0.05}$ & 1.69$^{+0.11}_{-0.06}$ & 1.80$^{+0.08}_{-0.08}$  \\
norm $\times$ $10^{-2}$ & 0.090$^{+0.06}_{-0.03}$ & 0.094$^{+0.05}_{-0.03}$ & 0.099$^{+0.095}_{-0.011}$ \\
$c_f$ & ... & 0.12$^{+0.69}_{-u}$ & ... \\
CTKcover & ... & ... & 0.22$^{+0.15}_{-u}$ \\
TORsigma & ... & ... & 6.86$^{+22.32}_{-6.66}$ \\
$\rm{N_{H,los}}$ $\times$ $10^{24}$ & 0.22$^{+0.05}_{-0.05}$ & 0.32$^{+0.07}_{-0.05}$ & 0.20$^{+0.02}_{-0.03}$ \\
$\rm{N_{H,avg}}$  $\times$ $10^{24}$ & $<$0.50 & 1.00(p) & ...  \\
$f_s$ $\times$ 10$^{-2}$ & 4.53$^{+3.02}_{-1.88}$ & 4.15$^{+1.01}_{-1.47}$ & 7.99$^{+0.03}_{-0.08}$ \\
$A_{0}$ & 1.00 (f) & ... & ... \\
$A_{90}$ & 1.00(f) & ... & ... \\
c$_{nus}$ & 1.04$^{+0.08}_{-0.07}$ & 1.04$^{+0.08}_{-0.07}$ & 1.04$^{+0.08}_{-0.07}$  \\
c$_{xmm}$ & 0.33$^{+0.05}_{-0.05}$ & 0.33$^{+0.05}_{-0.05}$ & 0.33$^{+0.05}_{-0.05}$  \\
\hline
F$\rm_{2-10\,keV}$ & 1.45$^{+0.87}_{-0.43}$ $\times$ 10$^{-12}$ & 1.45$^{+0.06}_{-0.14}$ $\times$ 10$^{-12}$ & 1.45$^{+1.05}_{-0.16}$ $\times$ 10$^{-12}$  \\
L$\rm_{2-10\,keV}$ & 6.17$^{+0.26}_{-0.26}$ $\times$ 10$^{41}$ & 6.85$^{+0.29}_{-0.29}$ $\times$ 10$^{41}$ & 4.45$^{+0.21}_{-0.32}$ $\times$ 10$^{41}$ \\
\hline
\multicolumn{4}{c}{\textbf{NGC 2273}}\\
\hline
$\chi^2$/dof & 195/161 & 194/168 & 207/168   \\
kT & 0.35$^{+0.18}_{-0.08}$  & 0.36$^{+0.11}_{-0.06}$ & 0.34$^{+0.09}_{-0.05}$ \\
$\Gamma$ & 1.47$^{+0.22}_{-u}$ & 1.46$^{+0.18}_{-u}$ & 1.16$^{+0.17}_{-0.09}$  \\
norm $\times$ $10^{-2}$ & 0.043$^{+0.049}_{-0.009}$ & 0.21$^{+0.11}_{-0.02}$ & 0.23$^{+0.04}_{-0.04}$ \\
$c_f$ & ... & 0.56$^{+0.25}_{-0.09}$ & ... \\
CTKcover & ... & ... & 0.03$^{+0.03}_{-0.02}$ \\
TORsigma & ... & ... & $<$1.00 \\
$\rm{N_{H,los}}$ $\times$ $10^{24}$ & 0.37$^{+0.11}_{-0.06}$ & 0.31$^{+0.11}_{-0.13}$ & 0.35$^{+0.05}_{-0.07}$ \\
$\rm{N_{H,avg}}$  $\times$ $10^{24}$ & 3.30$^{+u}_{-1.90}$ & 0.43$^{+0.23}_{-0.14}$ & ...  \\
$f_s$ $\times$ 10$^{-2}$ & 5.09$^{+0.03}_{-0.02}$ & 0.85$^{+0.22}_{-0.28}$ & 1.00(f) \\
$A_{0}$ & 1.00(f) & ... & ... \\
$A_{90}$ & 1.00(f) & ... & ... \\
c$_{nus}$ & 1.04$^{+0.10}_{-0.12}$ & 1.04$^{+0.12}_{-0.11}$ & 1.03$^{+0.11}_{-0.10}$  \\
c$_{xmm}$ & 0.94$^{+0.13}_{-0.12}$ & 0.95$^{+0.13}_{-0.12}$ & 0.94$^{+0.13}_{-0.11}$  \\
c$_{chandra}$ & 0.93$^{+0.20}_{-0.17}$ & 0.97$^{+0.17}_{-0.16}$ & 0.97$^{+0.17}_{-0.14}$  \\
\hline
F$\rm_{2-10\,keV}$ & 1.02$^{+0.06}_{-0.75}$ $\times$ 10$^{-12}$ & 1.02$^{+0.11}_{-0.24}$ $\times$ 10$^{-12}$ & 1.02$^{+1.29}_{-0.02}$ $\times$ 10$^{-12}$  \\
L$\rm_{2-10\,keV}$ & 2.31$^{+0.40}_{-0.40}$ $\times$ 10$^{41}$ & 1.07$^{+0.32}_{-0.32}$ $\times$ 10$^{42}$ & 3.98$^{+1.25}_{-0.69}$ $\times$ 10$^{42}$ \\
\hline
\multicolumn{4}{c}{\textbf{FRL 265}}\\
\hline
$\chi^2$/dof & 180/171 & 177/173 & 173/170   \\
kT & 0.18$^{+0.30}_{-0.07}$ & -- & 0.15$^{+0.56}_{-u}$ \\
$\Gamma$ & 1.97$^{+0.07}_{-0.07}$ & 1.87$^{+0.40}_{-0.15}$ & 1.80$^{+0.08}_{-0.06}$  \\
norm $\times$ $10^{-2}$ & 0.10$^{+0.02}_{-0.01}$ & 0.08$^{+0.30}_{-0.06}$ & 0.10$^{+0.70}_{-0.30}$ \\
$c_f$ & ... & 1.00(p) & ... \\
CTKcover & ... & ... & 0.36$^{+0.08}_{-0.05}$ \\
TORsigma & ... & ... & 8.01$^{+2.25}_{-7.18}$ \\
$\rm{N_{H,los}}$ $\times$ $10^{24}$ & 0.010$^{+0.004}_{-u}$ & $<$0.06 & 0.01$^{+0.23}_{-0.06}$ \\
$\rm{N_{H,avg}}$  $\times$ $10^{24}$ & 2.00$^{+2.95}_{-0.99}$ & 1.51$^{+0.34}_{-0.31}$ & ...  \\
$f_s$ $\times$ 10$^{-2}$ & -- & -- & -- \\
$A_{0}$ & 1.00(f) & ... & ... \\
$A_{90}$ & 1.00(f) & ... & ... \\
c$_{nus}$ & 1.06$^{+0.09}_{-0.08}$ & 1.06$^{+0.09}_{-0.08}$ & 1.05$^{+0.09}_{-0.08}$  \\
c$_{xmm}$ & 0.79$^{+0.07}_{-0.06}$ & 0.83$^{+0.07}_{-0.06}$ & 0.78$^{+0.07}_{-0.06}$  \\
\hline
F$\rm_{2-10\,keV}$ & 2.53$^{+0.18}_{-0.16}$ $\times$ 10$^{-12}$ & 2.59$^{+0.18}_{-0.15}$ $\times$ 10$^{-12}$ & 2.48$^{+-0.17}_{-2.48}$ $\times$ 10$^{-12}$  \\
L$\rm_{2-10\,keV}$ & 3.32$^{+0.20}_{-0.20}$ $\times$ 10$^{43}$ & 4.12$^{+2.59}_{-2.59}$ $\times$ 10$^{43}$ & 1.42$^{+0.73}_{-0.64}$ $\times$ 10$^{43}$ \\
\hline
\multicolumn{4}{c}{\textbf{MRK 231}}\\
\hline
$\chi^2$/dof & 556/539 & 557/541 & 558/540   \\
kT & -- & -- & -- \\
$\Gamma$ & 1.51$^{+0.07}_{-0.07}$ & 1.52$^{+0.21}_{-u}$ & 1.67$^{+0.10}_{-0.09}$  \\
norm $\times$ $10^{-2}$ & 0.023$^{+0.006}_{-0.004}$ & 0.025$^{+0.003}_{-0.002}$ & 0.043$^{+0.013}_{-0.012}$ \\
$c_f$ & ... & 0.30$^{+u}_{-0.18}$ & ... \\
CTKcover & ... & ... & $<$0.54 \\
TORsigma & ... & ... & 50.90$^{+u}_{-40.04}$  \\
$\rm{N_{H,los}}$ $\times$ $10^{24}$ & 0.06$^{+0.03}_{-0.02}$ & 0.10$^{+0.02}_{-0.02}$ & 0.07$^{+0.02}_{-0.02}$ \\
$\rm{N_{H,avg}}$  $\times$ $10^{24}$ & 0.24$^{+0.53}_{-0.15}$ & 1.62$^{+1.24}_{-u}$ & ...  \\
$f_s$ $\times$ 10$^{-2}$ & 1.00(F) & 1.00(f) & 1.00(f) \\
$A_{0}$ & 1.00(f) & ... & ... \\
$A_{90}$ & 1.00(f) & ... & ... \\
c$_{nus1}$ & 0.94$^{+0.10}_{-0.11}$ & 0.94$^{+0.10}_{-0.11}$ & 0.94$^{+0.10}_{-0.11}$  \\
c$_{nus2}$ & 1.00$^{+0.10}_{-0.11}$ & 1.00$^{+0.10}_{-0.11}$ & 1.00$^{+0.10}_{-0.11}$  \\
c$_{nu3}$ & 0.98$^{+0.11}_{-0.10}$ & 0.98$^{+0.11}_{-0.10}$ & 0.98$^{+0.11}_{-0.10}$  \\
c$_{nu4}$ & 1.04$^{+0.11}_{-0.10}$ & 1.04$^{+0.11}_{-0.10}$ & 1.04$^{+0.11}_{-0.10}$  \\
c$_{nu5}$ & 1.07$^{+0.10}_{-0.09}$ & 1.07$^{+0.10}_{-0.09}$ & 1.07$^{+0.10}_{-0.09}$  \\
c$_{nu6}$ & 0.90$^{+0.10}_{-0.09}$ & 0.90$^{+0.10}_{-0.09}$ & 0.90$^{+0.10}_{-0.09}$  \\
c$_{nu7}$ & 0.82$^{+0.10}_{-0.09}$ & 0.82$^{+0.10}_{-0.09}$ & 0.82$^{+0.10}_{-0.09}$  \\
c$_{nu8}$ & 0.91$^{+0.10}_{-0.09}$ & 0.91$^{+0.10}_{-0.09}$ & 0.91$^{+0.10}_{-0.09}$  \\
c$_{nu9}$ & 0.90$^{+0.10}_{-0.09}$ & 0.90$^{+0.10}_{-0.09}$ & 0.90$^{+0.10}_{-0.09}$  \\
c$_{nu10}$ & 1.25$^{+0.11}_{-0.09}$ & 1.25$^{+0.11}_{-0.09}$ & 1.25$^{+0.11}_{-0.09}$  \\
c$_{nu11}$ & 1.20$^{+0.10}_{-0.09}$ & 1.20$^{+0.10}_{-0.09}$ & 1.20$^{+0.10}_{-0.09}$  \\
\hline
F$\rm_{2-10\,keV}$ & 7.61$^{+4.20}_{-7.54}$ $\times$ 10$^{-13}$ & 7.37$^{+0.43}_{-0.47}$ $\times$ 10$^{-13}$ & 7.27$^{+0.30}_{-2.25}$ $\times$ 10$^{-13}$  \\
L$\rm_{2-10\,keV}$ & 4.78$^{+0.10}_{-0.10}$ $\times$ 10$^{42}$ & 5.12$^{+0.11}_{-0.11}$ $\times$ 10$^{42}$ & 2.95$^{+0.43}_{-0.46}$ $\times$ 10$^{43}$ \\
\hline
\multicolumn{4}{c}{\textbf{PG 1211+143}}\\
\hline
$\chi^2$/dof & 1616/1647 & 1654/1648 & 1662/1647   \\
kT & -- & -- & -- \\
$\Gamma$ & 2.42$^{+0.08}_{-0.06}$ & 2.22$^{+0.03}_{-0.06}$ & 2.17$^{+0.01}_{-0.03}$  \\
norm $\times$ $10^{-2}$ & 0.39$^{+0.07}_{-0.05}$ & 0.25$^{+0.02}_{-0.04}$ & 0.30$^{+0.03}_{-0.02}$ \\
$c_f$ & ... & 1.00(p) & ... \\
CTKcover & ... & ... & 0.30$^{+0.09}_{-0.05}$ \\
TORsigma & ... & ... & 84.00$^{+u}_{-7.42}$  \\
$\rm{N_{H,los}}$ $\times$ $10^{24}$ & 0.04$^{+0.01}_{-0.01}$ & 0.04$^{+0.01}_{-0.02}$ & 0.14$^{+0.05}_{-0.04}$ \\
$\rm{N_{H,avg}}$  $\times$ $10^{24}$ & 1.99$^{+0.21}_{-0.22}$ & 1.51$^{+0.31}_{-0.33}$ & ...  \\
$f_s$ $\times$ 10$^{-2}$ & 1.00(F) & 1.00(f) & 1.00(f) \\
$A_{0}$ & 3.89$^{+1.13}_{-0.85}$ & ... & ... \\
$A_{90}$ & 1.00(f) & ... & ... \\
c$_{nus1}$ & 0.98$^{+0.03}_{-0.03}$ & 0.99$^{+0.03}_{-0.03}$ & 0.99$^{+0.03}_{-0.03}$  \\
c$_{nus2}$ & 1.33$^{+0.04}_{-0.04}$ & 1.34$^{+0.04}_{-0.04}$ & 1.34$^{+0.05}_{-0.04}$  \\
c$_{nu3}$ & 1.32$^{+0.05}_{-0.04}$ & 1.32$^{+0.04}_{-0.04}$ & 1.32$^{+0.05}_{-0.04}$  \\
c$_{nu4}$ & 1.48$^{+0.04}_{-0.04}$ & 1.49$^{+0.04}_{-0.04}$ & 1.49$^{+0.04}_{-0.04}$  \\
c$_{nu5}$ & 1.41$^{+0.04}_{-0.04}$ & 1.41$^{+0.04}_{-0.04}$ & 1.41$^{+0.04}_{-0.04}$  \\
c$_{nu6}$ & 1.03$^{+0.03}_{-0.03}$ & 1.03$^{+0.03}_{-0.03}$ & 1.03$^{+0.03}_{-0.03}$  \\
c$_{nu7}$ & 0.99$^{+0.03}_{-0.03}$ & 0.99$^{+0.03}_{-0.03}$ & 0.99$^{+0.03}_{-0.03}$  \\
\hline
F$\rm_{2-10\,keV}$ & 3.52$^{+0.05}_{-0.07}$ $\times$ 10$^{-12}$ & 3.59$^{+0.06}_{-0.07}$ $\times$ 10$^{-12}$ & 3.64$^{+0.21}_{-0.51}$ $\times$ 10$^{-12}$  \\
L$\rm_{2-10\,keV}$ & 7.68$^{+0.07}_{-0.07}$ $\times$ 10$^{43}$ & 7.09$^{+0.06}_{-0.06}$ $\times$ 10$^{43}$ & 4.88$^{+0.11}_{-0.15}$ $\times$ 10$^{43}$ \\
\hline
\multicolumn{4}{c}{\textbf{Mrk 376}}\\
\hline
C-stat/dof & 201/199 & 198/201 & 210/205   \\
kT & 0.23$^{+0.03}_{-0.03}$  & 0.20$^{+0.04}_{-0.02}$ & 0.19$^{+0.02}_{-0.01}$ \\
$\Gamma$ & 1.90$^{+0.10}_{-0.10}$ & 1.64$^{+0.06}_{-0.08}$ & 1.70$^{+0.06}_{-0.06}$  \\
norm $\times$ $10^{-2}$ & 0.18$^{+0.03}_{-0.03}$ & 0.10$^{+0.01}_{-0.01}$ & 0.87$^{+0.64}_{-0.54}$ \\
$c_f$ & ... & 1.00(p) & ... \\
CTKcover & ... & ... & 0.45$^{+0.04}_{-0.05}$ \\
TORsigma & ... & ... & 1.50$^{+2.87}_{-0.74}$ \\
$\rm{N_{H,los}}$ $\times$ $10^{24}$ & 0.010$^{+0.001}_{-u}$ & 0.084$^{+0.017}_{-0.020}$ & 0.0056$^{+0.0007}_{-0.0005}$ \\
$\rm{N_{H,avg}}$  $\times$ $10^{24}$ & 2.46$^{+2.79}_{-2.33}$ & 1.11$^{+1.59}_{-0.78}$ & ...  \\
$f_s$ $\times$ 10$^{-2}$ & 1.00(f) & 1.00(f) & 1.00(f) \\
$A_{0}$ & $<$3.06 & ... & ... \\
$A_{90}$ & 1.64$^{+0.84}_{-1.39}$ & ... & ... \\
c$_{nus}$ & 1.01$^{+0.06}_{-0.06}$ & 1.01$^{+0.06}_{-0.06}$ & 1.01$^{+0.06}_{-0.06}$  \\
c$_{xrt}$ & 0.84$^{+0.08}_{-0.08}$ & 0.33$^{+0.09}_{-0.09}$ & 0.88$^{+0.09}_{-0.08}$  \\
\hline
F$\rm_{2-10\,keV}$ & 4.58$^{+0.15}_{-0.32}$ $\times$ 10$^{-12}$ & 4.56$^{+0.13}_{-0.39}$ $\times$ 10$^{-12}$ & 4.58$^{+0.12}_{-0.82}$ $\times$ 10$^{-12}$  \\
L$\rm_{2-10\,keV}$ & 2.93$^{+0.12}_{-0.12}$ $\times$ 10$^{43}$ & 3.33$^{+0.09}_{-0.09}$ $\times$ 10$^{43}$ & 2.62$^{+1.24}_{-1.83}$ $\times$ 10$^{44}$ \\
\hline
\multicolumn{4}{c}{\textbf{NGC 7378}}\\
\hline
C-stat/dof & 87/103 & 89/104 & 96/106   \\
kT & -- & -- & -- \\
$\Gamma$ & 2.22$^{+0.25}_{-0.22}$ & 1.98$^{+0.13}_{-0.05}$ & 1.80$^{+0.11}_{-0.06}$  \\
norm $\times$ $10^{-2}$ & 0.31$^{+0.15}_{-0.12}$ & 0.20$^{+0.08}_{-0.04}$ & 0.48$^{+0.21}_{-0.17}$ \\
$c_f$ & ... & 1.00(p) & ... \\
CTKcover & ... & ... & 0.46$^{+0.01}_{-0.03}$ \\
TORsigma & ... & ... & $<$1.17 \\
$\rm{N_{H,los}}$ $\times$ $10^{24}$ & 0.094$^{+0.023}_{-0.026}$ & 0.12$^{+0.04}_{-0.03}$ & 0.050$^{+0.029}_{-0.025}$ \\
$\rm{N_{H,avg}}$  $\times$ $10^{24}$ & 3.13$^{+4.61}_{-0.99}$ & 2.57$^{+2.00}_{-1.06}$ & ...  \\
$f_s$ $\times$ 10$^{-2}$ & 0.80$^{+0.64}_{-0.50}$ & 1.03$^{+0.74}_{-0.71}$ & 1.00(f) \\
$A_{0}$ & 7.11$^{+6.94}_{-5.47}$ & ... & ... \\
$A_{90}$ & 0.23$^{+2.00}_{-u}$ & ... & ... \\
c$_{nus}$ & 1.09$^{+0.09}_{-0.08}$ & 1.10$^{+0.09}_{-0.08}$ & 1.10$^{+0.09}_{-0.08}$  \\
c$_{xrt}$ & 0.55$^{+0.09}_{-0.08}$ & 0.55$^{+0.09}_{-0.09}$ & 0.53$^{+0.08}_{-0.08}$  \\
\hline
F$\rm_{2-10\,keV}$ & 2.80$^{+0.15}_{-0.22}$ $\times$ 10$^{-12}$ & 2.82$^{+0.05}_{-0.21}$ $\times$ 10$^{-12}$ & 2.87$^{+2.33}_{-0.72}$ $\times$ 10$^{-12}$  \\
L$\rm_{2-10\,keV}$ & 1.03$^{+0.04}_{-0.04}$ $\times$ 10$^{42}$ & 9.46$^{+0.37}_{-0.37}$ $\times$ 10$^{41}$ & 2.97$^{+0.78}_{-1.29}$ $\times$ 10$^{42}$ \\
\hline
\multicolumn{4}{c}{\textbf{2MASX J11462959+7421289}}\\
\hline
C-stat/dof & 107/96 & 108/96 & 116/96   \\
kT & -- & -- & -- \\
$\Gamma$ & 1.75$^{+0.17}_{-0.09}$ & 1.70$^{+0.18}_{-0.09}$ & 1.74$^{+0.08}_{-0.06}$  \\
norm $\times$ $10^{-2}$ & 0.083$^{+0.032}_{-0.012}$ & 0.070$^{+0.029}_{-0.012}$ & 0.86$^{+0.89}_{-0.44}$ \\
$c_f$ & ... & 1.00(p) & ... \\
CTKcover & ... & ... & 0.36$^{+0.05}_{-0.17}$ \\
TORsigma & ... & ... & 0.79$^{+1.71}_{-0.44}$ \\
$\rm{N_{H,los}}$ $\times$ $10^{24}$ & 0.010$^{+0.003}_{-u}$ & 0.013$^{+0.060}_{-0.010}$ & 0.045$^{+0.007}_{-0.010}$ \\
$\rm{N_{H,avg}}$  $\times$ $10^{24}$ & 0.20$^{+1.29}_{-u}$ & 0.11$^{+0.65}_{-u}$ & ...  \\
$f_s$ $\times$ 10$^{-2}$ & 1.00(f) & 1.00(f) & 1.00(f) \\
$A_{0}$ & 1.00(f) & ... & ... \\
$A_{90}$ & 1.00(f) & ... & ... \\
c$_{nus}$ & 1.03$^{+0.08}_{-0.08}$ & 1.03$^{+0.09}_{-0.08}$ & 1.03$^{+0.09}_{-0.08}$  \\
c$_{xrt}$ & 0.86$^{+0.16}_{-0.11}$ & 0.84$^{+0.18}_{-0.16}$ & 0.62$^{+0.12}_{-0.11}$  \\
\hline
F$\rm_{2-10\,keV}$ & 2.79$^{+0.03}_{-0.21}$ $\times$ 10$^{-12}$ & 2.75$^{+0.02}_{-0.46}$ $\times$ 10$^{-12}$ & 2.85$^{+0.12}_{-1.37}$ $\times$ 10$^{-12}$  \\
L$\rm_{2-10\,keV}$ & 2.27$^{+0.09}_{-0.09}$ $\times$ 10$^{43}$ & 2.08$^{+0.11}_{-0.11}$ $\times$ 10$^{43}$ & 2.62$^{+0.15}_{-0.09}$ $\times$ 10$^{43}$ \\
\hline
\multicolumn{4}{c}{\textbf{SWIFT J2006.5+5619}}\\
\hline
C-stat/dof & 84/74 & 108/96 & 116/96   \\
kT & -- & -- & -- \\
$\Gamma$ & 1.55$^{+0.15}_{-0.11}$ & 1.94$^{+0.17}_{-0.20}$ & 1.93$^{+0.12}_{-0.16}$  \\
norm $\times$ $10^{-2}$ & 0.070$^{+0.017}_{-0.024}$ & 0.19$^{+0.14}_{-0.10}$ & 0.22$^{+0.09}_{-0.08}$ \\
$c_f$ & ... & 1.00(p) & ... \\
CTKcover & ... & ... & 0.60$^{+u}_{-0.42}$ \\
TORsigma & ... & ... & 84.00(p) \\
$\rm{N_{H,los}}$ $\times$ $10^{24}$ & 0.15$^{+0.04}_{-0.03}$ & 0.31$^{+0.09}_{-0.11}$ & 0.19$^{+0.04}_{-0.05}$ \\
$\rm{N_{H,avg}}$  $\times$ $10^{24}$ & 0.20$^{+0.20}_{-0.16}$ & 3.24$^{+u}_{-1.46}$ & ...  \\
$f_s$ $\times$ 10$^{-2}$ & 1.00(f) & 1.00(f) & 1.00(f) \\
$A_{0}$ & 1.00(f) & ... & ... \\
$A_{90}$ & 1.00(f) & ... & ... \\
c$_{nus}$ & 0.99$^{+0.09}_{-0.08}$ & 0.99$^{+0.09}_{-0.08}$ & 0.99$^{+0.09}_{-0.08}$  \\
c$_{xrt}$ & 0.83$^{+0.19}_{-0.17}$ & 1.04$^{+0.18}_{-0.16}$ & 1.04$^{+0.21}_{-0.19}$  \\
\hline
F$\rm_{2-10\,keV}$ & 1.67$^{+0.06}_{-0.26}$ $\times$ 10$^{-12}$ & 1.80$^{+0.33}_{-1.46}$ $\times$ 10$^{-12}$ & 1.69$^{+0.11}_{-0.21}$ $\times$ 10$^{-12}$  \\
L$\rm_{2-10\,keV}$ & 1.51$^{+0.07}_{-0.07}$ $\times$ 10$^{43}$ & 2.69$^{+0.15}_{-0.15}$ $\times$ 10$^{43}$ & 2.82$^{+0.12}_{-0.11}$ $\times$ 10$^{43}$ \\
\hline
\multicolumn{4}{c}{\textbf{2MASX J06363227-2034532}}\\
\hline
C-stat/dof & 44/45 & 44/45 & 43/45   \\
kT & 0.23$^{+0.25}_{-0.08}$ & 0.22$^{+0.10}_{-0.06}$ & 0.20$^{+0.08}_{-0.05}$ \\
$\Gamma$ & 1.40$^{+1.2}_{-u}$ & 1.40$^{+1.2}_{-u}$ & 1.59$^{+0.28}_{-0.23}$  \\
norm $\times$ $10^{-2}$ & 0.11$^{+0.12}_{-0.09}$ & 0.17$^{+0.04}_{-0.03}$ & 0.15$^{+0.29}_{-0.05}$ \\
$c_f$ & ... & 0.10(p) & ... \\
CTKcover & ... & ... & 0.60$^{+u}_{-0.50}$ \\
TORsigma & ... & ... & 28.07$^{+29.03}_{-25.09}$ \\
$\rm{N_{H,los}}$ $\times$ $10^{24}$ & 0.92$^{+0.11}_{-0.11}$ & 1.48$^{+0.21}_{-0.17}$ & 0.86$^{+0.22}_{-0.10}$ \\
$\rm{N_{H,avg}}$  $\times$ $10^{24}$ & 0.01(p) & 0.01(p) & ...  \\
$f_s$ $\times$ 10$^{-2}$ & 1.09$^{+1.16}_{-0.51}$ & 0.78$^{+0.72}_{-0.74}$ & 7.85$^{+13.32}_{-0.05}$ \\
$A_{0}$ & 1.00(f) & ... & ... \\
$A_{90}$ & 1.00(f) & ... & ... \\
c$_{nus}$ & 1.00$^{+0.12}_{-0.12}$ & 1.00$^{+0.11}_{-0.08}$ & 1.02$^{+0.13}_{-0.12}$  \\
c$_{xrt}$ & 0.64$^{+0.28}_{-0.24}$ & 0.68$^{+0.25}_{-0.21}$ & 0.62$^{+0.24}_{-0.22}$  \\
\hline
F$\rm_{2-10\,keV}$ & 4.83$^{+1.39}_{-1.54}$ $\times$ 10$^{-13}$ & 4.82$^{+1.06}_{-2.40}$ $\times$ 10$^{-13}$ & 4.92$^{+2.23}_{-3.31}$ $\times$ 10$^{-13}$  \\
L$\rm_{2-10\,keV}$ & 5.03$^{+0.33}_{-0.33}$ $\times$ 10$^{43}$ & 7.94$^{+0.52}_{-0.52}$ $\times$ 10$^{43}$ & 5.37$^{+0.65}_{-0.72}$ $\times$ 10$^{43}$ \\
\hline
\multicolumn{4}{c}{\textbf{2MASX J09034285-7414170}}\\
\hline
C-stat/dof & 89/101 & 99/101 & 99/100   \\
kT & -- & -- & -- \\
$\Gamma$ & 2.35$^{+u}_{-0.53}$ & 1.73$^{+0.40}_{-0.15}$ & 1.73$^{+0.32}_{-0.22}$  \\
norm $\times$ $10^{-2}$ & 0.40$^{+0.67}_{-0.33}$ & 0.10$^{+0.30}_{-0.06}$ & 0.13$^{+0.21}_{-0.06}$ \\
$c_f$ & ... & 1.00(p) & ... \\
CTKcover & ... & ... & 0.30$^{+0.30}_{-u}$ \\
TORsigma & ... & ... & 84.00$^{+u}_{-37.92}$ \\
$\rm{N_{H,los}}$ $\times$ $10^{24}$ & 0.40$^{+0.22}_{-0.21}$ & 0.49$^{+0.21}_{-0.17}$ & 0.34$^{+0.17}_{-0.16}$ \\
$\rm{N_{H,avg}}$  $\times$ $10^{24}$ & 3.50$^{+5.69}_{-2.27}$ & 1.32$^{+0.74}_{-0.86}$ & ...  \\
$f_s$ $\times$ 10$^{-2}$ & 0.14$^{+1.22}_{-u}$ & 0.67$^{+1.38}_{-0.48}$ & 1.00(f) \\
$A_{0}$ & 9.89$^{+13.60}_{-u}$ & ... & ... \\
$A_{90}$ & 1.59$^{+u}_{-2.32}$ & ... & ... \\
c$_{nus}$ & 1.00$^{+0.12}_{-0.11}$ & 1.01$^{+0.13}_{-0.11}$ & 1.01$^{+0.13}_{-0.11}$  \\
c$_{xrt}$ & 0.74$^{+0.23}_{-0.20}$ & 0.83$^{+0.27}_{-0.24}$ & 0.84$^{+0.26}_{-0.24}$  \\
\hline
F$\rm_{2-10\,keV}$ & 1.02$^{+0.16}_{-0.18}$ $\times$ 10$^{-12}$ & 1.01$^{+0.07}_{-0.77}$ $\times$ 10$^{-12}$ & 1.02$^{+0.03}_{-0.73}$ $\times$ 10$^{-12}$  \\
L$\rm_{2-10\,keV}$ & 3.96$^{+0.40}_{-0.40}$ $\times$ 10$^{43}$ & 8.06$^{+0.58}_{-0.58}$ $\times$ 10$^{43}$ & 1.09$^{+0.40}_{-0.06}$ $\times$ 10$^{43}$ \\
\hline
\multicolumn{4}{c}{\textbf{2MASX J00091156-0036551}}\\
\hline
C-stat/dof & 53/48 & 53/50 & 53/49   \\
kT & -- & -- & -- \\
$\Gamma$ & 1.54$^{+0.25}_{-u}$ & 1.84$^{+0.40}_{-0.15}$ & 1.61$^{+0.21}_{-0.23}$  \\
norm $\times$ $10^{-2}$ & 0.05$^{+0.15}_{-0.06}$ & 0.09$^{+0.30}_{-0.06}$ & 0.12$^{+0.02}_{-0.05}$ \\
$c_f$ & ... & 1.00(p) & ... \\
CTKcover & ... & ... & 0.13$^{+0.24}_{-0.06}$ \\
TORsigma & ... & ... & $<$9.71 \\
$\rm{N_{H,los}}$ $\times$ $10^{24}$ & 0.29$^{+0.06}_{-0.09}$ & 0.45$^{+0.16}_{-0.16}$ & 0.27$^{+0.12}_{-0.09}$ \\
$\rm{N_{H,avg}}$  $\times$ $10^{24}$ & 0.01(p) & 1.80$^{+1.32}_{-0.92}$ & ...  \\
$f_s$ $\times$ 10$^{-2}$ & 1.23$^{+1.62}_{-1.23}$ & 0.67$^{+1.72}_{-0.48}$ & 1.00(f) \\
$A_{0}$ & 1.00(p) & ... & ... \\
$A_{90}$ & 0.98$^{+12.95}_{-u}$ & ... & ... \\
c$_{nus}$ & 0.88$^{+0.13}_{-0.12}$ & 1.01$^{+0.13}_{-0.11}$ & 0.88$^{+0.13}_{-0.12}$  \\
c$_{xrt}$ & 0.83$^{+0.26}_{-0.22}$ & 0.83$^{+0.24}_{-0.21}$ & 0.84$^{+0.23}_{-0.20}$  \\
\hline
F$\rm_{2-10\,keV}$ & 7.84$^{+1.00}_{-1.54}$ $\times$ 10$^{-13}$ & 6.95$^{+0.96}_{-3.66}$ $\times$ 10$^{-13}$ & 7.77$^{+10.58}_{-2.24}$ $\times$ 10$^{-13}$  \\
L$\rm_{2-10\,keV}$ & 3.17$^{+0.22}_{-0.22}$ $\times$ 10$^{43}$ & 3.90$^{+0.31}_{-0.31}$ $\times$ 10$^{43}$ & 7.26$^{+0.35}_{-0.36}$ $\times$ 10$^{43}$ \\
\hline\hline
\enddata
\tablecomments{\small
    $\chi^2$/dof: $\chi^2$ divided by degrees of freedom (total fit statistics). \\
    C-stat/dof: Cash statistics divided by degrees of freedom (total fit statistics). \\
    kT: \texttt{mekal} model temperature in units of keV. \\
    $\Gamma$: Power law photon index. \\
    norm: the main power-law normalization (in units of photons cm$^{-2}$ s$^{-1}$ keV$^{-1}$ $\times$ $10^{-2}$, measured at 1 keV). \\
    $c_f$: Covering factor of the torus. \\
    \texttt{CTKcover}: Covering factor of the inner ring of clouds. \\ \texttt{TORsigma}: Cloud dispersion factor, computed with \ux. \\
    $\rm{N_{H,los}}$: line-of-sight torus hydrogen column density, in units of $10^{24}$ cm$^{-2}$. \\
    $\rm{N_{H,avg}}$:  Average torus hydrogen column density, in units of $10^{24}$ cm$^{-2}$. \\
    $f_s$: Fraction of scattered continuum. \\
    $A_{90}$: Relative weighs of the \texttt{MYTS} and \texttt{MYTL} component for $\theta_{obs} = 90^{\circ}$.\\
    $A_{0}$: Relative weighs of the \texttt{MYTS} and \texttt{MYTL} component for $\theta_{obs} = 00^{\circ}$. \\
    $c$: The cross-normalization constant between the instruments. \\
    F$_{\rm 2-10\,keV}$: Observed flux in the 2--10\,keV band with units of erg cm$^{-2}$ s$^{-1}$. \\
    L$_{\rm 2-10\,keV}$: Intrinsic luminosity in the 2--10\,keV band with units of erg s$^{-1}$. \\
    $*$: Indicates the parameter was frozen to this value during fitting. \\
    u: The parameter is unconstrained. \\
    p: The parameter is pegged.}
\end{deluxetable*} 
\end{center}

\vspace{2cm}


\clearpage

\section{Supplementary Tables}\label{lumin:table}
We list in Tables~\ref{tab:type1} and~\ref{tab:type2} the derived bolometric luminosities, bolometric correction factors, and Eddington luminosities for the type~1 and type~2 AGN samples. The values illustrate the range of corrections applied from the intrinsic 2-10 keV luminosity, following the prescriptions described above.

\vspace{1cm}
\begin{table}[H]
\centering
\caption{Bolometric luminosity, correction factor, and Eddington luminosity for type~1 sources.}
\label{tab:type1}
\begin{tabular}{rrrr}
\toprule
 Source name & log$\rm{L_{bol}}$ &    $K_X$ &  log$\rm{L_{Edd}}$ \\
\midrule
   UGC 5101 & 44.46 & 14.36 &    46.46 \\
   IC 1198 & 45.21 & 18.37 &    45.62 \\
   MRK 231 & 44.64 & 14.95 &    46.05 \\
   PG 1211+143 & 44.90 & 16.09 &    46.71 \\
   MRK 376 & 45.88 & 28.64 &    46.33 \\
\bottomrule
\end{tabular}
\end{table}

\begin{table} [H]
\centering
\caption{Bolometric luminosity, correction factor, and Eddington luminosity for type~2 sources.}
\label{tab:type2}
\begin{tabular}{rrrr}
\toprule
 Source name & log$\rm{L_{bol}}$ &    $K_X$ &  log$\rm{L_{Edd}}$ \\
\midrule
   MRK 1073 & 44.27 & 12.12 &    45.89 \\
   NGC 7674 & 43.94 & 11.54 &    45.84 \\
   ESO 406$-$4 & 43.65 & 11.23 &    46.13 \\
   2MASX J03585442+1026033  & 44.66 & 13.47 &    45.86 \\
   M58 & 42.77 & 10.91 &    46.21 \\
   2MASX J09261742$-$8421330  & 45.15 & 17.03 &    45.21 \\
   ESO 234$-$50  & 42.69 & 10.90 &    44.15 \\
   NGC 2273 & 43.65 & 11.24 &    46.33 \\
   NGC 7378 & 43.52 & 11.15 &    43.60 \\
   2MASX J11462959+7421289 & 44.53 & 12.90 &    46.33 \\
   SWIFT J2006.5+5619  & 44.57 & 13.04 &    45.18 \\
   2MASX J06363227$-$2034532 & 44.90 & 14.85 &    46.55 \\
   2MASX J09034285$-$7414170 & 44.11 & 11.79 &    45.99 \\
   2MASX J00091156$-$0036551 & 45.07 & 16.23 &    46.65 \\
\bottomrule
\end{tabular}
\end{table}


\smallskip
\section{Best-fit spectra}\label{spectral fit 25}
\begin{figure*}
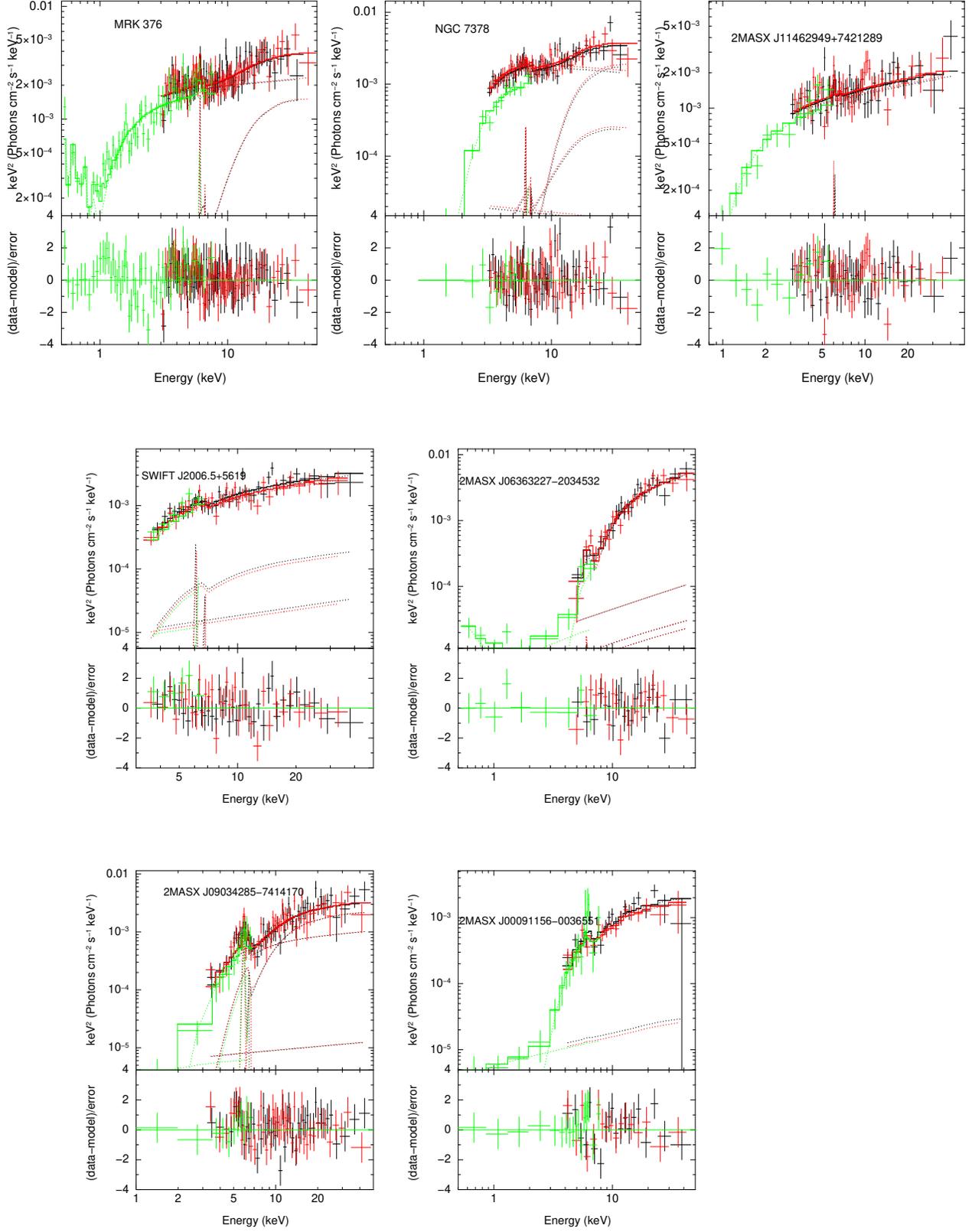

\centering
\hbox{
\hspace{1.2cm}
    \includegraphics[scale=0.26]{2mas170.eps}  
    \includegraphics[scale=0.26]{2mas551.eps}
    }
\caption{Best-fit unfolded spectra and residuals obtained with the \myt\ model for 25 sources. The source names are indicated in the plot legends.}\label{figure-38}
\end{figure*}

The best-fitted unfolded spectra and the data-to-model residuals of 25 sources are presented in the following section. The modeling is carried out using \myt\ and presented in Fig. \ref{figure-38}.
\clearpage

\newpage

\begin{figure*}
\centering
\hbox{
\hspace{-0.2cm}
    \includegraphics[scale=0.28]{MRK1073.eps}
    \hspace{-0.5cm}  
    \includegraphics[scale=0.28]{UGC5101.eps}
    \hspace{-0.5cm}
    \includegraphics[scale=0.28]{NGC7674.eps}}
\label{figure-30}
\end{figure*}
\begin{figure*}
\centering
\hbox{
\hspace{-0.2cm}
    \includegraphics[scale=0.28]{ESO362.eps}
    \hspace{-0.5cm}  
    \includegraphics[scale=0.28]{ESO406.eps}
    \hspace{-0.5cm}
    \includegraphics[scale=0.28]{2MFGC13496.eps}}
\label{figure-31}
\end{figure*}
\begin{figure*}
\centering
\hbox{
\hspace{-0.2cm}
    \includegraphics[scale=0.28]{2MAS033.eps}
    \hspace{-0.5cm}  
    \includegraphics[scale=0.28]{M58.eps}
    \hspace{-0.5cm}
    \includegraphics[scale=0.28]{3c371.eps}}
\label{figure-32}
\end{figure*}
\begin{figure*}
\centering
\hbox{
\hspace{-0.2cm}
    \includegraphics[scale=0.28]{IC1198.eps}
    \hspace{-0.5cm}  
    \includegraphics[scale=0.28]{2mas330.eps}
    \hspace{-0.5cm}
    \includegraphics[scale=0.28]{UGC12348.eps}}
\label{figure-33}
\end{figure*}

\begin{figure*}
\centering
\hbox{
\hspace{-0.2cm}
    \includegraphics[scale=0.28]{2MAS61.eps}
    \hspace{-0.5cm}  
    \includegraphics[scale=0.28]{ESO234.eps}
    \hspace{-0.5cm}
    \includegraphics[scale=0.28]{NGC2273.eps}}
\label{figure-34}
\end{figure*}

\begin{figure*}
\centering
\hbox{
\hspace{-0.2cm}
    \includegraphics[scale=0.28]{FRL265.eps}
    \hspace{-0.5cm}  
    \includegraphics[scale=0.28]{MRK231.eps}
    \hspace{-0.5cm}
    \includegraphics[scale=0.28]{PG1211.eps}}
\label{figure-35}
\end{figure*}

\begin{figure*}
\centering
\hbox{
\hspace{-0.2cm}
    \includegraphics[scale=0.28]{MRK376.eps}
    \hspace{-0.5cm}  
    \includegraphics[scale=0.28]{NGC7378.eps}
    \hspace{-0.5cm}
    \includegraphics[scale=0.28]{2mas1289.eps}}
\label{figure-36}
\end{figure*}

\begin{figure*}
\centering
\hbox{
\hspace{1.2cm}
    \includegraphics[scale=0.26]{swift5619.eps}  
    \includegraphics[scale=0.26]{2mas532.eps}
    }
\label{figure-37}
\end{figure*}

\clearpage

\bibliography{sample701}{}

\begin{thebibliography}{}
\expandafter\ifx\csname natexlab\endcsname\relax\def\natexlab#1{#1}\fi
\providecommand{\url}[1]{\href{#1}{#1}}
\providecommand{\dodoi}[1]{doi:~\href{http://doi.org/#1}{\nolinkurl{#1}}}
\providecommand{\doeprint}[1]{\href{http://ascl.net/#1}{\nolinkurl{http://ascl.net/#1}}}
\providecommand{\doarXiv}[1]{\href{https://arxiv.org/abs/#1}{\nolinkurl{https://arxiv.org/abs/#1}}}

\bibitem[{{Ajello} {et~al.}(2008){Ajello}, {Greiner}, {Sato}, {Willis}, {Kanbach}, {Strong}, {Diehl}, {Hasinger}, {Gehrels}, {Markwardt}, \& {Tueller}}]{2008ApJ...689..666A}
{Ajello}, M., {Greiner}, J., {Sato}, G., {et~al.} 2008, \apj, 689, 666, \dodoi{10.1086/592595}

\bibitem[{{Akylas} {et~al.}(2016){Akylas}, {Georgantopoulos}, {Ranalli}, {Gkiokas}, {Corral}, \& {Lanzuisi}}]{2016A&A...594A..73A}
{Akylas}, A., {Georgantopoulos}, I., {Ranalli}, P., {et~al.} 2016, \aap, 594, A73, \dodoi{10.1051/0004-6361/201628711}

\bibitem[{{Alexander} {et~al.}(2003){Alexander}, {Bauer}, {Brandt}, {Schneider}, {Hornschemeier}, {Vignali}, {Barger}, {Broos}, {Cowie}, {Garmire}, {Townsley}, {Bautz}, {Chartas}, \& {Sargent}}]{2003AJ....126..539A}
{Alexander}, D.~M., {Bauer}, F.~E., {Brandt}, W.~N., {et~al.} 2003, \aj, 126, 539, \dodoi{10.1086/376473}

\bibitem[{{Ananna} {et~al.}(2019){Ananna}, {Treister}, {Urry}, {Ricci}, {Kirkpatrick}, {LaMassa}, {Buchner}, {Civano}, {Tremmel}, \& {Marchesi}}]{2019ApJ...871..240A}
{Ananna}, T.~T., {Treister}, E., {Urry}, C.~M., {et~al.} 2019, \apj, 871, 240, \dodoi{10.3847/1538-4357/aafb77}

\bibitem[{{Arnaud}(1996)}]{1996ASPC..101...17A}
{Arnaud}, K.~A. 1996, in Astronomical Society of the Pacific Conference Series, Vol. 101, Astronomical Data Analysis Software and Systems V, ed. G.~H. {Jacoby} \& J.~{Barnes}, 17

\bibitem[{{Asmus} {et~al.}(2015){Asmus}, {Gandhi}, {H{\"o}nig}, {Smette}, \& {Duschl}}]{2015MNRAS.454..766A}
{Asmus}, D., {Gandhi}, P., {H{\"o}nig}, S.~F., {Smette}, A., \& {Duschl}, W.~J. 2015, \mnras, 454, 766, \dodoi{10.1093/mnras/stv1950}

\bibitem[{{Balokovi{\'c}} {et~al.}(2018){Balokovi{\'c}}, {Brightman}, {Harrison}, {Comastri}, {Ricci}, {Buchner}, {Gandhi}, {Farrah}, \& {Stern}}]{2018ApJ...854...42B}
{Balokovi{\'c}}, M., {Brightman}, M., {Harrison}, F.~A., {et~al.} 2018, \apj, 854, 42, \dodoi{10.3847/1538-4357/aaa7eb}

\bibitem[{{Boorman} {et~al.}(2025){Boorman}, {Gandhi}, {Buchner}, {Stern}, {Ricci}, {Balokovi{\'c}}, {Asmus}, {Harrison}, {Svoboda}, {Greenwell}, {Koss}, {Alexander}, {Annuar}, {Bauer}, {Brandt}, {Brightman}, {Civano}, {Chen}, {Farrah}, {Forster}, {Grefenstette}, {H{\"o}nig}, {Hill}, {Kammoun}, {Lansbury}, {Lanz}, {LaMassa}, {Madsen}, {Marchesi}, {Middleton}, {Mingo}, {Parker}, {Treister}, {Ueda}, {Urry}, \& {Zappacosta}}]{2025ApJ...978..118B}
{Boorman}, P.~G., {Gandhi}, P., {Buchner}, J., {et~al.} 2025, \apj, 978, 118, \dodoi{10.3847/1538-4357/ad8236}

\bibitem[{{Braito} {et~al.}(2004){Braito}, {Della Ceca}, {Piconcelli}, {Severgnini}, {Bassani}, {Cappi}, {Franceschini}, {Iwasawa}, {Malaguti}, {Marziani}, {Palumbo}, {Persic}, {Risaliti}, \& {Salvati}}]{2004A&A...420...79B}
{Braito}, V., {Della Ceca}, R., {Piconcelli}, E., {et~al.} 2004, \aap, 420, 79, \dodoi{10.1051/0004-6361:20040061}

\bibitem[{{Brightman} \& {Nandra}(2011)}]{2011MNRAS.413.1206B}
{Brightman}, M., \& {Nandra}, K. 2011, \mnras, 413, 1206, \dodoi{10.1111/j.1365-2966.2011.18207.x}

\bibitem[{{Brightman} {et~al.}(2015){Brightman}, {Balokovi{\'c}}, {Stern}, {Ar{\'e}valo}, {Ballantyne}, {Bauer}, {Boggs}, {Craig}, {Christensen}, {Comastri}, {Fuerst}, {Gandhi}, {Hailey}, {Harrison}, {Hickox}, {Koss}, {LaMassa}, {Puccetti}, {Rivers}, {Vasudevan}, {Walton}, \& {Zhang}}]{2015ApJ...805...41B}
{Brightman}, M., {Balokovi{\'c}}, M., {Stern}, D., {et~al.} 2015, \apj, 805, 41, \dodoi{10.1088/0004-637X/805/1/41}

\bibitem[{{Buchner} {et~al.}(2019{\natexlab{a}}){Buchner}, {Brightman}, {Nandra}, {Nikutta}, \& {Bauer}}]{Buchner2019}
{Buchner}, J., {Brightman}, M., {Nandra}, K., {Nikutta}, R., \& {Bauer}, F.~E. 2019{\natexlab{a}}, \aap, 629, A16, \dodoi{10.1051/0004-6361/201834771}

\bibitem[{{Buchner} {et~al.}(2019{\natexlab{b}}){Buchner}, {Brightman}, {Nandra}, {Nikutta}, \& {Bauer}}]{2019A&A...629A..16B}
---. 2019{\natexlab{b}}, \aap, 629, A16, \dodoi{10.1051/0004-6361/201834771}

\bibitem[{{Buchner} {et~al.}(2015){Buchner}, {Georgakakis}, {Nandra}, {Brightman}, {Menzel}, {Liu}, {Hsu}, {Salvato}, {Rangel}, {Aird}, {Merloni}, \& {Ross}}]{2015ApJ...802...89B}
{Buchner}, J., {Georgakakis}, A., {Nandra}, K., {et~al.} 2015, \apj, 802, 89, \dodoi{10.1088/0004-637X/802/2/89}

\bibitem[{{Burlon} {et~al.}(2011){Burlon}, {Ajello}, {Greiner}, {Comastri}, {Merloni}, \& {Gehrels}}]{2011ApJ...728...58B}
{Burlon}, D., {Ajello}, M., {Greiner}, J., {et~al.} 2011, \apj, 728, 58, \dodoi{10.1088/0004-637X/728/1/58}

\bibitem[{{Burrows} {et~al.}(2005){Burrows}, {Hill}, {Nousek}, {Kennea}, {Wells}, {Osborne}, {Abbey}, {Beardmore}, {Mukerjee}, {Short}, {Chincarini}, {Campana}, {Citterio}, {Moretti}, {Pagani}, {Tagliaferri}, {Giommi}, {Capalbi}, {Tamburelli}, {Angelini}, {Cusumano}, {Br{\"a}uninger}, {Burkert}, \& {Hartner}}]{2005SSRv..120..165B}
{Burrows}, D.~N., {Hill}, J.~E., {Nousek}, J.~A., {et~al.} 2005, \ssr, 120, 165, \dodoi{10.1007/s11214-005-5097-2}

\bibitem[{{Campitiello} {et~al.}(2020){Campitiello}, {Celotti}, {Ghisellini}, \& {Sbarrato}}]{2020A&A...640A..39C}
{Campitiello}, S., {Celotti}, A., {Ghisellini}, G., \& {Sbarrato}, T. 2020, \aap, 640, A39, \dodoi{10.1051/0004-6361/201936218}

\bibitem[{{Corral} {et~al.}(2011){Corral}, {Della Ceca}, {Caccianiga}, {Severgnini}, {Brunner}, {Carrera}, {Page}, \& {Schwope}}]{2011A&A...530A..42C}
{Corral}, A., {Della Ceca}, R., {Caccianiga}, A., {et~al.} 2011, \aap, 530, A42, \dodoi{10.1051/0004-6361/201015227}

\bibitem[{{Dasyra} {et~al.}(2011){Dasyra}, {Ho}, {Netzer}, {Combes}, {Trakhtenbrot}, {Sturm}, {Armus}, \& {Elbaz}}]{2011ApJ...740...94D}
{Dasyra}, K.~M., {Ho}, L.~C., {Netzer}, H., {et~al.} 2011, \apj, 740, 94, \dodoi{10.1088/0004-637X/740/2/94}

\bibitem[{{Decarli} {et~al.}(2008){Decarli}, {Dotti}, {Fontana}, \& {Haardt}}]{2008MNRAS.386L..15D}
{Decarli}, R., {Dotti}, M., {Fontana}, M., \& {Haardt}, F. 2008, \mnras, 386, L15, \dodoi{10.1111/j.1745-3933.2008.00451.x}

\bibitem[{{Duras} {et~al.}(2020){Duras}, {Bongiorno}, {Ricci}, {Piconcelli}, {Shankar}, {Lusso}, {Bianchi}, {Fiore}, {Maiolino}, {Marconi}, {Onori}, {Sani}, {Schneider}, {Vignali}, \& {La Franca}}]{2020A&A...636A..73D}
{Duras}, F., {Bongiorno}, A., {Ricci}, F., {et~al.} 2020, \aap, 636, A73, \dodoi{10.1051/0004-6361/201936817}

\bibitem[{{Elitzur} \& {Shlosman}(2006)}]{2006ApJ...648L.101E}
{Elitzur}, M., \& {Shlosman}, I. 2006, \apjl, 648, L101, \dodoi{10.1086/508158}

\bibitem[{{Evans} {et~al.}(2009){Evans}, {Beardmore}, {Page}, {Osborne}, {O'Brien}, {Willingale}, {Starling}, {Burrows}, {Godet}, {Vetere}, {Racusin}, {Goad}, {Wiersema}, {Angelini}, {Capalbi}, {Chincarini}, {Gehrels}, {Kennea}, {Margutti}, {Morris}, {Mountford}, {Pagani}, {Perri}, {Romano}, \& {Tanvir}}]{2009MNRAS.397.1177E}
{Evans}, P.~A., {Beardmore}, A.~P., {Page}, K.~L., {et~al.} 2009, \mnras, 397, 1177, \dodoi{10.1111/j.1365-2966.2009.14913.x}

\bibitem[{{Fruscione} {et~al.}(2006){Fruscione}, {McDowell}, {Allen}, {Brickhouse}, {Burke}, {Davis}, {Durham}, {Elvis}, {Galle}, {Harris}, {Huenemoerder}, {Houck}, {Ishibashi}, {Karovska}, {Nicastro}, {Noble}, {Nowak}, {Primini}, {Siemiginowska}, {Smith}, \& {Wise}}]{2006SPIE.6270E..1VF}
{Fruscione}, A., {McDowell}, J.~C., {Allen}, G.~E., {et~al.} 2006, in Society of Photo-Optical Instrumentation Engineers (SPIE) Conference Series, Vol. 6270, Observatory Operations: Strategies, Processes, and Systems, ed. D.~R. {Silva} \& R.~E. {Doxsey}, 62701V, \dodoi{10.1117/12.671760}

\bibitem[{{Fukazawa} {et~al.}(2011){Fukazawa}, {Hiragi}, {Mizuno}, {Nishino}, {Hayashi}, {Yamasaki}, {Shirai}, {Takahashi}, \& {Ohno}}]{2011ApJ...727...19F}
{Fukazawa}, Y., {Hiragi}, K., {Mizuno}, M., {et~al.} 2011, \apj, 727, 19, \dodoi{10.1088/0004-637X/727/1/19}

\bibitem[{{Gabriel} {et~al.}(2004){Gabriel}, {Denby}, {Fyfe}, {Hoar}, {Ibarra}, {Ojero}, {Osborne}, {Saxton}, {Lammers}, \& {Vacanti}}]{2004ASPC..314..759G}
{Gabriel}, C., {Denby}, M., {Fyfe}, D.~J., {et~al.} 2004, in Astronomical Society of the Pacific Conference Series, Vol. 314, Astronomical Data Analysis Software and Systems (ADASS) XIII, ed. F.~{Ochsenbein}, M.~G. {Allen}, \& D.~{Egret}, 759

\bibitem[{{Gandhi} {et~al.}(2017){Gandhi}, {Annuar}, {Lansbury}, {Stern}, {Alexander}, {Bauer}, {Bianchi}, {Boggs}, {Boorman}, {Brandt}, {Brightman}, {Christensen}, {Comastri}, {Craig}, {Del Moro}, {Elvis}, {Guainazzi}, {Hailey}, {Harrison}, {Koss}, {Lamperti}, {Malaguti}, {Masini}, {Matt}, {Puccetti}, {Ricci}, {Rivers}, {Walton}, \& {Zhang}}]{2017MNRAS.467.4606G}
{Gandhi}, P., {Annuar}, A., {Lansbury}, G.~B., {et~al.} 2017, \mnras, 467, 4606, \dodoi{10.1093/mnras/stx357}

\bibitem[{{Gaspari} \& {S{\k{a}}dowski}(2017)}]{2017ApJ...837..149G}
{Gaspari}, M., \& {S{\k{a}}dowski}, A. 2017, \apj, 837, 149, \dodoi{10.3847/1538-4357/aa61a3}

\bibitem[{{Gehrels} {et~al.}(2004){Gehrels}, {Chincarini}, {Giommi}, {Mason}, {Nousek}, {Wells}, {White}, {Barthelmy}, {Burrows}, {Cominsky}, {Hurley}, {Marshall}, {M{\'e}sz{\'a}ros}, {Roming}, {Angelini}, {Barbier}, {Belloni}, {Campana}, {Caraveo}, {Chester}, {Citterio}, {Cline}, {Cropper}, {Cummings}, {Dean}, {Feigelson}, {Fenimore}, {Frail}, {Fruchter}, {Garmire}, {Gendreau}, {Ghisellini}, {Greiner}, {Hill}, {Hunsberger}, {Krimm}, {Kulkarni}, {Kumar}, {Lebrun}, {Lloyd-Ronning}, {Markwardt}, {Mattson}, {Mushotzky}, {Norris}, {Osborne}, {Paczynski}, {Palmer}, {Park}, {Parsons}, {Paul}, {Rees}, {Reynolds}, {Rhoads}, {Sasseen}, {Schaefer}, {Short}, {Smale}, {Smith}, {Stella}, {Tagliaferri}, {Takahashi}, {Tashiro}, {Townsley}, {Tueller}, {Turner}, {Vietri}, {Voges}, {Ward}, {Willingale}, {Zerbi}, \& {Zhang}}]{2004ApJ...611.1005G}
{Gehrels}, N., {Chincarini}, G., {Giommi}, P., {et~al.} 2004, \apj, 611, 1005, \dodoi{10.1086/422091}

\bibitem[{{Gilli} {et~al.}(2007){Gilli}, {Comastri}, \& {Hasinger}}]{2007A&A...463...79G}
{Gilli}, R., {Comastri}, A., \& {Hasinger}, G. 2007, \aap, 463, 79, \dodoi{10.1051/0004-6361:20066334}

\bibitem[{{Guainazzi} {et~al.}(2005){Guainazzi}, {Matt}, \& {Perola}}]{2005A&A...444..119G}
{Guainazzi}, M., {Matt}, G., \& {Perola}, G.~C. 2005, \aap, 444, 119, \dodoi{10.1051/0004-6361:20053643}

\bibitem[{{Harrison} {et~al.}(2013){Harrison}, {Craig}, {Christensen}, {Hailey}, {Zhang}, {Boggs}, {Stern}, {Cook}, {Forster}, {Giommi}, {Grefenstette}, {Kim}, {Kitaguchi}, {Koglin}, {Madsen}, {Mao}, {Miyasaka}, {Mori}, {Perri}, {Pivovaroff}, {Puccetti}, {Rana}, {Westergaard}, {Willis}, {Zoglauer}, {An}, {Bachetti}, {Barri{\`e}re}, {Bellm}, {Bhalerao}, {Brejnholt}, {Fuerst}, {Liebe}, {Markwardt}, {Nynka}, {Vogel}, {Walton}, {Wik}, {Alexander}, {Cominsky}, {Hornschemeier}, {Hornstrup}, {Kaspi}, {Madejski}, {Matt}, {Molendi}, {Smith}, {Tomsick}, {Ajello}, {Ballantyne}, {Balokovi{\'c}}, {Barret}, {Bauer}, {Blandford}, {Brandt}, {Brenneman}, {Chiang}, {Chakrabarty}, {Chenevez}, {Comastri}, {Dufour}, {Elvis}, {Fabian}, {Farrah}, {Fryer}, {Gotthelf}, {Grindlay}, {Helfand}, {Krivonos}, {Meier}, {Miller}, {Natalucci}, {Ogle}, {Ofek}, {Ptak}, {Reynolds}, {Rigby}, {Tagliaferri}, {Thorsett}, {Treister}, \& {Urry}}]{2013ApJ...770..103H}
{Harrison}, F.~A., {Craig}, W.~W., {Christensen}, F.~E., {et~al.} 2013, \apj, 770, 103, \dodoi{10.1088/0004-637X/770/2/103}

\bibitem[{{Hern{\'a}ndez-Garc{\'\i}a} {et~al.}(2015){Hern{\'a}ndez-Garc{\'\i}a}, {Masegosa}, {Gonz{\'a}lez-Mart{\'\i}n}, \& {M{\'a}rquez}}]{2015A&A...579A..90H}
{Hern{\'a}ndez-Garc{\'\i}a}, L., {Masegosa}, J., {Gonz{\'a}lez-Mart{\'\i}n}, O., \& {M{\'a}rquez}, I. 2015, \aap, 579, A90, \dodoi{10.1051/0004-6361/201526127}

\bibitem[{{Jaffe} {et~al.}(2004){Jaffe}, {Meisenheimer}, {R{\"o}ttgering}, {Leinert}, {Richichi}, {Chesneau}, {Fraix-Burnet}, {Glazenborg-Kluttig}, {Granato}, {Graser}, {Heijligers}, {K{\"o}hler}, {Malbet}, {Miley}, {Paresce}, {Pel}, {Perrin}, {Przygodda}, {Schoeller}, {Sol}, {Waters}, {Weigelt}, {Woillez}, \& {de Zeeuw}}]{2004Natur.429...47J}
{Jaffe}, W., {Meisenheimer}, K., {R{\"o}ttgering}, H.~J.~A., {et~al.} 2004, \nat, 429, 47, \dodoi{10.1038/nature02531}

\bibitem[{{Jana} {et~al.}(2023){Jana}, {Chatterjee}, {Chang}, {Nandi}, {Rubinur}, {Kumari}, {Naik}, {Safi-Harb}, \& {Ricci}}]{2023MNRAS.524.4670J}
{Jana}, A., {Chatterjee}, A., {Chang}, H.-K., {et~al.} 2023, \mnras, 524, 4670, \dodoi{10.1093/mnras/stad2140}

\bibitem[{{Kammoun} {et~al.}(2020){Kammoun}, {Miller}, {Koss}, {Oh}, {Zoghbi}, {Mushotzky}, {Barret}, {Behar}, {Brandt}, {Brenneman}, {Kaastra}, {Lohfink}, {Proga}, \& {Stern}}]{2020ApJ...901..161K}
{Kammoun}, E.~S., {Miller}, J.~M., {Koss}, M., {et~al.} 2020, \apj, 901, 161, \dodoi{10.3847/1538-4357/abb29f}

\bibitem[{{Koss} {et~al.}(2017){Koss}, {Trakhtenbrot}, {Ricci}, {Lamperti}, {Oh}, {Berney}, {Schawinski}, {Balokovi{\'c}}, {Baronchelli}, {Crenshaw}, {Fischer}, {Gehrels}, {Harrison}, {Hashimoto}, {Hogg}, {Ichikawa}, {Masetti}, {Mushotzky}, {Sartori}, {Stern}, {Treister}, {Ueda}, {Veilleux}, \& {Winter}}]{2017ApJ...850...74K}
{Koss}, M., {Trakhtenbrot}, B., {Ricci}, C., {et~al.} 2017, \apj, 850, 74, \dodoi{10.3847/1538-4357/aa8ec9}

\bibitem[{{Koss} {et~al.}(2016){Koss}, {Assef}, {Balokovi{\'c}}, {Stern}, {Gandhi}, {Lamperti}, {Alexander}, {Ballantyne}, {Bauer}, {Berney}, {Brandt}, {Comastri}, {Gehrels}, {Harrison}, {Lansbury}, {Markwardt}, {Ricci}, {Rivers}, {Schawinski}, {Trakhtenbrot}, {Treister}, \& {Urry}}]{2016ApJ...825...85K}
{Koss}, M.~J., {Assef}, R., {Balokovi{\'c}}, M., {et~al.} 2016, \apj, 825, 85, \dodoi{10.3847/0004-637X/825/2/85}

\bibitem[{{Koss} {et~al.}(2022){Koss}, {Ricci}, {Trakhtenbrot}, {Oh}, {den Brok}, {Mej{\'\i}a-Restrepo}, {Stern}, {Privon}, {Treister}, {Powell}, {Mushotzky}, {Bauer}, {Ananna}, {Balokovi{\'c}}, {B{\"a}r}, {Becker}, {Bessiere}, {Burtscher}, {Caglar}, {Congiu}, {Evans}, {Harrison}, {Heida}, {Ichikawa}, {Kamraj}, {Lamperti}, {Pacucci}, {Ricci}, {Riffel}, {Rojas}, {Schawinski}, {Temple}, {Urry}, {Veilleux}, \& {Williams}}]{2022ApJS..261....2K}
{Koss}, M.~J., {Ricci}, C., {Trakhtenbrot}, B., {et~al.} 2022, \apjs, 261, 2, \dodoi{10.3847/1538-4365/ac6c05}

\bibitem[{{LaMassa} {et~al.}(2019){LaMassa}, {Yaqoob}, {Boorman}, {Tzanavaris}, {Levenson}, {Gandhi}, {Ptak}, \& {Heckman}}]{2019ApJ...887..173L}
{LaMassa}, S.~M., {Yaqoob}, T., {Boorman}, P.~G., {et~al.} 2019, \apj, 887, 173, \dodoi{10.3847/1538-4357/ab552c}

\bibitem[{{Leighly} {et~al.}(2014){Leighly}, {Terndrup}, {Baron}, {Lucy}, {Dietrich}, \& {Gallagher}}]{2014ApJ...788..123L}
{Leighly}, K.~M., {Terndrup}, D.~M., {Baron}, E., {et~al.} 2014, \apj, 788, 123, \dodoi{10.1088/0004-637X/788/2/123}

\bibitem[{{Marchesi} {et~al.}(2018){Marchesi}, {Ajello}, {Marcotulli}, {Comastri}, {Lanzuisi}, \& {Vignali}}]{2018ApJ...854...49M}
{Marchesi}, S., {Ajello}, M., {Marcotulli}, L., {et~al.} 2018, \apj, 854, 49, \dodoi{10.3847/1538-4357/aaa410}

\bibitem[{{Marchesi} {et~al.}(2019){Marchesi}, {Ajello}, {Zhao}, {Marcotulli}, {Balokovi{\'c}}, {Brightman}, {Comastri}, {Cusumano}, {Lanzuisi}, {La Parola}, {Segreto}, \& {Vignali}}]{2019ApJ...872....8M}
{Marchesi}, S., {Ajello}, M., {Zhao}, X., {et~al.} 2019, \apj, 872, 8, \dodoi{10.3847/1538-4357/aafbeb}

\bibitem[{{Marchesi} {et~al.}(2022){Marchesi}, {Zhao}, {Torres-Alb{\`a}}, {Ajello}, {Gaspari}, {Pizzetti}, {Buchner}, {Bertola}, {Comastri}, {Feltre}, {Gilli}, {Lanzuisi}, {Matzeu}, {Pozzi}, {Salvestrini}, {Sengupta}, {Silver}, {Tombesi}, {Traina}, {Vignali}, \& {Zappacosta}}]{2022ApJ...935..114M}
{Marchesi}, S., {Zhao}, X., {Torres-Alb{\`a}}, N., {et~al.} 2022, \apj, 935, 114, \dodoi{10.3847/1538-4357/ac80be}

\bibitem[{{Matt} {et~al.}(2006){Matt}, {Bianchi}, {de Rosa}, {Grandi}, \& {Perola}}]{2006A&A...445..451M}
{Matt}, G., {Bianchi}, S., {de Rosa}, A., {Grandi}, P., \& {Perola}, G.~C. 2006, \aap, 445, 451, \dodoi{10.1051/0004-6361:20054013}

\bibitem[{{Mej{\'\i}a-Restrepo} {et~al.}(2022){Mej{\'\i}a-Restrepo}, {Trakhtenbrot}, {Koss}, {Oh}, {den Brok}, {Stern}, {Powell}, {Ricci}, {Caglar}, {Ricci}, {Bauer}, {Treister}, {Harrison}, {Urry}, {Ananna}, {Asmus}, {Assef}, {B{\"a}r}, {Bessiere}, {Burtscher}, {Ichikawa}, {Kakkad}, {Kamraj}, {Mushotzky}, {Privon}, {Rojas}, {Sani}, {Schawinski}, \& {Veilleux}}]{2022ApJS..261....5M}
{Mej{\'\i}a-Restrepo}, J.~E., {Trakhtenbrot}, B., {Koss}, M.~J., {et~al.} 2022, \apjs, 261, 5, \dodoi{10.3847/1538-4365/ac6602}

\bibitem[{{Murphy} \& {Yaqoob}(2009)}]{2009MNRAS.397.1549M}
{Murphy}, K.~D., \& {Yaqoob}, T. 2009, \mnras, 397, 1549, \dodoi{10.1111/j.1365-2966.2009.15025.x}

\bibitem[{{Nasa High Energy Astrophysics Science Archive Research Center (Heasarc)}(2014)}]{2014ascl.soft08004N}
{Nasa High Energy Astrophysics Science Archive Research Center (Heasarc)}. 2014, {HEAsoft: Unified Release of FTOOLS and XANADU}, Astrophysics Source Code Library, record ascl:1408.004.
\newblock \doeprint{1408.004}

\bibitem[{{Nenkova} {et~al.}(2008){Nenkova}, {Sirocky}, {Nikutta}, {Ivezi{\'c}}, \& {Elitzur}}]{2008ApJ...685..160N}
{Nenkova}, M., {Sirocky}, M.~M., {Nikutta}, R., {Ivezi{\'c}}, {\v{Z}}., \& {Elitzur}, M. 2008, \apj, 685, 160, \dodoi{10.1086/590483}

\bibitem[{{Oda} {et~al.}(2017){Oda}, {Tanimoto}, {Ueda}, {Imanishi}, {Terashima}, \& {Ricci}}]{2017ApJ...835..179O}
{Oda}, S., {Tanimoto}, A., {Ueda}, Y., {et~al.} 2017, \apj, 835, 179, \dodoi{10.3847/1538-4357/835/2/179}

\bibitem[{{Osorio-Clavijo} {et~al.}(2022){Osorio-Clavijo}, {Gonz{\'a}lez-Mart{\'\i}n}, {S{\'a}nchez}, {Esparza-Arredondo}, {Masegosa}, {Victoria-Ceballos}, {Hern{\'a}ndez-Garc{\'\i}a}, \& {D{\'\i}az}}]{2022MNRAS.510.5102O}
{Osorio-Clavijo}, N., {Gonz{\'a}lez-Mart{\'\i}n}, O., {S{\'a}nchez}, S.~F., {et~al.} 2022, \mnras, 510, 5102, \dodoi{10.1093/mnras/stab3752}

\bibitem[{{Panagiotou} {et~al.}(2021){Panagiotou}, {Walter}, \& {Paltani}}]{2021A&A...653A.162P}
{Panagiotou}, C., {Walter}, R., \& {Paltani}, S. 2021, \aap, 653, A162, \dodoi{10.1051/0004-6361/202140379}

\bibitem[{{Peterson} {et~al.}(2004){Peterson}, {Ferrarese}, {Gilbert}, {Kaspi}, {Malkan}, {Maoz}, {Merritt}, {Netzer}, {Onken}, {Pogge}, {Vestergaard}, \& {Wandel}}]{2004ApJ...613..682P}
{Peterson}, B.~M., {Ferrarese}, L., {Gilbert}, K.~M., {et~al.} 2004, \apj, 613, 682, \dodoi{10.1086/423269}

\bibitem[{{Pfeifle} {et~al.}(2022){Pfeifle}, {Ricci}, {Boorman}, {Stalevski}, {Asmus}, {Trakhtenbrot}, {Koss}, {Stern}, {Ricci}, {Satyapal}, {Ichikawa}, {Rosario}, {Caglar}, {Treister}, {Powell}, {Oh}, {Urry}, \& {Harrison}}]{2022ApJS..261....3P}
{Pfeifle}, R.~W., {Ricci}, C., {Boorman}, P.~G., {et~al.} 2022, \apjs, 261, 3, \dodoi{10.3847/1538-4365/ac5b65}

\bibitem[{{Pizzetti} {et~al.}(2025){Pizzetti}, {Torres-Alb{\`a}}, {Marchesi}, {Buchner}, {Cox}, {Zhao}, {Neal}, {Sengupta}, {Silver}, \& {Ajello}}]{2025ApJ...979..170P}
{Pizzetti}, A., {Torres-Alb{\`a}}, N., {Marchesi}, S., {et~al.} 2025, \apj, 979, 170, \dodoi{10.3847/1538-4357/ad9c64}

\bibitem[{{Reeves} {et~al.}(2018){Reeves}, {Lobban}, \& {Pounds}}]{2018ApJ...854...28R}
{Reeves}, J.~N., {Lobban}, A., \& {Pounds}, K.~A. 2018, \apj, 854, 28, \dodoi{10.3847/1538-4357/aaa776}

\bibitem[{{Ricci} {et~al.}(2010){Ricci}, {Beckmann}, {Audard}, \& {Courvoisier}}]{2010A&A...518A..47R}
{Ricci}, C., {Beckmann}, V., {Audard}, M., \& {Courvoisier}, T.~J.~L. 2010, \aap, 518, A47, \dodoi{10.1051/0004-6361/200912509}

\bibitem[{{Ricci} {et~al.}(2015){Ricci}, {Ueda}, {Koss}, {Trakhtenbrot}, {Bauer}, \& {Gandhi}}]{2015ApJ...815L..13R}
{Ricci}, C., {Ueda}, Y., {Koss}, M.~J., {et~al.} 2015, \apjl, 815, L13, \dodoi{10.1088/2041-8205/815/1/L13}

\bibitem[{{Ricci} {et~al.}(2017){Ricci}, {Trakhtenbrot}, {Koss}, {Ueda}, {Delvecchio}, {Treister}, {Schawinski}, {Paltani}, {Oh}, {Lamperti}, {Berney}, {Gandhi}, {Ichikawa}, {Bauer}, {Ho}, {Asmus}, {Beckmann}, {Soldi}, {Balokovi{\'c}}, {Gehrels}, \& {Markwardt}}]{2017ApJS..233...17R}
{Ricci}, C., {Trakhtenbrot}, B., {Koss}, M.~J., {et~al.} 2017, \apjs, 233, 17, \dodoi{10.3847/1538-4365/aa96ad}

\bibitem[{{Ricci} {et~al.}(2021){Ricci}, {Privon}, {Pfeifle}, {Armus}, {Iwasawa}, {Torres-Alb{\`a}}, {Satyapal}, {Bauer}, {Treister}, {Ho}, {Aalto}, {Ar{\'e}valo}, {Barcos-Mu{\~n}oz}, {Charmandaris}, {Diaz-Santos}, {Evans}, {Gao}, {Inami}, {Koss}, {Lansbury}, {Linden}, {Medling}, {Sanders}, {Song}, {Stern}, {U}, {Ueda}, \& {Yamada}}]{2021MNRAS.506.5935R}
{Ricci}, C., {Privon}, G.~C., {Pfeifle}, R.~W., {et~al.} 2021, \mnras, 506, 5935, \dodoi{10.1093/mnras/stab2052}

\bibitem[{{Risaliti} {et~al.}(2007){Risaliti}, {Elvis}, {Fabbiano}, {Baldi}, {Zezas}, \& {Salvati}}]{2007ApJ...659L.111R}
{Risaliti}, G., {Elvis}, M., {Fabbiano}, G., {et~al.} 2007, \apjl, 659, L111, \dodoi{10.1086/517884}

\bibitem[{{Saha} {et~al.}(2022){Saha}, {Markowitz}, \& {Buchner}}]{2022MNRAS.509.5485S}
{Saha}, T., {Markowitz}, A.~G., \& {Buchner}, J. 2022, \mnras, 509, 5485, \dodoi{10.1093/mnras/stab3250}

\bibitem[{{Sambruna} {et~al.}(1999){Sambruna}, {Eracleous}, \& {Mushotzky}}]{1999ApJ...526...60S}
{Sambruna}, R.~M., {Eracleous}, M., \& {Mushotzky}, R.~F. 1999, \apj, 526, 60, \dodoi{10.1086/307981}

\bibitem[{{Sengupta} {et~al.}(2023){Sengupta}, {Marchesi}, {Vignali}, {Torres-Alb{\`a}}, {Bertola}, {Pizzetti}, {Lanzuisi}, {Salvestrini}, {Zhao}, {Gaspari}, {Gilli}, {Comastri}, {Traina}, {Tombesi}, {Silver}, {Pozzi}, \& {Ajello}}]{2023A&A...676A.103S}
{Sengupta}, D., {Marchesi}, S., {Vignali}, C., {et~al.} 2023, \aap, 676, A103, \dodoi{10.1051/0004-6361/202245646}

\bibitem[{{Silver} {et~al.}(2023){Silver}, {Torres-Alb{\`a}}, {Zhao}, {Marchesi}, {Pizzetti}, {Cox}, \& {Ajello}}]{2023AandA...675A..65S}
{Silver}, R., {Torres-Alb{\`a}}, N., {Zhao}, X., {et~al.} 2023, \aap, 675, A65, \dodoi{10.1051/0004-6361/202345980}

\bibitem[{{Silver} {et~al.}(2022){Silver}, {Torres-Alb{\`a}}, {Zhao}, {Marchesi}, {Pizzetti}, {Cox}, {Ajello}, {Cusumano}, {La Parola}, \& {Segreto}}]{2022ApJ...940..148S}
---. 2022, \apj, 940, 148, \dodoi{10.3847/1538-4357/ac9bf8}

\bibitem[{{Str{\"u}der} {et~al.}(2001){Str{\"u}der}, {Briel}, {Dennerl}, {Hartmann}, {Kendziorra}, {Meidinger}, {Pfeffermann}, {Reppin}, {Aschenbach}, {Bornemann}, {Br{\"a}uninger}, {Burkert}, {Elender}, {Freyberg}, {Haberl}, {Hartner}, {Heuschmann}, {Hippmann}, {Kastelic}, {Kemmer}, {Kettenring}, {Kink}, {Krause}, {M{\"u}ller}, {Oppitz}, {Pietsch}, {Popp}, {Predehl}, {Read}, {Stephan}, {St{\"o}tter}, {Tr{\"u}mper}, {Holl}, {Kemmer}, {Soltau}, {St{\"o}tter}, {Weber}, {Weichert}, {von Zanthier}, {Carathanassis}, {Lutz}, {Richter}, {Solc}, {B{\"o}ttcher}, {Kuster}, {Staubert}, {Abbey}, {Holland}, {Turner}, {Balasini}, {Bignami}, {La Palombara}, {Villa}, {Buttler}, {Gianini}, {Lain{\'e}}, {Lumb}, \& {Dhez}}]{2001A&A...365L..18S}
{Str{\"u}der}, L., {Briel}, U., {Dennerl}, K., {et~al.} 2001, \aap, 365, L18, \dodoi{10.1051/0004-6361:20000066}

\bibitem[{{Tanimoto} {et~al.}(2019){Tanimoto}, {Ueda}, {Odaka}, {Kawaguchi}, {Fukazawa}, \& {Kawamuro}}]{2019ApJ...877...95T}
{Tanimoto}, A., {Ueda}, Y., {Odaka}, H., {et~al.} 2019, \apj, 877, 95, \dodoi{10.3847/1538-4357/ab1b20}

\bibitem[{{Tanimoto} {et~al.}(2020){Tanimoto}, {Ueda}, {Odaka}, {Ogawa}, {Yamada}, {Kawaguchi}, \& {Ichikawa}}]{2020ApJ...897....2T}
---. 2020, \apj, 897, 2, \dodoi{10.3847/1538-4357/ab96bc}

\bibitem[{{Tanimoto} {et~al.}(2022){Tanimoto}, {Ueda}, {Odaka}, {Yamada}, \& {Ricci}}]{2022ApJS..260...30T}
{Tanimoto}, A., {Ueda}, Y., {Odaka}, H., {Yamada}, S., \& {Ricci}, C. 2022, \apjs, 260, 30, \dodoi{10.3847/1538-4365/ac5f59}

\bibitem[{{Teng} {et~al.}(2014){Teng}, {Brandt}, {Harrison}, {Luo}, {Alexander}, {Bauer}, {Boggs}, {Christensen}, {Comastri}, {Craig}, {Fabian}, {Farrah}, {Fiore}, {Gandhi}, {Grefenstette}, {Hailey}, {Hickox}, {Madsen}, {Ptak}, {Rigby}, {Risaliti}, {Saez}, {Stern}, {Veilleux}, {Walton}, {Wik}, \& {Zhang}}]{2014ApJ...785...19T}
{Teng}, S.~H., {Brandt}, W.~N., {Harrison}, F.~A., {et~al.} 2014, \apj, 785, 19, \dodoi{10.1088/0004-637X/785/1/19}

\bibitem[{{Tombesi} {et~al.}(2015){Tombesi}, {Mel{\'e}ndez}, {Veilleux}, {Reeves}, {Gonz{\'a}lez-Alfonso}, \& {Reynolds}}]{2015Natur.519..436T}
{Tombesi}, F., {Mel{\'e}ndez}, M., {Veilleux}, S., {et~al.} 2015, \nat, 519, 436, \dodoi{10.1038/nature14261}

\bibitem[{{Torres-Alb{\`a}} {et~al.}(2023){Torres-Alb{\`a}}, {Marchesi}, {Zhao}, {Cox}, {Pizzetti}, {Sengupta}, {Ajello}, \& {Silver}}]{2023A&A...678A.154T}
{Torres-Alb{\`a}}, N., {Marchesi}, S., {Zhao}, X., {et~al.} 2023, \aap, 678, A154, \dodoi{10.1051/0004-6361/202345947}

\bibitem[{{Torres-Alb{\`a}} {et~al.}(2021){Torres-Alb{\`a}}, {Marchesi}, {Zhao}, {Ajello}, {Silver}, {Ananna}, {Balokovi{\'c}}, {Boorman}, {Comastri}, {Gilli}, {Lanzuisi}, {Murphy}, {Urry}, \& {Vignali}}]{2021ApJ...922..252T}
---. 2021, \apj, 922, 252, \dodoi{10.3847/1538-4357/ac1c73}

\bibitem[{{Treister} {et~al.}(2009){Treister}, {Urry}, \& {Virani}}]{2009ApJ...696..110T}
{Treister}, E., {Urry}, C.~M., \& {Virani}, S. 2009, \apj, 696, 110, \dodoi{10.1088/0004-637X/696/1/110}

\bibitem[{{Ueda} {et~al.}(2014){Ueda}, {Akiyama}, {Hasinger}, {Miyaji}, \& {Watson}}]{2014ApJ...786..104U}
{Ueda}, Y., {Akiyama}, M., {Hasinger}, G., {Miyaji}, T., \& {Watson}, M.~G. 2014, \apj, 786, 104, \dodoi{10.1088/0004-637X/786/2/104}

\bibitem[{{Verner} {et~al.}(1996){Verner}, {Ferland}, {Korista}, \& {Yakovlev}}]{1996ApJ...465..487V}
{Verner}, D.~A., {Ferland}, G.~J., {Korista}, K.~T., \& {Yakovlev}, D.~G. 1996, \apj, 465, 487, \dodoi{10.1086/177435}

\bibitem[{{Weisskopf} {et~al.}(2000){Weisskopf}, {Tananbaum}, {Van Speybroeck}, \& {O'Dell}}]{2000SPIE.4012....2W}
{Weisskopf}, M.~C., {Tananbaum}, H.~D., {Van Speybroeck}, L.~P., \& {O'Dell}, S.~L. 2000, in Society of Photo-Optical Instrumentation Engineers (SPIE) Conference Series, Vol. 4012, X-Ray Optics, Instruments, and Missions III, ed. J.~E. {Truemper} \& B.~{Aschenbach}, 2--16, \dodoi{10.1117/12.391545}

\bibitem[{{Williams} {et~al.}(2022){Williams}, {Pahari}, {Baldi}, {McHardy}, {Mathur}, {Beswick}, {Beri}, {Boorman}, {Aalto}, {Alberdi}, {Argo}, {Dullo}, {Fenech}, {Green}, {Knapen}, {Mart{\'\i}-Vidal}, {Moldon}, {Mundell}, {Muxlow}, {Panessa}, {P{\'e}rez-Torres}, {Saikia}, {Shankar}, {Stevens}, \& {Uttley}}]{2022MNRAS.510.4909W}
{Williams}, D.~R.~A., {Pahari}, M., {Baldi}, R.~D., {et~al.} 2022, \mnras, 510, 4909, \dodoi{10.1093/mnras/stab3310}

\bibitem[{{Willingale} {et~al.}(2013){Willingale}, {Starling}, {Beardmore}, {Tanvir}, \& {O'Brien}}]{2013MNRAS.431..394W}
{Willingale}, R., {Starling}, R.~L.~C., {Beardmore}, A.~P., {Tanvir}, N.~R., \& {O'Brien}, P.~T. 2013, \mnras, 431, 394, \dodoi{10.1093/mnras/stt175}

\bibitem[{{Wilms} {et~al.}(2000){Wilms}, {Allen}, \& {McCray}}]{2000ApJ...542..914W}
{Wilms}, J., {Allen}, A., \& {McCray}, R. 2000, \apj, 542, 914, \dodoi{10.1086/317016}

\bibitem[{{Yamada} {et~al.}(2020){Yamada}, {Ueda}, {Tanimoto}, {Oda}, {Imanishi}, {Toba}, \& {Ricci}}]{2020ApJ...897..107Y}
{Yamada}, S., {Ueda}, Y., {Tanimoto}, A., {et~al.} 2020, \apj, 897, 107, \dodoi{10.3847/1538-4357/ab94b1}

\bibitem[{{Yamada} {et~al.}(2024){Yamada}, {Kawamuro}, {Mizumoto}, {Ricci}, {Ogawa}, {Noda}, {Ueda}, {Enoto}, {Kokubo}, {Minezaki}, {Sameshima}, {Horiuchi}, \& {Mizukoshi}}]{2024ApJS..274....8Y}
{Yamada}, S., {Kawamuro}, T., {Mizumoto}, M., {et~al.} 2024, \apjs, 274, 8, \dodoi{10.3847/1538-4365/ad5961}

\bibitem[{{Yaqoob}(2012)}]{2012MNRAS.423.3360Y}
{Yaqoob}, T. 2012, \mnras, 423, 3360, \dodoi{10.1111/j.1365-2966.2012.21129.x}

\bibitem[{{Yaqoob} {et~al.}(2015){Yaqoob}, {Tatum}, {Scholtes}, {Gottlieb}, \& {Turner}}]{2015MNRAS.454..973Y}
{Yaqoob}, T., {Tatum}, M.~M., {Scholtes}, A., {Gottlieb}, A., \& {Turner}, T.~J. 2015, \mnras, 454, 973, \dodoi{10.1093/mnras/stv2021}

\bibitem[{{Zhao} {et~al.}(2020){Zhao}, {Marchesi}, {Ajello}, {Balokovi{\'c}}, \& {Fischer}}]{2020ApJ...894...71Z}
{Zhao}, X., {Marchesi}, S., {Ajello}, M., {Balokovi{\'c}}, M., \& {Fischer}, T. 2020, \apj, 894, 71, \dodoi{10.3847/1538-4357/ab879d}

\bibitem[{{Zhao} {et~al.}(2021){Zhao}, {Marchesi}, {Ajello}, {Cole}, {Hu}, {Silver}, \& {Torres-Alb{\`a}}}]{2021A&A...650A..57Z}
{Zhao}, X., {Marchesi}, S., {Ajello}, M., {et~al.} 2021, \aap, 650, A57, \dodoi{10.1051/0004-6361/202140297}

\end{thebibliography}
\bibliographystyle{aasjournal}



\end{document}